\RequirePackage{fix-cm}
\documentclass[twocolumn]{svjour3}          
\smartqed  
\usepackage{graphicx}
\usepackage{flushend}
\usepackage[a-1b]{pdfx}   
\usepackage{balance}

\usepackage{color}
\usepackage[labelfont=bf]{caption}
\usepackage{subcaption}
\usepackage{hyperref}
\usepackage{fixltx2e}
\usepackage{makecell}
\usepackage{etoolbox,xspace}
\usepackage{algorithmicx}
\usepackage{algorithm}
\usepackage{algpseudocode}
\algtext*{EndWhile}
\algtext*{EndIf}
\algtext*{EndFunction}
\algtext*{EndFor}
\algrenewcommand\algorithmicrequire{\textbf{Input:}}
\algrenewcommand\algorithmicensure{\textbf{Output:}}

\usepackage{amsthm}

\usepackage{tabularx}
\usepackage{ragged2e}  
\usepackage{booktabs}  
\usepackage{textcomp}
\usepackage{url}
\usepackage{comment}
\usepackage{enumitem}
\usepackage{amsmath,amsfonts}
\usepackage{graphicx}
\usepackage{textcomp}
\usepackage{xcolor}
\newcolumntype{Y}{>{\RaggedRight\arraybackslash}X}

\newcommand{\eat}[1]{}
\newcommand{\submit}[1]{}
\newcommand{\techrep}[1]{#1}

\newcommand{\squishlist}{
 \begin{list}{$\bullet$}
  { \setlength{\itemsep}{0pt}
     \setlength{\parsep}{1pt}
     \setlength{\topsep}{1pt}
     \setlength{\partopsep}{0pt}
     \setlength{\leftmargin}{1em}
     \setlength{\labelwidth}{1em}
     \setlength{\labelsep}{0.5em} } }

\newcommand{\squishend}{
  \end{list}
}

\definecolor{americanrose}{rgb}{1.0, 0.01, 0.24}
\definecolor{airforceblue}{rgb}{0.36, 0.54, 0.66}
\definecolor{ao(english)}{rgb}{0.0, 0.5, 0.0}
\definecolor{ao}{rgb}{0.0, 0.0, 1.0}

\newcommand{\rev}[1]{\textcolor{black}{#1}}

\usepackage{xspace}
\newcommand{\stitle}[1]{\vspace{2mm}\noindent{\bf #1:}}
\newcommand{\ru}{{\sc RU}\xspace}
\newcommand{\sru}{{\sc strongRU}\xspace}
\newcommand{\wru}{{\sc weakRU}\xspace}

\newcommand{\dee}{\mathcal{D}}
\newcommand{\dist}{\xi}
\newcommand{\alg}{\mathsf{A}}
\newcommand{\qu}{\mathbf{q}}
\newcommand{\ex}{\mathbf{x}}
\newcommand{\ti}{\mathbf{t}}
\newcommand{\sdt}{\mathsf{SRU}}
\newcommand{\wdt}{\mathsf{WRU}}

\newcommand{\pe}{\mathbb{P}}

\newcommand{\eps}{\varepsilon}

\newcommand{\at}[1]{{\tt \small #1}\xspace}

\usepackage{bbm, dsfont}
\begin{document}

\title{Reliability Evaluation of Individual Predictions: A Data-centric Approach\thanks{This research was support by NSF 2107290. The authors would like to thank the reviewers and the meta-reviewer for their constructive feedback. We would also like to thank Dr. Saravanan Thirumuruganathan and Prof. H. V. Jagadish for their valuable insights.}
}


\author{Nima Shahbazi         \and
        Abolfazl Asudeh 
}


\institute{N. Shahbazi \at
              University of Illinois Chicago \\
              \email{nshahb3@uic.edu}           
           \and
           A. Asudeh \at
              University of Illinois Chicago \\
              \email{asudeh@uic.edu}
}

\date{Received: date / Accepted: date}

\maketitle

\begin{abstract}
Machine learning models only provide probabilistic guarantees on the expected loss of random samples from the distribution represented by their training data. As a result, a model with high accuracy, may or may not be reliable for predicting an individual query point.
To address this issue, XAI aims to provide explanations of individual predictions, while approaches such as conformal predictions, probabilistic predictions, and prediction intervals count on the model's certainty in its prediction to identify unreliable cases.

Conversely, instead of relying on the model itself, we look for insights in the training data. That is, following the fact a model's performance is limited to the data it has been trained on, we ask ``{\em is a model trained on a given data set, fit for making a specific prediction?}''. Specifically, we argue that a model's prediction is not reliable if (i) there were not enough similar instances in the training set to the query point, and (ii) if there is a high fluctuation (uncertainty) in the vicinity of the query point in the training set.

Using these two observations, we propose data-centric reliability measures for individual predictions and develop novel algorithms for efficient and effective computation of the reliability measures during inference time. 
The proposed algorithms learn the necessary components of the measures from the data itself and are sublinear, which makes them scalable to very large and multi-dimensional settings.
Furthermore, an estimator is designed to enable no-data access during the inference time.
We conduct extensive experiments using multiple real and synthetic data sets and different tasks, which reflect a consistent correlation between distrust values and model performance.
\end{abstract}

\vspace{-2em}
\section{Introduction}\label{sec:intro}
\vspace{-1em}
\stitle{Motivation} 
Interpretability is a necessity for data scientists who develop predictive models for critical decision-making.
In such settings, it is important to provide additional means to support the following question:
{\em is an individual prediction of the model reliable for decision-making?}
To further motivate this, let us use Example~\ref{ex-0}:

\begin{example}\label{ex-0}
Using data-driven predictive models is prevalent for loan approval~\cite{sheikh2020approach}.
Consider a data science company that has developed a predictive model tailored for bankers, aiming to predict the chance of repayment by a prospective loan applicant. 
Indeed, such models can be beneficial to help financial institutes make wise loan application decisions.
Suppose the model predicts the queried individual has a poor chance of repaying the loan in case of approval.
The data science company and the bankers are aware and concerned about the critiques surrounding such models~\cite{butler2020racial,blanchard2008lenders,montoya2020bad}.
In particular, a major question the banker faces is whether or not they should rely on the prediction outcome to take action for this case.
Therefore, the data science company would like to provide additional means alongside the model itself to help with the reliability question regarding individual predictions i.e. although the model demonstrates to be accurate on average, is it reliable for this individual prediction as well?
Furthermore, 
for the cases in which the banker finds the model prediction unreliable, what evidence could be provided for them?
\end{example} 

\vspace{-3mm}
\stitle{Novelty}
There has been extensive research on trustworthy AI~\cite{wing2021trustworthy,kentour2021analysis,liu2021trustworthy,singh2021trustworthy,moskovitch2022reliability} to address the above issues. For example, explainable AI~\cite{harradon2018causal,ribeiro2016should,gunning2019darpa} provides simple explanations and rules that justify the model's prediction. On the other hand, approaches such as probabilistic predictions~\cite{zadrozny2001obtaining,zadrozny2002transforming,platt1999probabilistic,niculescu2005predicting}, conformal predictions~\cite{angelopoulos2021gentle,shafer2008tutorial}, and prediction intervals~\cite{khosravi2010lower,pearce2018high,chatfield93predictionintervals} utilize the model's confidence in its prediction to identify instances deemed less reliable.

In contrast, rather than relying on the model itself, we seek insights within the training data used for drawing the prediction.
Acknowledging that a model's accuracy is limited to the data on which it was trained, we pose the question: ``{\em Is a model trained on a specific dataset suitable for making a specific prediction?}''

\begin{figure}[!t]
    \centering
    \includegraphics[width=.5\textwidth]{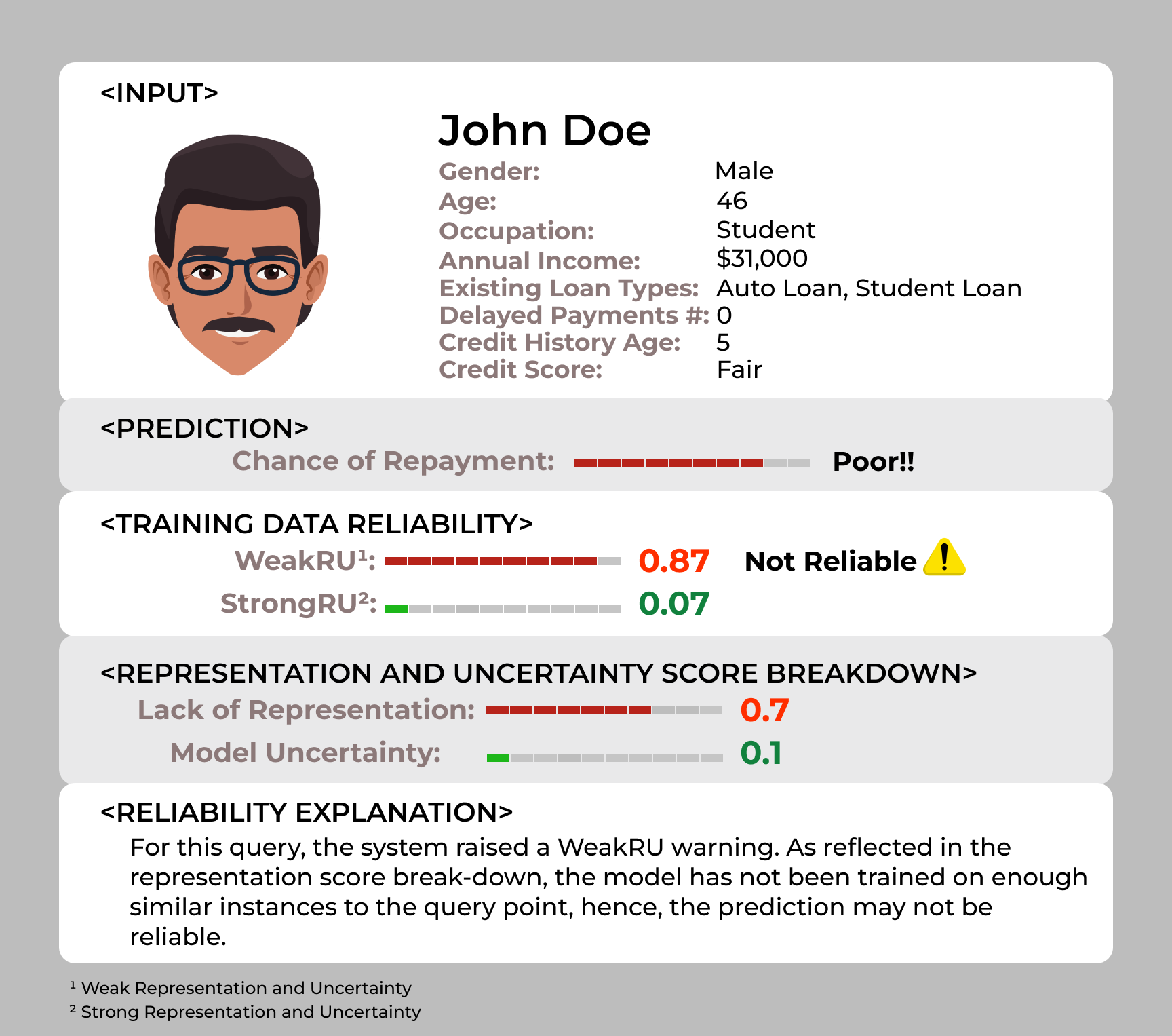}
    \caption{Illustration of a mock interface for individual prediction reliability evaluation -- Example~\ref{ex-0} (Part 2).}
    \label{fig:high_distrust_mock_interface}
    \vspace{-5mm}
\end{figure}

\stitle{Technical Highlights}
We argue that, irrespective of the choice of model and its details, {\em the prediction for a query point is not reliable if the model is not trained on instances similar to the query point or if there is a high variance in the vicinity of the query point in the training set.}
Therefore, we introduce two data-centric reliability measures based on the Representation and Uncertainty ({\bf \ru}) around the query point, called {\bf \sru} and {\wru}.
The \ru measures are defined based on two components: 
\begin{itemize}[leftmargin=*]
    \item {\em Representativeness:} Predictive models provide only probabilistic guarantees on the \underline{average} loss over the distribution represented by the data set used for training them. As a result, their predictions are not distribution generalizable~\cite{kulynych2022you}.
    Consequently, if the query point is {\em not represented} by the data, the guarantees may not hold, hence one cannot rely on the prediction outcome.
    \item {\em Uncertainty:} When the query point belongs to an {\em uncertain} neighborhood with high variance on the target values, the model prediction may not be reliable.
\end{itemize}
\wru is a warning that is raised if the individual prediction is problematic at least based on one of the two components.
That is, if the query point belongs to an uncertain region or it is not represented by the data.
\sru, on the other hand, is a conservative but strong warning that is only raised when the query point fails based on both of the two components. That is, it both belongs to uncertain regions and is also not well-represented in the data.
While \wru is a weaker yellow flag (e.g., Figure~\ref{fig:high_distrust_mock_interface}) warning that can be ignored for less critical decisions, \sru is a red flag, which if raised, the model outcome should be ignored or at least considered with extra caution.
\color{black}

\noindent{\bf Example \ref{ex-0} (part 2):}
\ru measures {raise warning} when the fitness of the data set used for drawing a prediction is questionable, helping the banker to be cautious when taking action. 
Besides, these measures provide {quantitative justification} to support the banker's action when they decide to ignore a prediction outcome that is not trustworthy.
Suppose the \wru for the query point is low.
Besides the score, our system specifies that, for example, lack of representation is the issue, reflected by the low representation score.
The banker can then argue to ignore a model outcome for this case, justifying that the model has been built using a data set that fails to represent the given case.

To see a concrete example, let us consider the mock interface in Figure~\ref{fig:high_distrust_mock_interface}, showing an individual prediction with ``poor chance or repayment''. However, the \at{$\langle$training data reliability$\rangle$} section raises a warning signal.
Next, the \at{$\langle$RU score breakdown$\rangle$} reveals that the model has not seen sufficient samples similar to this individual. That is, the prediction suffers from a lack of representation.
Based on this analysis, the banker can reject the model prediction and use the \ru values along with the \at{$\langle$reliability explanation$\rangle$} section to justify their action.
\hfill$\square$ \vspace{1mm}

While being agnostic to the choice of the uncertainty and lack of representation components, we propose an implementation based on the $k$-vicinity of a query point. In particular, given the radius of the $k$-vicinity and its uncertainty, we develop functions that return probabilities indicating the lack of representation and uncertainty. We propose methods to {\em learn} the probabilities {\em from the data set itself}.
We devise proper indexing and algorithms that enable sublinear query processing that {\em scales} to large data sets.

\stitle{Positioning in the context of existing work}
Our work differs from existing literature including model-centric uncertainty quantification, local interpretation techniques, and data coverage in several radical ways:
\begin{itemize}[leftmargin=*]
    \item We offer a {\em data}-centric measure, a {quantitative reliability warning} that measures whether a query point is in the {scope of use of a data}. 
    Unlike model-centric uncertainty quantification techniques, our techniques reveal a property of the data set that regardless of the constructed model this property stands still.
    \item Although model-centric techniques such as~\cite{khosravi2010lower,pearce2018high,chatfield93predictionintervals} guarantee a user-specified assurance level of error, this error is still computed over the entire data and consequently {\em may fail to focus the error on local regions} in data, representing, for example, minority populations in social applications. 
    On the other hand, the local fidelity of our techniques satisfies equal treatment for every query point.
    \item While some model-centric uncertainty quantification techniques \cite{khosravi2010lower,pearce2018high} claim considering lack of representation as a source of uncertainty, as we observed in our experiments, they fail to capture the associated uncertainty for such query points in sparse regions.
    This failure originates from the perfect sampling assumption in development and production data which may not hold in practice. Our measures however properly capture such cases and directly target the lack of representation of a query point.  
    \item The literature on data coverage~\cite{asudeh2019assessing,asudeh2021identifying,lin2020identifying} only focuses on representation, and hence fails to capture uncertainty. Additionally, they only return a binary signal of whether to trust the outcome of the model for a query point or not which practically is not very informative. Whereas our proposed measures target both sources of uncertainty and representation and return a quantitative value that is easily interpretable.
    \item Unlike techniques in interpretable machine learning~\cite{molnar2020interpretable} {\em justify} (advocate) individual predictions, our technique questions those that it finds unreliable.
\end{itemize}

\vspace{-3mm}
\stitle{Summary of contributions}
In summary, our contributions in this paper include the following:
\begin{enumerate}[leftmargin=*]
    \item We propose data set \ru measures to raise warnings when the fitness of a data set for an individual prediction is questionable. To the best of our knowledge, we are the first to propose data-centric \ru measures, a property associated with data sets.
    \item Our proposal is a quantitative measure based on two components: the query's lack of representation and uncertainty in the data set.
    The proposed measures can be extended to different data types and are independent of the model and prediction task (classification and regression). The measures are also agnostic to the choice of metric or approach for computing the two components.
    Proposing quantitative probabilistic outcomes, our measures are interpretable for the users since beyond the scores, the uncertainty and lack of representation components provide an explanation to justify them.
    \item We propose novel algorithms based on the $k$-vicinity of a query point to compute the query's lack of representation and uncertainty.
    In particular, we ``learn'' the measurements from the data set itself. We also propose proper preprocessing and algorithms that enable sub-linear query answering that scales to very large and high-dimensional data sets.
    Furthermore, to enable no-data access during the query time, we build regression models to accurately estimate parameters needed to compute \ru measures. We design an exponential search strategy for constructing large enough samples for training the estimators.
    \item We conduct comprehensive experiments on multiple synthetic and real-world data sets with various scales and dimensions, on different prediction tasks {(regression and binary/multi-class classification including text classification and image processing)}, using several models (such as Logistic Regression, DNN, Random Forest, etc.), and distance measures to (i) {\em validate} the effectiveness and consistency of the \ru measures, (ii) evaluate the {\em efficiency and scalability} of our algorithms and (iii) evaluate the {\em existing works}.
\end{enumerate}

Our extensive proof-of-concept experiments verify a consistent correlation between \ru values and ML performance metrics on a variety of tasks, data sets, and ML algorithms. For tuples that have higher \ru values (meaning they are less reliable w.r.t. to our measures), an ML model is more likely to fail to capture the truth and make a correct decision. 

\noindent{\bf How to use?}  As demonstrated in our experiments, when \ru values for a query point are high, one should \underline{discard} or at least \underline{not rely on} the individual prediction for critical decisions.
We would like to reiterate that our proposal in this paper is complementary to the existing literature and {\em should be used \underline{alongside other} \underline{techniques} and potential approaches for trustworthy AI}.
\section{Preliminaries}\label{sec:pre}
\subsection{Data Model}\label{sec:pre:data}
Consider a data set $\dee$ with $n$ tuples, each consisting of $d$ (observation) attributes $\ex=\langle x_1, x_2, \cdots,x_d\rangle$
and a target attribute $y$, also known as label attribute\footnote{The measures and the algorithms proposed in this paper extend for data sets with multiple target attributes.
In such cases, each measure is defined per each target attribute.}.
The observation attributes $\ex$ are called the {\em input space} and the target attribute is called the {\em output space}.
We assume the data set is used for training a prediction model $h$, as we shall further explain in \S~\ref{sec:pre:model}.
Prediction models assume that $\dee$ is a set of iid (independent and identically distributed random) samples\footnote{We would like to note that our proposal does not make the iid assumption, which can be violated in practice, especially in the presence of issues such as sampling bias.
We raise a warning when the data set is not fit to draw a specific prediction. As a result, in such cases, the warnings will be raised more frequently for the query points that are not represented by data.}, drawn from an (unknown) underlying distribution $\dist$.
Attribute values may be discrete ordinal, continuous-valued, or non-ordinal categorical.
Throughout the paper, we assume ordinal attributes are normalized in the range $[0,1]$, with values drawn from the set of rational or real numbers.
For non-ordinal attributes, we assume one-hot encoded representations.
We use $\ti^j$ to refer to the 
$j$-th tuple in the data set $\dee$ and its values of the observation attributes in particular. Similarly, we use $y^j$ to refer to the value of the target attribute of $\ti^j$.
For every tuple $\ti\in\dee$, we use the notation $t_i$ to show the value of $\ti$ on attribute $x_i\in \ex$.
\vspace{-5mm}
\subsection{Query and Prediction Model}\label{sec:pre:model}
The goal of prediction is to guess the target value $y$ of a query point based on the observations on $\ex$. 
In other words, given a point $\qu=\langle q_1, q_2,\cdots,q_d\rangle$, the goal is to predict the value of the target attribute of $\qu$.
We consider the prediction model $h: \mathbbm{R}^d\rightarrow \mathbbm{R}$ as a function that predicts the target value of $\qu$ as $h(\qu)$.
When $y$ is categorical, the task is classification, while regression is considered when $y$ is continuous.

The underlying assumption is that $\qu$ is drawn from the same distribution $\dist$ from which $\dee$ has been generated.
Now, consider the Cartesian product of the input and output space $\ex \times y$, and fix the hypothesis universe $\mathcal{H}$ of prediction functions.
A learning algorithm $\alg$ takes as input the set of samples in the data set $\dee$ and finds a specific function $h = \alg(\dee)$ by minimizing the empirical risk (maximizing the empirical accuracy or minimizing empirical loss) over $\dee$.
Empirical accuracy for classification is computed as the sum of samples in $\dee$ for which the true label is the same as the predicted label:

\begin{align}
    \hspace{20mm}\max \sum\limits_{j=1}^n \mathbbm{1}\Big(y^j==h(\ti^j)\Big)
\end{align}

The equivalent objective for regression is to minimize the empirical error between the target variable and predicted values. Sum of Squares Error (SSE) is the de-facto error measure for regression:
\begin{align}
    \hspace{22mm}\min \sum\limits_{j=1}^n \Big(y^j - h(\ti^j)\Big)^2
\end{align}

Having a prediction model $h$ trained by maximizing its empirical accuracy over the sample points in $\dee$, the model is then used to predict the value of {\em unseen} target attribute of each query point $\qu$, observed {\em after} model deployment, as $h(\qu)$. 
A central question at this point is whether a decision-maker should rely on the model prediction (at least for critical decisions).
In the next section, we propose data-centric measures generated to answer this concern.

\begin{figure*}[!ht] 
    \begin{subfigure}[t]{0.32\linewidth}
        	\centering
        	\includegraphics[width=\textwidth]{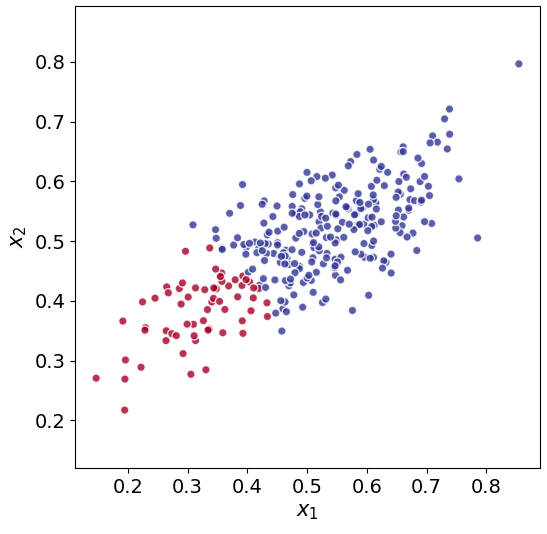} 
        \vspace{-5mm}
        	\caption{the data set $\dee$ generated using a Gaussian distribution where $x_1$ and $x_2$ are positively correlated}
            \label{fig:ex1:1}
    \end{subfigure}
    \hfill
    \begin{subfigure}[t]{0.32\linewidth}
        \centering
        	\includegraphics[width =\textwidth]{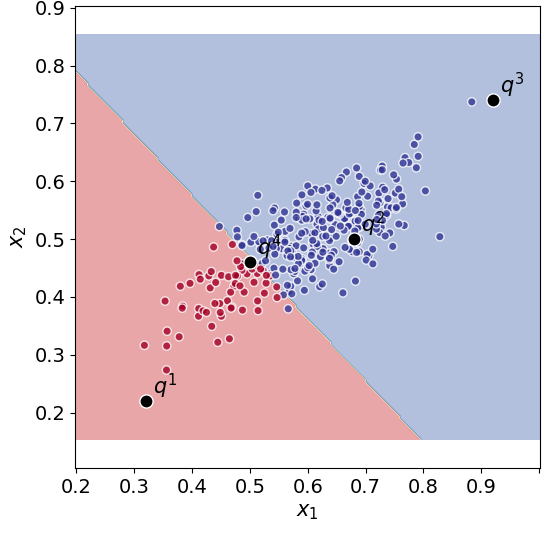} 
        \vspace{-5mm}
        	\caption{the decision boundary of learned model $h$ and query points $\qu^1$ to $\qu^4$}
            \label{fig:ex1:2}
    \end{subfigure}
    \hfill
    \begin{subfigure}[t]{0.32\linewidth}
        	\centering
        	\includegraphics[width =\textwidth]{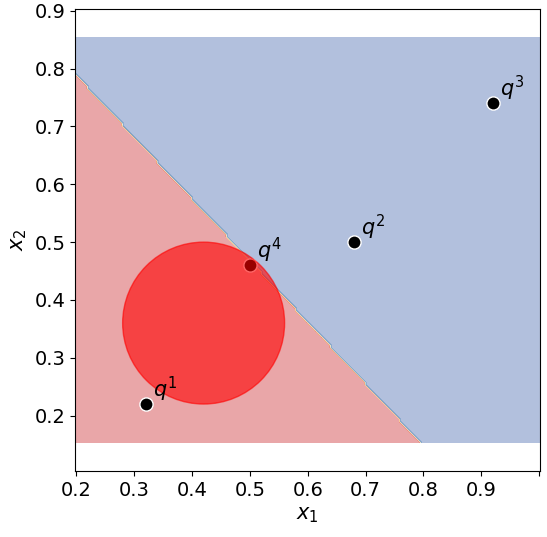}
        \vspace{-5mm}
        	\caption{ground-truth boundary, overlaid on the model decision boundary and query points}
            \label{fig:ex1:3}
    \end{subfigure}
    \submit{\vspace{-7mm}}
    \caption{a toy example (Example~\ref{ex-1}) representing a binary classification task.}
    \label{fig:ex1}
    \vspace{-5mm}
\end{figure*}   
\vspace{-1em}
\section{Reliability\&Uncertainty (\ru) Measures}\label{sec:measures}
Not every data set is fit for all data science tasks~\cite{sun2019mithralabel,asudeh2019assessing}.
An essential requirement for a learning algorithm is that its training data $\dee$ should represent the underlying distribution $\dist$.
Even if so, the trained model guarantees to perform well only {\em on average} over the query points drawn from $\dist$, not necessarily on a specific query point.
To further explain this, let us provide some background from the machine learning theory.

Let $\mathcal{L}$ be the loss function used by the learning algorithm.
Considering the underlying distribution $\dist$, the optimal model $h^*\in\mathcal{H}$ is the one with the minimum expected loss for a random sample drawn from $\dist$:
\begin{align}
    \mathcal{E}^*_\mathcal{H} = \inf_{h\in\mathcal{H}}\mathbb{E}\Big[\mathcal{L}\big(h(\ex), y\big)\Big]
    = \inf_{h\in\mathcal{H}}\int_\ex \mathcal{L}\big(h(\ex), y\big) d_\dist(\ex)
\end{align}

Let $h_n$ be the model generated with the algorithm $\alg$ over a data set $\dee$ with $n$ samples drawn from $\dist$. 
Given the values $\epsilon,\delta>0$, the {\em sample complexity}~\cite{vapnik1999overview} of $\alg$ is the minimum value of $n$ such that
\begin{align}
   \hspace{22mm} \pe_\dist\big(\mathcal{E}(h_n)-\mathcal{E}^* >\epsilon\big)\leq \delta
\end{align}

If the sample complexity of $\alg$ for given values of $\epsilon$ and $\delta$ is unbounded, the function space is {\em not learnable}.
The interesting immediate question is whether a distribution-free function is learnable.
In other words, is there a learning algorithm $\alg$ such that its sample complexity is bounded, independent of the underlying distribution?
Unfortunately, following the so-called ``no free lunch'' theorem~\cite{kakade2003sample}, the answer to the above question is negative. 

In summary,
a trained model $h$ only provides probabilistic guarantee on the {\em expected} loss on random samples from the {\em underlying distribution} $\dist$ represented by the data set $\dee$.  
While ML models guarantee to perform well on average over the query points that follow $\dist$, our objective is {\em use-case base}, i.e., on {\em a single query point} -- as opposed to the average performance of the model over a set of samples.
A model that performs well on {\em majority} of samples drawn from $\dist$ will have a high performance on average. Still, it does not necessarily mean it will perform well on the {\em minorities} and outlier points~\cite{asudeh2021identifying}.
To further observe this, we present an example that leads to the design of our measures:
\vspace{-5mm}
\subsection{A Toy Example}\label{sec:measure:toy}

As the running example in this section, let us consider the following classification task:

\begin{example}\label{ex-1}
Consider a binary classification task where the input space is $\ex=\langle x_1, x_2\rangle$ and the output space is the binary label $y$ with values $\{-1$ (red) $,+1$ (blue)$\}$.
Suppose the underlying data distribution $\dist$ follows a 2D Gaussian, where $x_1$ and $x_2$ 
are positively correlated as shown in Figure~\ref{fig:ex1:1}.
The figure shows the data set $\dee$ drawn independently from the distribution $\dist$, along with their labels as their colors.
Using $\dee$, the prediction model $h$ is constructed as shown in Figure~\ref{fig:ex1:2}. 
The decision boundary is specified in the picture; while any point above the line is predicted as +1, a query point below it is labeled as -1.
The classifier has been evaluated using a test set that is an iid sample set drawn from the underlying data set $\dist$. The accuracy on the test set is high (above 90\%), and hence, the model gets deployed for predicting the outcome of unseen query points.
We cherry-picked four query points, $\qu^1$ to $\qu^4$, that are also included in Figure~\ref{fig:ex1:2}. Using $h$ for prediction, $h(\qu^1)=-1$, $h(\qu^2)=+1$,  $h(\qu^3)=+1$, and $h(\qu^4)=-1$.
Figure~\ref{fig:ex1:3} adds the ground-truth boundary to the search space, revealing the true label of the query points: every point inside the red circle has the true label $-1$ while any point outside of it is $+1$.
Looking at the figure, $y^1=+1$ while the model predicted it as $h(\qu^1)=-1$.
\end{example}

Let us take a closer look at the four query points in this example and their placement w.r.t. the tuples in $\dee$ used for training $h$. 
$\qu^2$ belongs to a {\em dense region} with many training tuples in $\dee$ surrounding it. Besides, all of the tuples in its vicinity have the same label $y=+1$. As a result, one can expect that the model's outcome $h(\qu^2)=+1$ should be a reliable prediction.
Similar to $\qu^2$, $\qu^4$ also belongs to a dense region in $\dee$; however, $\qu^4$ belongs to an {\em uncertain region}, where some of the tuples in its vicinity have a label $y=+1$, and some others have the label $y=-1$. Considering the uncertainty in the vicinity of $\qu^4$, one cannot confidently rely on the outcome of the model $h$. 
On the other hand, the neighbors of $\qu^1$ (resp. $\qu^3$) are not uncertain, all having the label $y=-1$ (resp. $y=+1$).
However, the query points $\qu^1$ and $\qu^3$ are not well represented by $\dee$, as those would be {\em outlier} w.r.t. $\dee$. In other words, $\qu^1$ and $\qu^3$ are unlikely to be generated according to the underlying distribution $\dist$, represented by $\dee$. As a result, following the no-free-lunch theorem, one cannot expect the outcome of model $h$ to be reliable for these points.
\rev{Note that, as we observed in our experiments, model-centric techniques such as prediction intervals and conformal prediction fail to detect $\qu^1$ and $\qu^3$ as not trustworthy.}

Looking at the ground-truth boundaries in Figure~\ref{fig:ex1:3}, $h$ luckily predicted the outcome for $\qu^3$ correctly, but it was not fortunate to predict the $y^1$ correctly.
Nevertheless, 
since the model has not reliably been trained for these outlier points, 
its outcome {\em may or may not} be accurate for these query points, hence is not trustworthy.

\subsection{Strong and Weak \ru Measures}\label{sec:measure:main}
From Example~\ref{ex-1}, we observe that the outcome of a model $h$, trained using a data set $\dee$ is not reliable for a query point $\qu$, if:
\begin{itemize}[leftmargin=*]
    \item {\em Lack of representation:} $\qu$ is not well-presented by $\dee$. In other words, $\qu$ is an outlier w.r.t. the tuples in $\dee$. In such cases, the model has not seen ``enough'' samples similar to $\qu$ to reliably learn and predict the outcome of $\qu$.
    \item {\em Lack of certainty:} $\qu$ belongs to an uncertain region, where different tuples of $\dee$ in the vicinity of $\qu$ have different target values. In a classification context, that means the tuples have different labels (similar to $\qu^4$ in Example~\ref{ex-1}). Similarly, in a regression setting,
    $\qu$ belongs to a high-fluctuating area, where tuples in the vicinity of $\qu$ have a wide range of values on the target variable.
\end{itemize}

We design the data-centric \ru measures based on these two observations.
In order to identify if a query suffers from uncertainty or lack of representation, one could use a deterministic approach using a fixed threshold. Then if the number of similar samples to (resp. label fluctuation in vicinity of) $\qu$ is larger than the threshold it is considered as unrepresented (resp. uncertain).
This approach, however, would be misleading since two numbers close to the threshold could be treated very differently. Also, all points on each side of the threshold would be considered equally represented (resp., certain). Instead, we consider {\em a randomized approach}, widely popular in the literature, including~\cite{dwork2012fairness}.
That is, instead of using fixed thresholds, a Bernoulli variable (a biased coin) is used that 
assigns $\qu$ as unrepresented (resp., uncertain) based on the number of samples similar to it (resp., its neighborhood uncertainty).
We represent the probability of the Bernoulli variables for lack of representation or uncertainty components as $\pe_o$ and $\pe_u$, respectively.

Note that the two Bernoulli variables $\pe_o$ and $\pe_u$ are independent from each other.
That simply follows the argument that after specifying the number of similar samples to $\qu$ whether or not it should be considered as unrepresented does not depend on the uncertainty in the neighborhood of $\qu$.
We will further discuss this in \S~\ref{sec:discussions:indep}.
Before formally defining the \ru measures, we would like to emphasize that our definitions are agnostic and independent from how $\pe_o$ and $\pe_u$ are computed. We still shall provide the details of how to compute these probabilities in \S~\ref{sec:dev}.

\begin{definition}[\sru]\label{def:sdt}
The {\em strong} representation and uncertainty measure is a probabilistic measure that considers the outcome of a model for a query point $\qu$ untrustworthy if $\qu$ is not represented by $\dee$ {\bf and} it belongs to an uncertain region. 
Formally, the strong representation and uncertainty measure is:
\begin{align} \label{eq:strong}
    \nonumber
    \sdt(\qu) &= \pe\big((\qu \mbox{ is outlier}) \wedge (\qu \mbox{ is in uncertain region})\big) \\
    &= \pe_o(\qu) \times \pe_u(\qu)
\end{align}
\end{definition}

\sru raises the warning signal only when the query point fails on {\em both} conditions of being represented by $\dee$ and not belonging to an uncertain region. 
For instance, in Example~\ref{ex-1} none of the query points fail both on representation and on uncertainty; hence neither has a high \sru score.
On the other hand, 
a high \sru score for a query point $\qu$ {\em provides a strong warning signal} that one should perhaps reject the model outcome and not consider it for decision-making.

\sru is a strong signal that raises warning only for the fearfully-concerning cases that fail both on representation and uncertainty.
However, as observed in Example~\ref{ex-1} a query points failing {\em at least} one of these conditions may also not be reliable, at least for critical decision making.
We define the weak representation and uncertainty measures to raise a warning for such cases.

\begin{definition}[\wru]\label{def:wdt}
The {\em weak} representation and uncertainty measure is a probabilistic measure that considers the outcome of a model for a query point $\qu$ untrustworthy if $\qu$ is not represented by $\dee$ {\bf or} it belongs to an uncertain region.
Formally, the weak representation and uncertainty measure is computed as follows:
\begin{align} \label{eq:weak} 
    \nonumber
    \wdt(\qu) &= \pe\big((\qu \mbox{ is outlier}) \vee (\qu \mbox{ is in uncertain region})\big) \\
    &= \pe_o(\qu) + \pe_u(\qu) - \pe_o(\qu) \times \pe_u(\qu)
\end{align}
\end{definition}
\vspace{-3em}
\section{Implementation of the measures}\label{sec:dev}
\subsection{Lack of Representation Oracle}\label{sec:dev:po}
The first component of the \ru measures identifies if the data set $\dee$ misses to represent the query point $\qu$.
The oracle returns the probabilistic measure $\pe_o$, indicating if $\qu$ is an outlier in $\dee$.
Different techniques have been proposed to identify the outliers and the anomalies~\cite{breunig2000lof,ramaswamy2000,ester1996density,liu2008isolation,chandola2009anomaly} of a data set.
The \ru measures proposed in this paper are agnostic to the choice of the outlier detection technique, and alternative approaches that can compute $\pe_o$ are equally applicable.

Still, in this section we provide a new approach for computing the probability $\pe_o$, indicating if $\qu$ is an outlier.
In particular, we follow the existing work \cite{asudeh2021identifying,ester1996density,ramaswamy2000,breunig2000lof} by considering the $k$ nearest neighbors of $\qu$ in $\dee$ for studying if it is an outlier.

\begin{figure*}[!htb]
    \begin{minipage}[t]{0.32\linewidth}
        \centering
        \includegraphics[width=\textwidth]{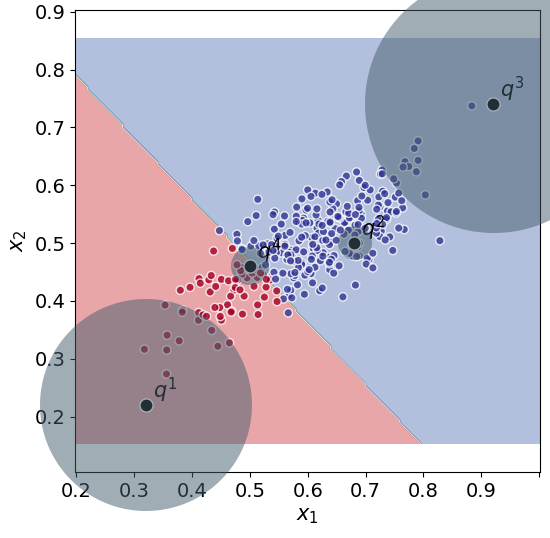}
    \vspace{-7mm}
    \caption{illustration of the $k$-vicinity ($k=10$) of $\qu^1$ to $\qu^4$ in Example~\ref{ex-1}.}
    \label{fig:ex1:4}
    \end{minipage}
    \hfill
    \begin{minipage}[t]{0.32\linewidth}
        \centering
        \vspace{-42mm}\includegraphics[width=\textwidth]{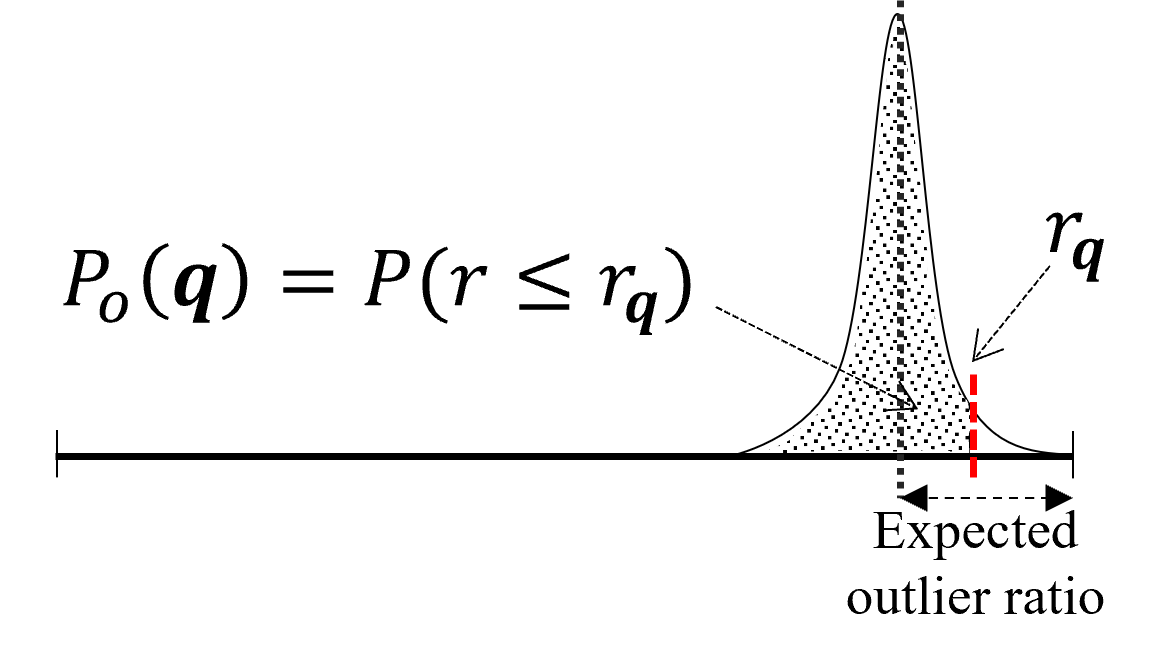}
    \vspace{-4mm}
    \caption{computation of of $\pe_o(\qu)$ using the ratio of tuples in $\dee$ with smaller $k$-vicinity radius than $\Delta_k(\qu,\dee)$.}
    \label{fig:or}
    \end{minipage}
    \hfill
    \begin{minipage}[t]{0.32\linewidth}
        \centering
        \vspace{-50mm}
        \begin{tabular}{|c|c|c|c|c|}
        \hline
        id&$x_1$&$x_2$&2-NN&$\rho$ \\ \hline
        $t^1$&0.61&0.58&$\{t^5$,$t^6\}$&0.191 \\ \hline
        $t^2$&0.32&0.77&$\{t^7$,$t^9\}$&0.161 \\ \hline
        $t^3$&0.79&0.41&$\{t^1$,$t^5\}$&0.247 \\ \hline
        $t^4$&0.13&0.9&$\{t^7$,$t^{10}\}$&0.082 \\ \hline
        $t^5$&0.74&0.44&$\{t^1$,$t^3\}$&0.191 \\ \hline
        $t^6$&0.55&0.64&$\{t^1$,$t^9\}$&0.183 \\ \hline
        $t^7$&0.18&0.85&$\{t^4$,$t^{10}\}$&0.147 \\ \hline
        $t^8$&0.93&0.12&$\{t^3$,$t^5\}$&0.372 \\ \hline
        $t^9$&0.38&0.71&$\{t^2$,$t^6\}$&0.183 \\ \hline
        $t^{10}$&0.05&0.92&$\{t^4$,$t^7\}$&0.147 \\ \hline
    \end{tabular}
    \vspace{2mm}
    \caption{the data set in Example~\ref{ex-2}.}
    \label{fig:ex:2:1}
    \end{minipage}
    \vspace{-5mm}
\end{figure*}

Given a distance metric $\Delta$, let $\rho_\qu=\Delta_k(\qu,\dee)$ be the distance of the $k$-th nearest tuple in $\dee$ to $\qu$. 
Considering euclidean\footnote{Please note that while we use euclidean distance for the explanation and examples in the paper, our metrics and algorithms are agnostic to the choice of the distance measure, and those equally work for other ones. We evaluate the impact of the choice of the distance measure (using multiple well-known measures) in our experiments. Our experiments show consistent results across different measures.} distance measure for $\Delta$, $\rho_\qu$ is the radius of the $k$-vicinity of $\qu$, the tight hyper-sphere (circle in 2D) centered at $\qu$ that includes exactly $k$ tuples from $\dee$.
For example, Figure~\ref{fig:ex1:4} shows the $k$-vicinity of the query points $\qu^1$ to $\qu^4$ in Example~\ref{ex-1}. 
It is easy to see that smaller values of $\rho_\qu$ correspond to denser $k$-vicinities around $\qu$, meaning that the data set $\dee$ is 
more representative of the query point. 
We use this observation to develop the lack of representation component $\pe_o$.
That is, we consider the $k$-vicinity of $\qu$ and the value of $\rho_\qu$ to identify whether or not $\qu$ is represented by $\dee$.

In particular, we would like to develop the function $O:\mathbb{R}\rightarrow [0,1]$ that given the value of $\rho_\qu$ returns the probability $\pe_o(\qu)$.
That is, $\pe_o(\qu) = O(\Delta_k(\qu,\dee))$.
The function $O$ takes a distance value as the input and returns a probability indicating if the query point with that $k$-vicinity radius is not represented by $\dee$.
It is clear that as the distance values increase, the probability $\pe_o$ should monotonically increase as well. 
However, translating the distances to the probabilities is unclear and may vary from one data set to another.

Our idea is to {\em learn the function} $O$ using the tuples in the data set $\dee$.
Specifically, we note that the probability of sampling an outlier tuple according to the underlying distribution $\dist$ is low, and hence most of the tuples in $\dee$ are not outliers.
Therefore, the comparison between $\rho_\qu$ and the $k$-vicinity radii of the tuples in $\dee$ can reveal if $\qu$ is an outlier.
As a result, instead of directly translating the distance values to probabilities, we can first identify the {\em rank of $\rho_\qu$} in comparison with other tuples in $\dee$ and use this information to specify if $\qu$ is an outlier.
For example, if $\rho_\qu$ is smaller than more than half of $k$-vicinity radii of the tuples in $\dee$, one can conclude that $\qu$ is not an outlier. On the other hand, if $\rho_\qu$ is larger than the $k$-vicinity radii of all tuples in $\dee$, it should be an outlier.

Besides, it is often the case in practice that data sets are associated with information such as outlier ratio, showing approximately what percentage of its samples are outliers.
We use such information to develop the function $O$.

In particular, since the ratio of the outliers in $\dee$ is often an estimation by the experts, we consider a Normal distribution $\mathcal{N}(\mu,\sigma)$, where the user-specified outlier ratio is $(1-\mu)$ and $\sigma$ is the standard deviation specifying the outlier ratio estimation variance. Figure~\ref{fig:or} demonstrates such a distribution as a bell curve centered at one minus the expected outlier ratio.
Recall that instead of directly using the value of $k$-vicinity radius to decide if a point is an outlier or not, we use the relative position of this value to compute the probability. 
That is, we define the probability distribution on the {\em ratio of outliers in $\dee$}. 

To do so, we first compute $\Gamma_\dee$, the multi-set (including duplicate values) of $k$-vicinity radii of the tuples in $\dee$. Now,
let $r_\qu$ be the percentage of values in $\Gamma_\dee$ that are not larger than $\Delta_k(\qu,\dee)$:
\begin{align}
    \hspace{18mm}r_\qu=\frac{|\{r\in\Gamma_\dee | r \leq \Delta_k(\qu,\dee)\}|}{n}
\end{align}

Using the value of $r_\qu$, 
the query point $\qu$ is an outlier if its $k$-vicinity radius falls within the range of outlier radii. In particular, suppose $r$ is the boundary of outlier values in $\Gamma_\dee$. Then $\qu$ is an outlier if $r_q \geq r$. Following this argument, the function $O$ can use the probability distribution $\mathcal{N}(\mu,\sigma)$ to compute the probability $\pe_o(\qu)$. 
As shown in Figure~\ref{fig:or},  $\pe_o(\qu)$ is the probability that the outlier boundary $r$ is less than or equal to $r_\qu$, i.e. 
$\pe_o(\qu) = \pe(r\leq r_\qu)$.

Converting the values to the standard-Normal distribution and using the $Z$-table:
\vspace{-3mm}
\begin{align} \label{eq:po}
    \pe_o(\qu) = \pe(r\leq r_\qu) = \mathcal{Z}\Big(\frac{r_\qu - \mu}{\sigma}\Big)
\end{align}

To further elaborate on how $\pe_o(\qu)$ is computed, let us consider the following example:

\begin{example}\label{ex-2}
Consider the 2D data set $\dee$ with $n=10$ tuples shown in the table of Figure~\ref{fig:ex:2:1}.
In addition to the tuple values on $x_1$ and $x_2$, the table also includes the $k$-NN ($k=2$) of the tuples and the radius $\rho$ of their $k$-vicinity.
Let the outlier ratio of the data set be 20\% ($\mu = 1- 0.2= 0.8$) with a standard deviation of $\sigma=0.1$.
Now consider the query point $\qu:\langle 0.81,0.76\rangle$. The 2-NN of $\qu$ are $\{t_1, t_6\}$, and $\rho_\qu = 0.286$. Looking at the last column of Figure~\ref{fig:ex:2:1}, only $t^8$ has a larger $k$-vicinity radius than $\rho_\qu$, i.e., for 90\% of tuples the $k$-vicinity radius is smaller than $\rho_\qu$.  Therefore, using Equation~\ref{eq:po},
$\pe_o(\qu) = \mathcal{Z} ((0.9-0.8)/0.1) = 0.84
$.
\end{example}

Computing Equation~\ref{eq:po} requires (i) computing $\Delta_k(\qu,\dee)$, which requires finding the $k$-NN of $\qu$, and (ii) computing the value of $r_\qu$.
The baseline approach for computing these values makes a linear pass over $\dee$ to identify the $k$-NN of $\qu$. Besides, it requires $O(n^2)$ to compute the multi-set of $k$-vicinity radii $\Gamma_\dee$ for the tuples in $\dee$ and then it needs $O(n)$ to make a pass over $\Gamma_\dee$ to compute $r_\qu$.

However, given the interactive nature of query answering for ML systems and potentially large size of $\dee$, we are interested in designing an algorithm that runs in a {\em sublinear} time to $n$. Theoretically speaking, finding the $k$ nearest neighbors of a point can be done in $O(\log n)$ using $k$-voronoi diagrams\footnote{$k$-voronoi diagram is a partitioning of the query space into convex cells where the $k$-NN of all points in each cell is the same set of tuples.}~\cite{chazelle1987improved,agarwal1998constructing,lee1982k,bohler2013complexity}, while constructing the $k$-voronoi cells takes polynomial time for a constant number of dimensions~\cite{agarwal1998constructing,edelsbrunner1986voronoi} ($O(k^2n\log n)$ for 2D~\cite{lee1982k}).
Besides, practically efficient algorithms have been proposed~\cite{hautamaki2004,wu2002improved,suguna2010improved}, construct data structures in preprocessing time that enables identifying $k$-NN is near-logarithmic time. 
We rely on the off-the-shelf techniques for finding the $k$-NN of a query point.

During the preprocessing time, we first construct the $k$-NN data structure. Next, for every tuple in $\dee$, we identify its $k$-vicinity radius and add it to the list $\Gamma_\dee$.
Finally, to quickly identify the value of $r_\qu$ in query time, we sort the list $\Gamma_D$.

\begin{algorithm}[t]
\caption{}\label{alg:po}
\begin{algorithmic}[1]
\Require {data set $\dee$; $k$; expected outlier ratio $\tau$; standard deviation $\sigma$; query point $\qu$}
\Function{preprocess$_o$}{$\dee$,$k$}
    \State $M\gets$ build the $k$-NN index of $\dee$; 
    ~$\Gamma\gets [~]$
    \techrep{
    \For{$t\in\dee$}
        \State $V\gets$ $k$-NN$(t)$
        \State add $\max_{t'\in V} \Delta(t,t')$ to $\Gamma$
    \EndFor
    }
    \submit{
    \State {\bf for} $t\in\dee$ {\bf do} add $\max_{t'\in k-\mbox{NN}(t)} \Delta(t,t')$ to $\Gamma$
    }
    \State {\bf return} $M$, {\bf sort}$(\Gamma)$
\EndFunction
\Function{$\pe_o$}{$\qu$}
    \State $\rho_\qu\gets \max \Delta(\qu,t'), \forall {t'\in k\mbox{-NN}(\qu,M)}$ \label{line:8}
    \State $r_\qu\gets\frac{1}{n}\mbox{\sc BinarySearch}(\rho_\qu,\Gamma)$ 
    \State {\bf return} $\mathcal{Z}(\frac{r_\qu-\mu}{\sigma})$
\EndFunction
\end{algorithmic}
\end{algorithm}

The preprocessing algorithm and the function for identifying $\pe_o(\qu)$ are provided in Algorithm~\ref{alg:po}.
To compute $r_\qu$ for a query point $\qu$, the algorithm first finds the $k$-vicinity of $\qu$ and identifies the tuple in $k$-vicinity with maximum distance from $\qu$.
Next, it applies a binary search on the sorted list $\Gamma$ to identify the number of cells in $\Gamma$ that have a value not larger than $\qu$, and use it to compute $r_\qu$. At last, it uses Equation~\ref{eq:po} and returns the value of $\pe_o(\qu)$.

Let $T_{c_n}$ be the time to construct the $k$-NN index. Also, let $T_{q_n}\simeq O(\log n)$ be the time to identify the $k$-NN of a query point, using the constructed index.
The preprocessing function constructs the $k$-NN index, identifies the $k$-NN of each tuple in $\dee$, and finally spends $O(n \log n)$ to sort the list $\Gamma$.
Therefore, the total preprocessing time is $O\big(T_{c_n} + n^2)$.
Computing $\pe_o(\qu)$ requires $T_{q_n}$ to identify the $k$-NN of $\qu$ and $O(\log n)$ for the binary search. As a result, the time to compute $\pe_o(\qu)$ is $O(T_{q_n}+\log n)\simeq O(\log n)$.

\vspace{-4em}
\subsection{Lack of Certainty Oracle}\label{sec:dev:pe}
After the lack of representation oracle, we now turn our attention to the uncertainty oracle that, given the query point, the data set $\dee$, and the target variable $y$, returns $\pe_u$, the probabilistic measure that indicates if $\qu$ belongs to an uncertain region.
There has been extensive research, and there exist different metrics for computing uncertainty, namely, entropy, Gini impurity, Brier score, and probability calibration~\cite{shannon1948mathematical,breiman2017classification,brier1950verification,pakdaman2015,gebel2009multivariate}. 
Indeed, we are agnostic to the choice of the technique for developing the uncertainty oracle, and any method that can compute the probabilistic measure $\pe_u$ is equally applicable.
Even so, in the rest of this section, we provide a development of the uncertainty oracle, following the technique proposed in \S~\ref{sec:dev:po}.
Similar to \S~\ref{sec:dev:po}, we use the $k$-vicinity of a query point $\qu$ as the region for studying uncertainty.

Binary classification is among the most popular ML tasks.
A straightforward approach for developing the uncertainty oracle for such cases is to use the {\em Shannon entropy} ($\mathcal{H}$)~\cite{shannon1948mathematical}. Let $v_1,\cdots,v_\ell$ be the set of possible values for a target variable $y$. 
Known as a measure of uncertainty, the entropy of the random variable $y$ is
\begin{align}\label{eq:uncertainty}
\hspace{18mm}\mathcal{H}(y) = -\sum_{i=1}^\ell \pe(v_\ell) \log \pe(v_\ell)
\end{align}
Higher entropy values refer to higher uncertainty, while values close to zero indicate a high certainty in the value of $y$.
For a binary variable $y$, the maximum value of entropy is one, and it refers to the cases where the probability of each value is $0.5$.
Using entropy to measure uncertainty for binary classification, we consider the set of tuples $V_k(\qu)\subseteq \dee$ in the $k$-vicinity of the query point $\qu$ and compute $\pe_u(\qu)$ as the entropy among them.
Let $p_1$ be the ratio of tuples in $V_k(\qu)$ with label $1$.
Then,
\begin{align}
    \hspace{10mm}\pe_u(\qu) = -p_1 \log p_1 - (1-p_1) \log (1-p_1)
\end{align}
Entropy can also be used for non-binary classification. However, when the cardinality of $y$ is larger than 2, entropy is not bounded by $1$ anymore. 
Therefore, instead of using the absolute value of the entropy, we use the comparison between the uncertainty value of the query point vs. the uncertainty values in $k$-vicinities of the tuples in $\dee$ to identify if $\qu$ belongs to an uncertain region. Intuitively, if the uncertainty in the neighborhood of $\qu$ is smaller than a large portion of the other points in the training set, it is considered safe. 
Specifically, suppose $r_{u}$ is the expected ratio of the tuples in $\dee$ that belong to uncertain regions. Then, we consider a Normal distribution $\mathcal{N}(\mu_u,\sigma_u)$ where $\mu_u=(1-r_u)$ and $\sigma_u$ is the standard deviation of uncertain ratio estimation. 
During the preprocessing, we construct $\Gamma_{u_\dee}$, the sorted list of uncertainty values for the tuples in $\dee$.
Then, given a query point $\qu$, we first compute $\mathcal{H}_\qu(y)$, the uncertainty in the $k$-vicinity of $\qu$ using Equation~\ref{eq:uncertainty}.
Next, applying a binary search on $\Gamma_{u_\dee}$, we compute $r_{u_\qu}$, the ratio of uncertainty values in $\Gamma_{u_\dee}$ that are not larger than $r_{u_\qu}$. Finally, converting the values to standard-Normal distribution, $\pe_u(\qu)$ is computed as following:
\begin{align} \label{eq:pu}
    \hspace{12mm}\pe_u(\qu) = \pe(r\leq r_{u_\qu}) = \mathcal{Z}\Big(\frac{r_{u_\qu} - \mu_u}{\sigma_u}\Big)
\end{align}

The residual sum of squares (RSS) is a popular measure for regression. In regression trees~\cite{breiman2017classification}, for example, the objective is to split the search space into regions with high certainty, where the RSS values in each region are minimized, i.e., the certainty in each region is maximized.
We also use RSS for measuring $\pe_u(\qu)$ for the regression tasks.
Let $V_k(\qu)\subseteq \dee$ be the set of tuples in the $k$-vicinity of $\qu$.
Also, let $m_{u_\qu}$ be the average of $y$ values in $V_k(\qu)$. That is, $m_{u_\qu} = (\sum_{\mathbf{t}^i\in V_k(\qu)} y^i)/k$.
Then the uncertainty around $\qu$ is computed as
\begin{align}
    \hspace{16mm}rss_\qu(y) = \sum_{\mathbf{t}^i\in V_k(\qu)} (y^i - m_{u_\qu})^2
\end{align}
The process for computing $\pe_u(\qu)$ is the same as the one for classification, with the only difference being that RSS should be used for computing uncertainty (instead of entropy).
The pseudo-code of the function $\pe_u(\qu)$, along with preprocessing steps, are provided in Algorithm~\ref{alg:pe}. Following a similar procedure as of Algorithm~\ref{alg:po}, the time to compute $\pe_u(\qu)$ is $O(T_{q_n}+\log n)\simeq O(\log n)$.
Using the functions $\pe_o$ and $\pe_u$, it takes a (near) logarithmic time to compute the uncertainty measures SRU and WRU.

\vspace{-5mm}

\begin{algorithm}[t]
\caption{}\label{alg:pe}
\begin{algorithmic}[1]
\Require {data set $\dee$; $k$; expected uncertainty ratio $r_u$; standard deviation $\sigma_u$; query point $\qu$; $k$-NN index $M$}
\Function{uncertainty}{$V$}
    \If{$y$ is categorical {\small \tt /*classification*/}}
        \State $r_\ell\gets \left|\{t^i\in V|y^i=v_\ell\}\right|/k,~\forall v_\ell\in Dom(y)$
        \State {\bf return} $-\sum_{i=1}^\ell r_\ell \log(r_\ell)$
    \EndIf
        \State $m_u \gets (\sum_{\mathbf{t}^i\in V} y^i)/k$
        \State {\bf return} $\sum_{\mathbf{t}^i\in V} (y^i - m_{u})^2$
\EndFunction
\Function{preprocess$_u$}{$\dee$,$k$}
    \State $\Gamma_u\gets [~]$
    \State {\bf for} $t\in\dee$ {\bf do} add {\sc uncertainty}($k$-NN$(t)$) to $\Gamma_u$
    \State {\bf return} {\bf sort}$(\Gamma_u)$
\EndFunction
\Function{$\pe_u$}{$\qu$}
    \State $u_\qu\gets${\sc uncertainty}($k$-NN$(\qu,M)$)\label{line:12}
    \State $r_\qu\gets\mbox{\sc BinarySearch}(u_\qu,\Gamma_u)/n$
    \State {\bf return} $\mathcal{Z}(\frac{r_\qu-\mu_u}{\sigma_u})$
\EndFunction
\State {\bf function} {\sc SRU}($\qu$) {\bf return} $\pe_o(\qu)\times\pe_u(\qu)$
\Function{WRU}{$\qu$}
\State $p_1\gets \pe_o(\qu); p_2\gets \pe_u(\qu)$
    \State {\bf return} $p_1 + p_2- p_1\times p_2$
\EndFunction
\end{algorithmic}
\end{algorithm}

\subsection{No Data Access During the Query Time}\label{sec:sampling}
During the query-answering phase,
Algorithms~\ref{alg:pe} and ~\ref{alg:po} require to compute the $k$-vicinity radius and the entropy with the $k$-NN of a query point $q$. Although off-the-shelf $k$-NN indices are used to find this information, one could view it as requiring to access the training data after preprocessing.
Our practical approach to address this is to {\em learn} these values.
That is, to create two models that take a query point as the input, returning the $k$-vicinity radii and the entropy values. Creating these models requires sampling from the query space\footnote{We refer to the space of valid queries as the query space. Specifically, let $dom_i$ be the cardinality of the feature $x_i$. Then the query space is $\Pi_{i=1}^n dom_i$.}, i.e., to generate a large-enough training set with observations being i.i.d samples from the query space, while the target variables are the $k$-vicinity radius and entropy.
On the positive side, one can generate an arbitrarily large training set by generating i.i.d sample queries and then computing the target values using their $k$-NN.
On the flip side, however, as proven in~\cite{asudeh2021identifying}, the theoretical upper bound on the number of samples needed is exponential. In other words, theoretically speaking, the size of the training set may need to be exponential in the number of dimensions $d$ for adversarial cases, in order to guarantee a given error $\eps$. Fortunately, as we observe in our experiments, the theoretical upper bound is not tight, and in practice, the training set size is much smaller.

We still need to specify the proper training set size for our learning tasks. To do so, we design an {\em exponential search} algorithm as follows:
the algorithm starts by setting the sample set size $N_s$ to an initial value ($O(n)$). 
It then collects $N_s$ i.i.d samples $\mathcal{S}$ from the query space and finds the $k$-vicinity of each sample $s_i$ and identifies the $k$-vicinity radius\footnote{The process to learn the entropy values is the same as learning the $k$-vicinity radii.} of $s_i$ as $\rho_{i}\gets \max \Delta(s_i,t'), \forall {t'\in k\mbox{-NN}(s_i)}$.
Next, the algorithm builds a regression model $\mathcal{M}$ using $\mathcal{S}$ as the training set.
After building the model, the algorithm checks if $\mathcal{M}$ has the error of at most $\eps$, for a user-specified error $\eps$. 
To check this, the model uses the test set $\mathcal{T}$. If $error>\eps$, the algorithm {\em doubles the sample size} and repeats the process until it reaches the right sample size for $N_s$.

Let $Reg_\rho$ and $Reg_U$ be the trained regression models that return the $k$-vicinity radius and the $k$-NN entropy of a query point $\qu$, respectively.
Then, the only changes in the proposed algorithms are (i) replace Line~\ref{line:8} of Algorithm~\ref{alg:po} with $\rho_\qu\gets Reg_\rho(\qu)$ and (ii) replace Line~\ref{line:8} of Algorithm~\ref{alg:pe} with $u_\qu\gets Reg_U(\qu)$.
\section{Discussions}\label{sec:discussions}
\subsection{Assumptions and limitations}
Let us begin our discussions by providing a synopsis of the assumptions underlying our approach's development and delve into the limitations it faces:

\begin{itemize}[leftmargin=*]
    \item \ru measures are data-centric and and complementary to the the model-centric approaches for uncertainty quantification, such as conformal predictions, prediction intervals, and prediction probabilities, as well as the explainable AI literature.
    The \ru measures should be considered alongside existing model-centric approaches in order to consider the influence of the model as well.
    \item 
    Our data model defines a data set in a tabular manner, over a set of numeric attributes. For non-tabular data such as text, image, audio, etc., we rely on the existence of proper vector representations (aka embeddings), where each object is represented as a high-dimensional vector. We demonstrate this on an image data set in our experiments. The accuracy of our metrics is limited to the accuracy of the embeddings.
    \item In our development of lack of representation and uncertainty components, we assume the outlier ratio, the uncertainty ratio, and the variance values are provided as input, while considering a Normal distribution.
    \item To enable no-access to the training data at the inference time, we are required to learn the essential values via an exponential search across the query space with a $1-\eps$ guarantee. This could become computationally expensive in the preprocessing step for small values of $\eps$ as a large number of samples are required.
\end{itemize}

In the remainder of this section, we present solutions aimed at tackling the limitations concerning hyper-parameters within our approach.

\vspace{-5mm}
\subsection{Parameter Tuning}
Similar to many other techniques in data mining and ML, the \ru measures require parameter tuning. In implementing \ru measures, we take the neighborhood size $k$ in $k$-NN, along with the outlier ratio $c$ of the training samples, the uncertainty ratio $u$, and the standard deviation for uncertainty and outlier distributions as hyper-parameters.
The techniques proposed in this paper are agnostic to the choice of parameter tuning. Nevertheless, we present some heuristics for tuning these parameters in this section.

\subsubsection{Tuning Neighborhood Size and Outlier Ratio Parameters}

The first parameter to determine is $k$: the number of tuples in $\dee$ that specify the vicinity of the queried point. The second parameter is the outlier ratio $c$, which estimates the percentage of the tuples in the data set that are outliers.  

To jointly tune $c$ and $k$ for a data set $\dee$, we adopt a technique proposed in \cite{9006151} for tuning the parameters of the local outlier factor (LOF)~\cite{breunig2000lof} algorithm.
However, instead of choosing top $\left \lfloor{cn}\right \rfloor$ points with the highest LOF scores, we select top $\left \lfloor{cn}\right \rfloor$ with the highest $k$-vicinity radii.

We define a grid of values for $c$ and $k$. For each combination, we calculate the $k$-vicinity radius for all tuples in $\dee$, choose the top $\left \lfloor{cn}\right \rfloor$ tuples as the outliers, and the top $\left \lfloor{cn}\right \rfloor$ of remaining tuples as the inliers. The inliers are chosen in this manner because we are only interested in the tuples that are most similar to the outliers.

\noindent For each $c$ and $k$, now we have a list of $k$-vicinity radii for outliers and a list for inliers and we calculate mean ($\mu_{out}(c,k), \mu_{in}(c,k)$) and variance ($\sigma^2_{out}(c,k), \sigma^2_{in}(c,k)$) over the log of the values in each list. We define the standardized difference in mean log $k$-vicinity radii between the outliers and the inliers as

$$T_{c,k}=\frac{\mu_{out}(c,k)-\mu_{in}(c,k)}{\sqrt{\left\lfloor{cn}\right\rfloor^{-1}(\sigma^2_{out}(c,k)+\sigma^2_{in}(c,k))}}$$

If $c$ is known, it is enough to find $k^{*}_{c}=$ arg max\textsubscript{k} $T_{c,k}$ that maximizes the standardized difference between the outliers and inliers for the corresponding $c$. 
Otherwise, we assume that $k$-vicinity radii form a random sample following a Normal distribution with the mean $\mu_{out}(c)$ and variance $\sigma^2_{out}(c)$ for outliers, and one with mean $\mu_{in}(c)$ and variance $\sigma^2_{in}(c)$ for the inliers. 
Then given a value of $c$, $T_{c,k}$ approximately follows a non-central $t$ distribution with degrees of freedom $\mathsf{df}_c=2\left \lfloor{cn}\right \rfloor-2$ and the non-centrality parameter:

$$\mathsf{ncp}_c=\frac{\mbox{$\mu_{out}(c)$-$\mu_{in}(c)$}}{\sqrt{\left\lfloor{cn}\right\rfloor^{-1}(\sigma^2_{out}(c)+\sigma^2_{in}(c))}}$$

We cannot directly compare the largest standardized difference $T_{c,k^{*}_{c}}$ across different values of $c$ because $T_{c,k}$ follows different non-central $t$ distributions depending on $c$. 
Instead, we can compare the quantiles that correspond to $T_{c,k^{*}_{c}}$ in each respective non-central distribution so that the comparison is on the same scale. 
To do so, we define $c_{opt}=$ arg max\textsubscript{$c$} $P(z<T_{c,k^{*}_{c}};\mathsf{df}_c;\mathsf{ncp}_c)$, where the random variable $z$ follows a non-central $t$ distribution with $\mathsf{df}_c$ degrees of freedom and $\mathsf{ncp}_c$ non-centrality parameter. 
$c_{opt}$ is where $T_{c,k^{*}_{c}}$ is the largest quantile in the corresponding $t$ distribution as compared to the others.

\subsubsection{Tuning Uncertainty Ratio Parameter}
\begin{figure}[!tbh]
    \centering
    \includegraphics[width=.4\textwidth]{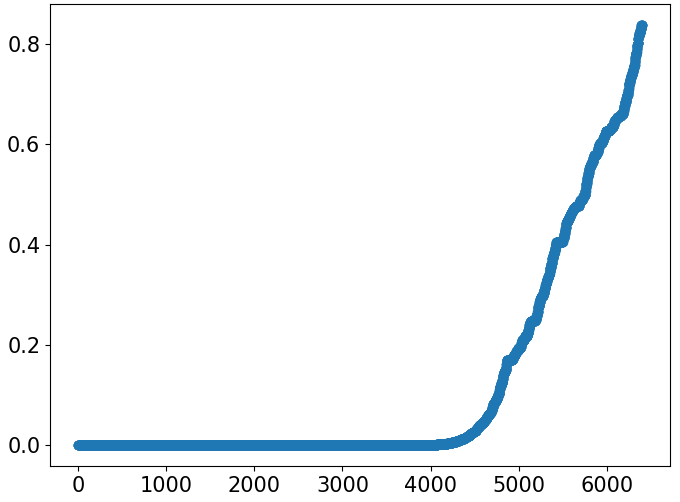}
    \vspace{-4mm}\caption{regression: sorted uncertainty values for {\it RN}}
    \vspace{-5mm}
    \label{fig:uncertainty_ratio_regression}
\end{figure}

\begin{figure}[!tbh]
    \centering
    \includegraphics[width=.4\textwidth]{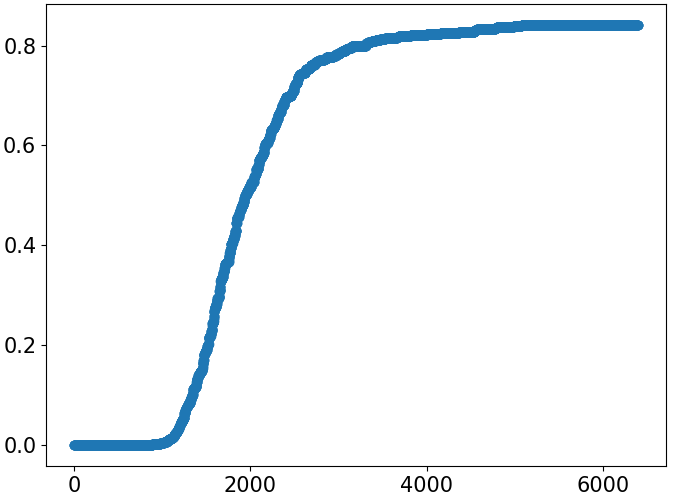}
    \vspace{-4mm}\caption{classification: sorted uncertainty values for multi-class classification data set shown in Figure \ref{fig:ext_1}}
    \vspace{-3mm}
    \label{fig:uncertainty_ratio_classification}
\end{figure}

The next parameter we need to tune is the uncertainty ratio $u$, which estimates what percentage of data belongs to uncertain regions.
Similar to the outliers ratios that help us transform the $k$-vicinity radii to probabilities, the expected
uncertainty ratio $u$ helps us transform an uncertainty value in a $k$-vicinity to a probability.
Consider the array $U_\dee=\{u_1,\cdots,u_n\}$, where $u_i$ is the uncertainty in the $k$-vicinity of $t_i\in \dee$. 
We use the distribution of values $U_\dee$ for identifying $u$.
To explain the intuition behind this, let us consider a classification task. While the uncertainty for the tuples far from the decision boundary should be low,
our experiments verify the uncertainty sharply increases as one gets close to the boundary. 
As a result, looking at the distribution of uncertainty values, one should be able to identify an estimation of $u$ by finding the sharp slope in the distribution of uncertainty values.
For example, Figures~\ref{fig:uncertainty_ratio_regression} and \ref{fig:uncertainty_ratio_classification} highlight our experiment results for two different settings for regression and classification. In Figure~\ref{fig:uncertainty_ratio_regression}, one can visually confirm that the sharp slope happens around 5000 with an uncertainty value of around 2. Similarly, in Figures~\ref{fig:uncertainty_ratio_classification}, the sharp slope happens around 2000, with an uncertainty around 0.4.
Following this intuition, we calculate the 
$k$-vicinity uncertainty for each tuple in $\dee$, and create the reverse cumulative distribution $\mathsf{V}:[0,1]\rightarrow \mathbbm{R}$ such that,
for every value $r$, the ratio of tuples in $\dee$ with an uncertainty value larger than $\mathsf{V}(r)$ is $r$. For example, $\mathsf{V}(0.1)$ returns the value $u_{0.1}$ such that the uncertainty for $10\%$ of tuples is larger than it.
We then identify the knee of this function (the sharp decrease in $\mathsf{V}(r)$) as the estimated uncertainty ratio. 
As a rule of thumb in our experiments, we observe that the knee falls around 10-15\%.

\vspace{-5mm}
\subsection{Independence of \ru measure components}\label{sec:discussions:indep}
Lack of representation and lack of certainty are the two components that are used in the definition of \ru measures (Definitions~\ref{def:sdt} and \ref{def:wdt}). In particular, we use the radius of $k$-vicinity of the query point $\qu$, $\Delta_k(\qu,\dee)$, and define a Bernoulli random variable $\pe_{o}$ on whether $q$ is an outlier in $\dee$.
Similarly, we use the entropy in the vicinity of $q$ to define the Bernoulli random variable $\pe_{u}$ on whether $\qu$ belongs to an uncertain region.
Note that $\pe_{o}$ only depends on $\Delta_k(\qu,\dee)$ and does not change based on the entropy in the vicinity of $\qu$.

Still, one can argue that in practice, there can be a correlation between uncertainty and sparsity of sub-spaces for a specific case. 
First, in our experiments, this correlation was very minimal, if not zero.
Nevertheless, it is important to note that
such correlation only impacts how frequently a similar {pair of independent Bernoulli variables are sampled}.
To further clarify this, let us consider a toy example with a bag of paired coins, where the coins are biased. Suppose the paired coins have a positive correlation, i.e., if the first coin has a higher chance for the head, the other one has a high chance of being biased the same way.
Now let us select a pair of coins from the bag. 
Suppose, the pairs are correlated, one has a probability of 0.8 and the other a probability of 0.75 for the head.
Still, the pair of coins are independent of each other since the second coin will have a probability of 0.75 for the head, independent of the first coin. 
In our case, $\pe_o$ and $\pe_u$ can be modeled as paired independent coins, where their probabilities may (or may not) be correlated. However, after selecting a pair (by selecting a query point), the two variables are independent since whether a point is an outlier only depends on the density of points in its neighborhood not the variance in their target values (uncertainty).

\vspace{-3mm}
\section{Experiments}
\label{sec:exp}

We conduct comprehensive experiments on multiple synthetic and real-world data sets of diverse sizes and dimensions using a variety of models (Logistic Regression, K-Nearest-Neighbor, Artificial Neural Networks, Deep Neural Networks, ElasticNet, Random Forest, and SVM), distance measures (Chebyshev, Manhattan, and Euclidean), {and tasks (regression and binary/multi-class classification including text classification and image processing)} to validate the effectiveness and consistency of our proposal and evaluate the efficiency and scalability of our algorithms.
In our proof of concept experiments, we follow a similar evaluation approach to the ones conducted in the existing literature on the reliability of individual predictions~\cite{bosnic2008comparison,petersen2009generic} where the correlation of the reliability scores and the prediction error over a test set is evaluated.
{We also demonstrate the failure of existing work such as Conformal Prediction, Prediction Probabilities (for cases that are not represented by the data), and data coverage (for the cases that belong to the uncertain regions) and how our proposed measures perform superior in capturing the prediction unreliability associated with point queries.}

\vspace{-5mm}
\subsection{Experiments Setup} \label{sec:exp:setup}
The experiments were conducted using a 2.5 GHz Quad-Core Intel Core i7 processor, 16 GB memory, and running macOS. The algorithms were implemented in Python.

\subsubsection{Motivating Use-cases}\label{sec:usecases} 
We motivate our experiments based on the following example data science tasks:
\begin{itemize}
    \item {\bf Regression:} we consider three regression use-cases where (i) a GIS application requires to estimate the land altitude of a point $p=\langle long,lat\rangle$, (ii) a real-estate agency would like to predict house sale prices for investment, and (iii) a jewelry app would like to predict the price of diamonds based on their attributes. The \ru measures will serve as warning signals when the altitude or price predictions are not reliable.
    \item {\bf Classification:} We consider two classification use-cases where (i) a banking application that would like to predict the payment type of its credit-card holders and (ii) an employment application that needs to predict if an employee's salary is above 50K or not. In addition to class labels, the \ru scores are provided as a reliability analysis of the predictions.
    \item {\bf Text and Image Classification:} Last but not least, we consider two applications on text and image data as examples of unstructured data. In particular, we consider (i) classifying an aviation article and (ii) a handwritten digit recognition from images. Similar to the previous cases, we will use \ru measures to identify when predicted labels are not reliable.
\end{itemize}

Considering these use cases for our non-synthetic experiments, the measure of success for \ru measures is to see if the prediction reliability values are indeed aligned with these scores.
That is, the evaluation is successful if the model predictions for queries with higher \ru scores are worse with a higher probability. In other words, there should be a high correlation between the \ru values and the model performance metrics.

We will use standard metrics for the model performances: Accuracy, F1, FNR, FPR for classification, and residual sum of squares (RSS) for regression.

\vspace{-2mm}
\subsubsection{Data Sets}\label{sec:datasets}
For evaluation purposes, we used (i) a collection of synthetic data sets and (ii) {\em seven real-world data sets} for regression and binary/multi-class classification including text and image classification. 

\stitle{Challenge}
In the evaluation of our \ru measures, we needed to generate samples with different \ru values. However, 
since the tuples with high \ru values are unlikely to be drawn from the underlying distribution $\dist$, it is challenging to collect enough samples (as a test set) to evaluate the effectiveness of our measures.
A comprehensive evaluation requires query points drawn uniformly from the query space to cover different parts of it.
To achieve this, we need to have access to a ground truth
oracle that for {\em any} given sample taken from the query space returns the value of {\em target variable}.
However, finding a real-world data set
in a context where the ground truth oracle exists (publicly) is challenging.
To overcome this challenge, we take three directions: first, to have full control of the shape and complexity of the ground truth labels over different data sets, we generate synthetic data; second, we find a real-world data set and a third party (public) service that provides access to ground truth labels; and third, we find a very large data set that contains samples from different parts of the query space and apply sub-sampling on it. Next, we remove the outliers from each sample (detecting the outliers using the Local Outlier Factor (LOF) algorithm) and split each cleaned sample into train and test sets, and add the outliers to the test set to cover larger parts of the query space.
A downside of this approach is that it further reduces the presentation of points from under-represented regions in the training set,  which may impact the model performance for those regions. Alternatively, one can partition the data in two halves and use one for the training set (without removing the outliers from it), while using the other solely for generating the test set. Indeed the training set size in such an approach is smaller. We tried this approach in \S~\ref{exp:additional:train-test} and observed consistent results between the two approaches.

\begin{figure}[!tb]
\centering
\frame{\includegraphics[width=0.48\textwidth]{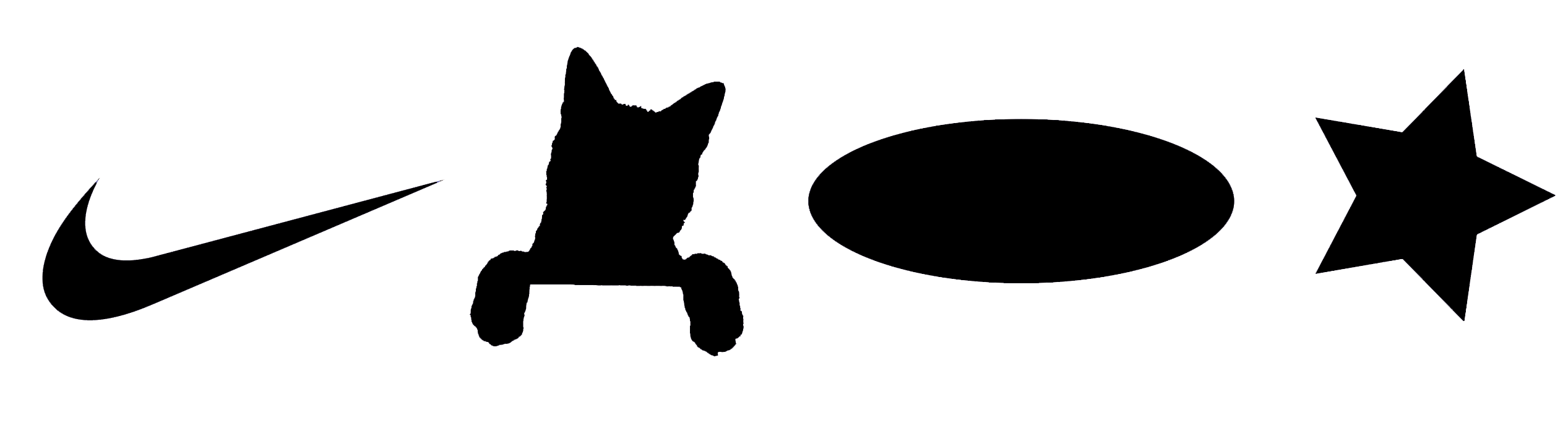}}
\vspace{-6mm}
\caption{shapes used as the ground truth in creating the synthetic data sets {\it SYN}}
\label{fig:exp:syn-ground-truth}
\vspace{-5mm}
\end{figure}

\stitle{Real Data Sets} 
We use multiple real data sets, as briefly explained in the following:
\begin{enumerate}[leftmargin=*]
\item\textbf{3D Road Network (RN) Data set~\cite{kaul2013building}} is a benchmark data set for {\bf regression} that was constructed by adding elevation information to a 2D road network in North Jutland, Denmark. It includes 434,874 records with attributes \at{Latitude}, \at{Longitude}, and \at{Altitude}.
We took 30 samples of size 10,000 from {\it RN} data set and generated 30 data sets and repeated each experiment 30 times, using different data sets.
To address the evaluation challenge for the {\it RN} data set, we generated a uniform sample of 6,400 points $\langle x_1, x_2\rangle$ in the range [0, 1].  We then transform the uniform samples back to the same space as the points in {\it RN}.
To obtain the ground-truth labels for the query points, we used an {\em off-the-shelf API}\footnote{https://api.open-elevation.com/} that given every coordinate $\langle\at{Latitude},\at{Longitude}\rangle$ in the data space, it yields the corresponding $\at{Altitude}$.

\item\textbf{House Sales in King County (HS) Data set~\cite{harlfoxem_2016}} is a {\bf regression} data set for house sale prices for King County (Seattle). It includes houses sold between May 2014 and May 2015. It includes 21,614 records having 21 attributes with 2 categorical and 16 continuous types. Given attributes such as {\tt no. of bedrooms, square footage, floors,} etc., the task is to predict the {\tt price} of the house.
We took 30 samples of size 10,000 from {\it HS} data set, 
generated 30 data sets, and repeated each experiment 30 times, using different data sets.
To address the evaluation challenge for {\it HS} data set, for each sample, we removed the outliers and then split the data set into train and test sets. We then added the outliers back to the test set. Although with {\it HS} we can not measure the \ru values for the whole query space, we believe the findings can still confirm the effectiveness of our measures.

\item\textbf{Diamond (DI) Data set \cite{agrawal_2017}} is a {\bf regression} data set for predicting the price of diamond given some visual properties. This data set has 53,941 records with 14 attributes, 6 of which are continuous and 3 categorical. We used a similar approach to {\it HS} data set for utilizing {\it DI} in our experiments.

\item\textbf{Default of Credit Card Clients (DCC) Data set~\cite{YEH20092473}} is a data set for {\bf classification} that was constructed from payment data in October 2005 from an important bank in Taiwan.
The data set is a binary class with default payment (Yes = 1, No = 0), as the response variable. Among the 30,000 records, 6,636 (22.12\%) are cardholders with default payments. The data set has 23 features (9 categorical and 14 continuous) including \at{credit line}, \at{age}, \at{gender}, \at{education}, \at{history of payment}, \at{amount of bill statement}, \at{amount of the previous statement}, etc.
Since it was not feasible for us to devise a function that can produce the ground truth for {\it DCC}, we took a sample of size 15,000 from the data set and then split it into two sets of train (5,000 tuples) and test (10,000 tuples) and used the test set as a substitute for the uniform sample over the query space. Following the same procedure, we generated 30 data sets and repeated each experiment 30 times, using different data sets. Similar to {\it HS}, we can not measure the \ru values for the whole query space in {\it DCC}, yet the findings can still confirm the effectiveness of our measures.

\item\textbf{Adult (AD) Data set \cite{kohavi1996scaling}} is a well-known benchmark data set for {\bf classification} tasks predicting whether income exceeds \$50K annually based on census data. This data set has 32,561 records with 14 attributes, 6 of which are continuous and 8 categorical. We used a similar approach to the {\it HS} data set for utilizing {\it AD} in our experiments.

\item\textbf{Real-sim (RS) Data set \cite{mccallum}} is based on {\it SRAA}\cite{mccallum_sraa} data set, preprocessed for SVMlin project \cite{sindhwani2006large}. The data set is designed for {\bf text classification} tasks and is based on UseNet articles of four discussion groups on simulated auto racing, simulated aviation, real autos, and real aviation. The task is to separate real data from simulated data.
{\it RS} is a sparse data set and has 72,309 records with 20,958 attributes with continuous values. We used a similar approach to {\it HS} data set for utilizing {\it RS} in our experiments.

\item\textbf{Gisette (GS) Data set \cite{guyon2004result}} is a handwritten digit recognition data set based on the popular {\it MNIST} data set \cite{lecun-mnisthandwrittendigit-2010} for {\bf image classification} to separate highly confusible digits `4' and `9'. The digits have been size-normalized and centered in a fixed-size image of $28\times28$ pixels. 
This data set has 6,000 records with 5,000 attributes with continuous values. We used a similar approach to {\it HS} data set for utilizing {\it GS} in our experiments.
\end{enumerate}
\vspace{-1em}
\stitle{Synthetic (SYN) Data Sets}
To fully investigate the relationship between the \ru measures and the model performance, we generated a collection of 60 data sets and repeated each experiment 60 times, using different data sets.
Each data set is a random sample following a 2D Gaussian distribution with $\mu=[0,0]$ and $\Sigma= 
\begin{bmatrix} 6 & 4\\ 3 & 1\end{bmatrix}$ over the input space $\ex=\langle x_1, x_2\rangle$ where $x_1$ and $x_2$ are positively correlated and the output space is the binary label $y$ with values $\{-1,+1\}$. To create the binary classes for each data set, we randomly moved the samples over each shape in Figure \ref{fig:exp:syn-ground-truth} in a way that the sample and shape have an intersection. As a result, each shape is the ground truth for 15 data sets but with different placements. A data point belongs to the $-1$ class, if it falls into the corresponding shape, otherwise, it belongs to the $+1$ class.
To address what we discussed in the evaluation challenge, we create a uniform sample of size 6,400 over [0, 1] and will label the points w.r.t. each shape and its placement in the space, generating a total of 60 uniform samples corresponding to each data set. 
In particular, following Example~\ref{ex-1}, we consider a binary classification task over the observation variables $x_1$ and $x_2$. 
We chose a 2D setting for visualization purposes.
All continuous values used are normalized in the range $[0,1]$, using $(v_i-\min)/(\max - \min)$ and the non-ordinal ones are one-hot encoded using scikit-learn OneHotEncoder.

\stitle{Default values}
To evaluate the performance of our algorithms under different settings, we vary the value of a parameter, while fixing the value of the other ones. The parameters that are varied among our performance evaluation experiments include $n$ (number of points), $k$ (neighborhood size), $c$ outlier ratio, and $d$ number of dimensions.
The default value for neighborhood size $k$ is 10. The outlier ratio $c$ is set to 0.1 suggesting that a mean $\mu=0.9$ is chosen for outlier distribution with a standard deviation $\sigma=0.1$. We adopt a technique proposed in \cite{9006151} to jointly tune $k$ and $c$ parameters for a given data set. The tuning procedure for these two parameters alongside uncertainty ratio parameter $u$ is discussed in detail in \S~\ref{sec:discussions}. 
The default value for $d$ (number of attributes) is 2 for {\it SYN} and {\it RN} data sets, while it is 20, 18, 9, 14, 6,000, and 20,958 for {\it DCC}, {\it HS}, {\it DI}, {\it AD}, {\it GS} and {\it RS} respectively.
The default value of $n$ (size of data set) for the {\it SYN}, {\it RN}, {\it HS}, {\it DCC}, {\it DI}, {\it AD}, {\it GS} and {\it RS} data sets are 1,000, 10,000, 10,000, 5,000, 43,150, and 32,560, 6,000 and 72,309.
The uncertainty ratio $u$ is set to 0.1, therefore, a mean $\mu_u=0.9$ is chosen for uncertainty distribution with a standard deviation $\sigma_u=0.1$.

\vspace{-5mm}
\subsection{Proof of Concept}\label{exp:effectiveness}
We start our experiments by evaluating the effectiveness of \ru measures across different data sets, ML models, and different parameters. 
Since the \ru measures are model-independent, we perform the effectiveness validation experiments for both classification and regression tasks. For the classification tasks, we use {\it SYN}, {\it DCC}, {\it AD}, {\it RS} and {\it GS} data sets, and for the regression tasks, we employ {\it RN}, {\it HS} 
 and {\it DI} data sets. To demonstrate the effectiveness of the \ru measures
we first provide a visual validation, using one of the 2D {\it SYN} data sets.
We then present a comprehensive validation over all our data sets by providing the correlation between the \ru values and the performance of an ML model's prediction on the same data. 
To do so, we deliver the results as bar graphs in which the $x$-axis is a bucketization of the ranges of the \ru measures and the $y$-axis is the ML model's evaluation score. Each bar represents a value corresponding to a measure of accuracy/error i.e. {\it Accuracy, F1 score, FPR} and {\it FNR} of the ML model for all the tuples that have a \ru value in the same range as the bar.

\begin{figure*}[!htb] 
\centering
    \begin{subfigure}[t]{0.23\linewidth}
        \centering
        \includegraphics[width=\textwidth]{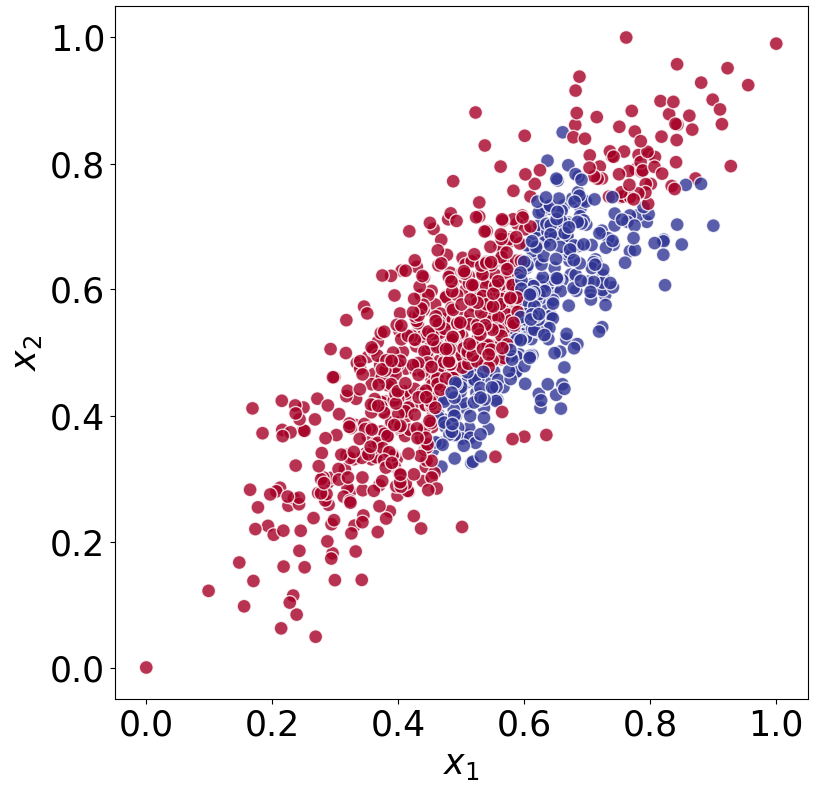}
        \vspace{-5mm}\caption{a data set $\dee$ in SYN}
        \label{fig:exp-cat-train}
    \end{subfigure}
    \hfill
    \begin{subfigure}[t]{0.23\linewidth}
        \centering
        \includegraphics[width=\textwidth]{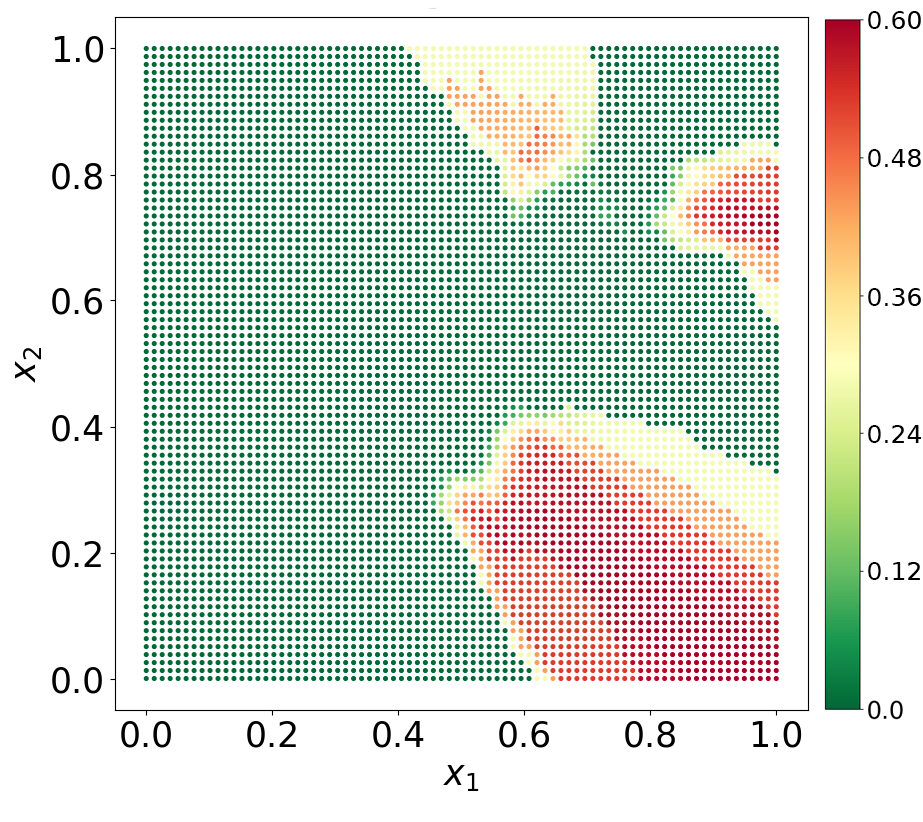}
        \vspace{-5mm}\caption{query space colored based on \sru values with regards to $\dee$ in Fig.~\ref{fig:exp-cat-train}}
        \label{fig:exp-cat-sdt}
    \end{subfigure}
    \hfill
    \begin{subfigure}[t]{0.23\linewidth}
        \centering
        \includegraphics[width=\textwidth]{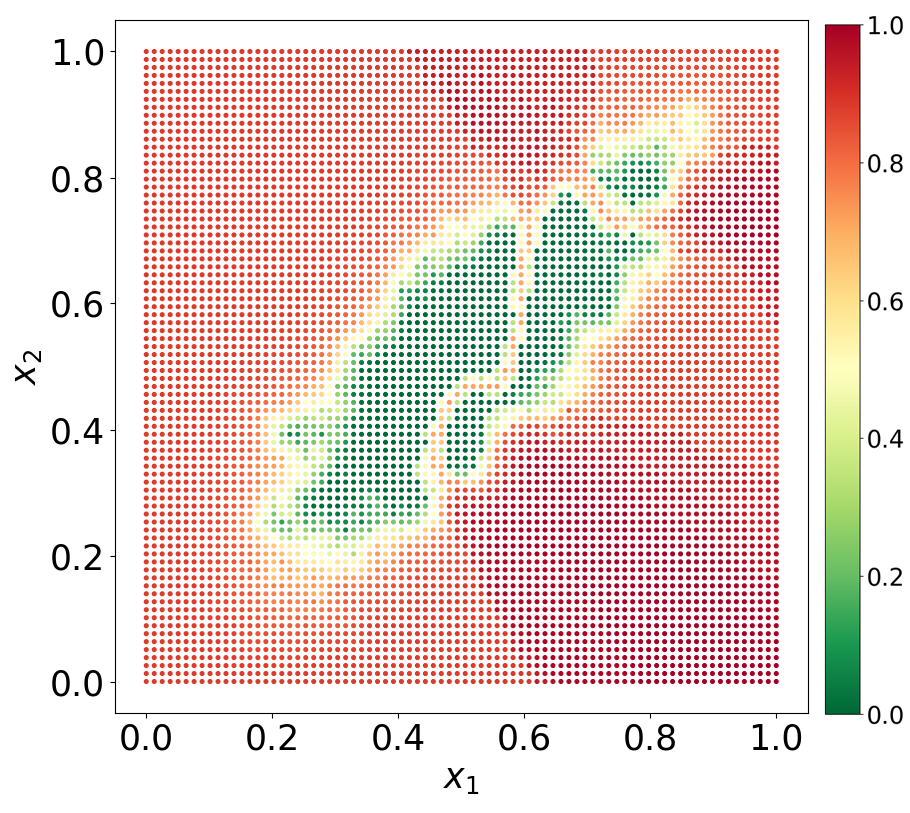}
        \vspace{-5mm}\caption{query space colored based on \wru values with regards to $\dee$ in Fig.~\ref{fig:exp-cat-train}}
        \label{fig:exp-cat-wdt}
    \end{subfigure}
    \hfill
        \begin{subfigure}[t]{0.23\linewidth}
        \centering
        \includegraphics[width=\textwidth]{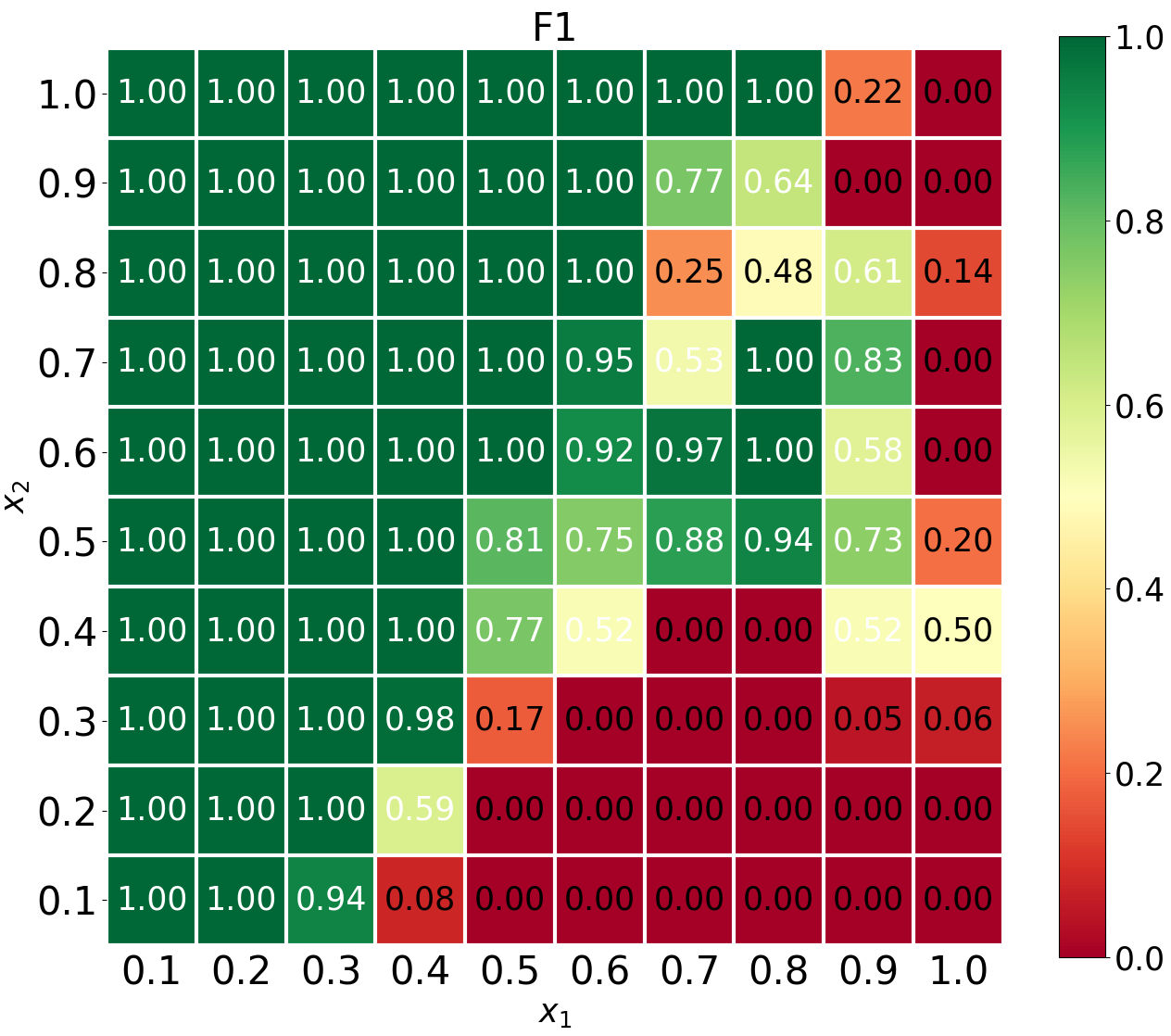}
        \vspace{-5mm}\caption{LR model's F1 score over discretized $\dee$ in Fig.~\ref{fig:exp-cat-train}}
        \label{fig:exp-cat-heatmap-f1}
    \end{subfigure}

    \begin{subfigure}[t]{0.23\linewidth}
        \centering
        \vspace{-38mm}
        \includegraphics[width=\textwidth]{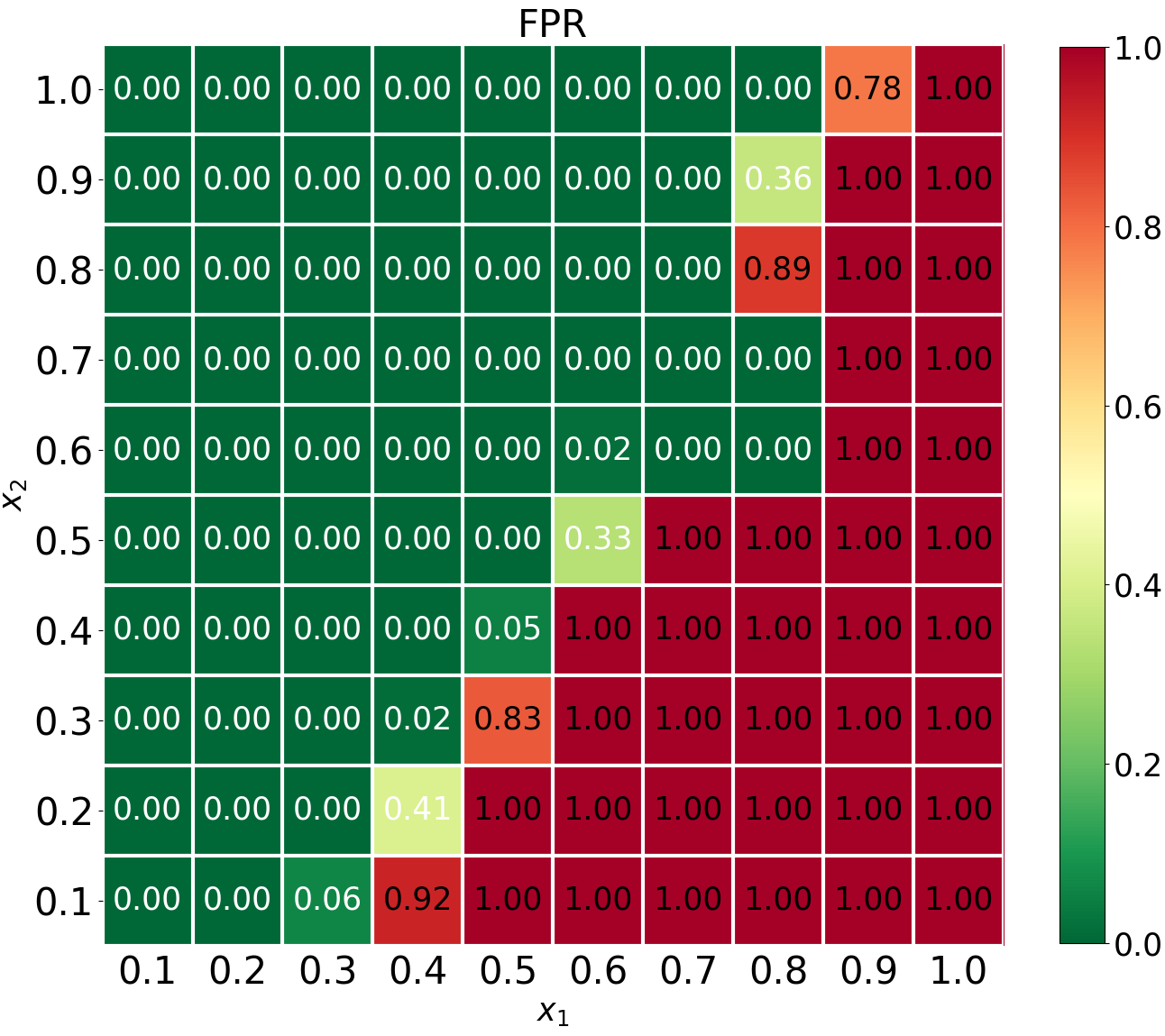}
        \vspace{-5mm}\caption{LR model's FPR score over discretized $\dee$ in Fig.~\ref{fig:exp-cat-train}}
        \label{fig:exp-cat-heatmap-fpr}
    \end{subfigure}\hfill
        \begin{subfigure}[t]{0.23\linewidth}
        \centering
        \includegraphics[width=\textwidth]{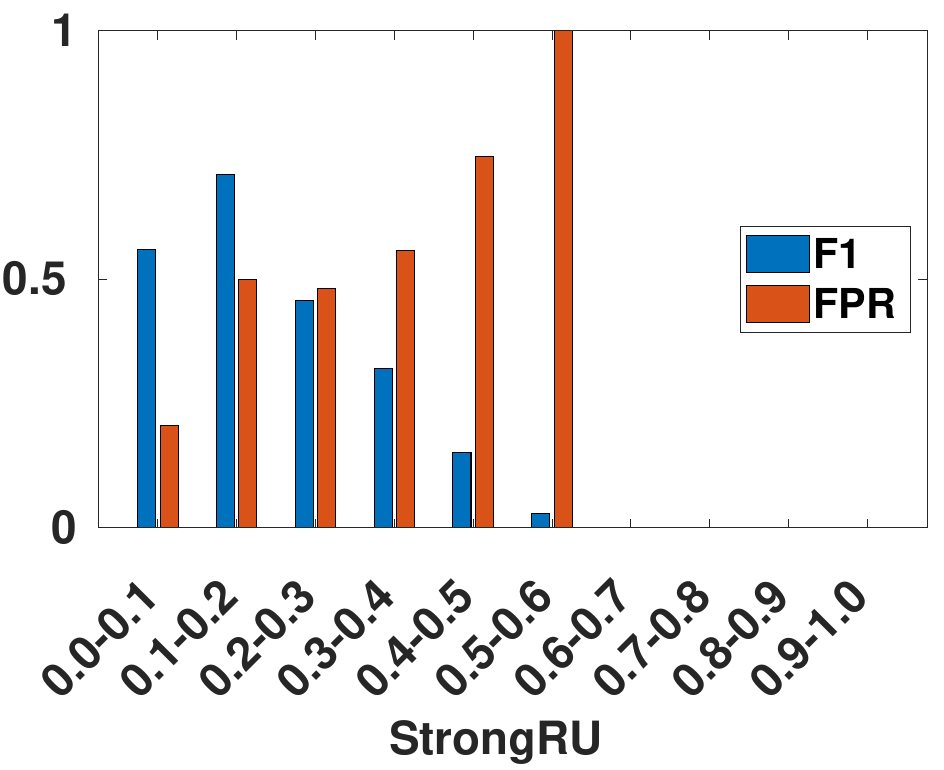}
        \caption{effectiveness of \sru over $\dee$ in Fig.~\ref{fig:exp-cat-train}\footnotemark}
        \label{fig:exp-cat-bar-sdt}
    \end{subfigure}
    \hfill
    \begin{subfigure}[t]{0.23\linewidth}
        \centering
        \includegraphics[width=\textwidth]{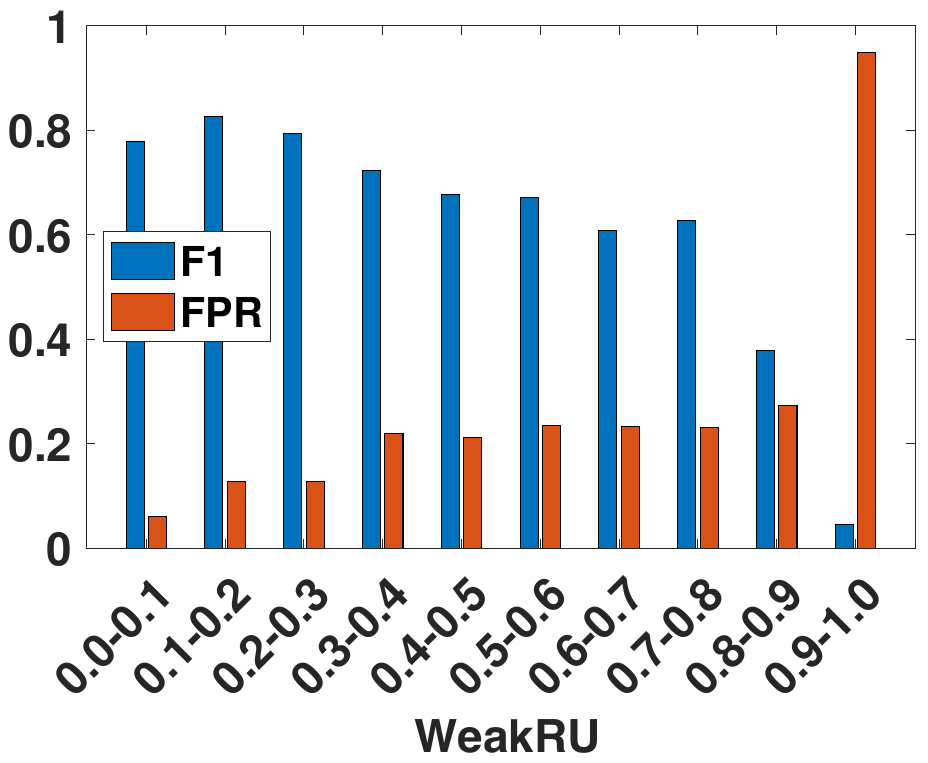}
        \caption{effectiveness of \wru over $\dee$ in Fig.~\ref{fig:exp-cat-train}}
        \label{fig:exp-cat-bar-wdt}
    \end{subfigure}\hfill
    \begin{subfigure}[t]{0.23\linewidth}
        \centering
        \includegraphics[width=\textwidth]{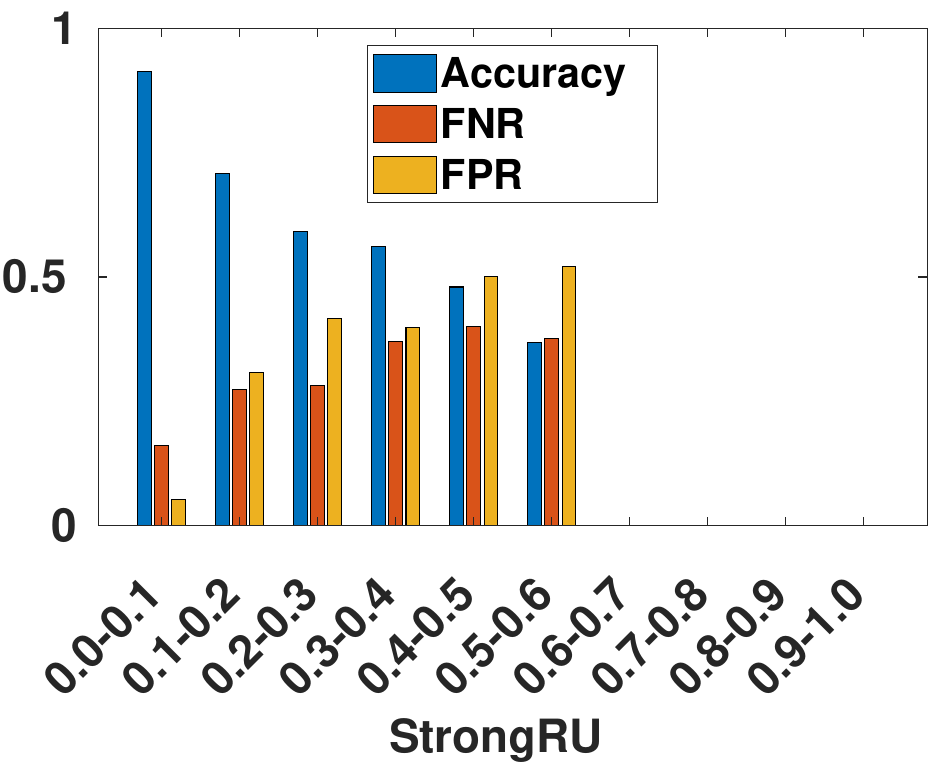}
        \caption{average effectiveness of \sru over SYN data sets}
        \label{fig:exp-class-bar-syn-sdt}
    \end{subfigure}
    
    \hfill
    \begin{subfigure}[t]{0.23\linewidth}
        \centering
        \includegraphics[width=\textwidth]{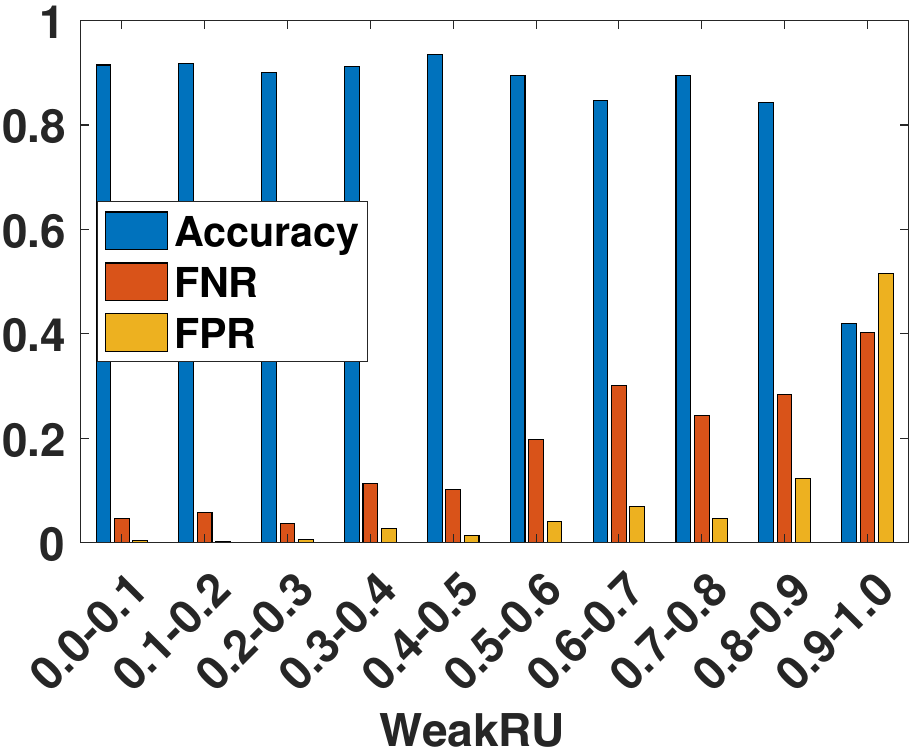}
        \vspace{-5mm}\caption{average effectiveness of \wru over SYN data sets}
        \label{fig:exp-class-bar-syn-wdt}
    \end{subfigure}
    \hfill
    \begin{subfigure}[t]{0.23\linewidth}
        \centering
        \includegraphics[width=\textwidth]{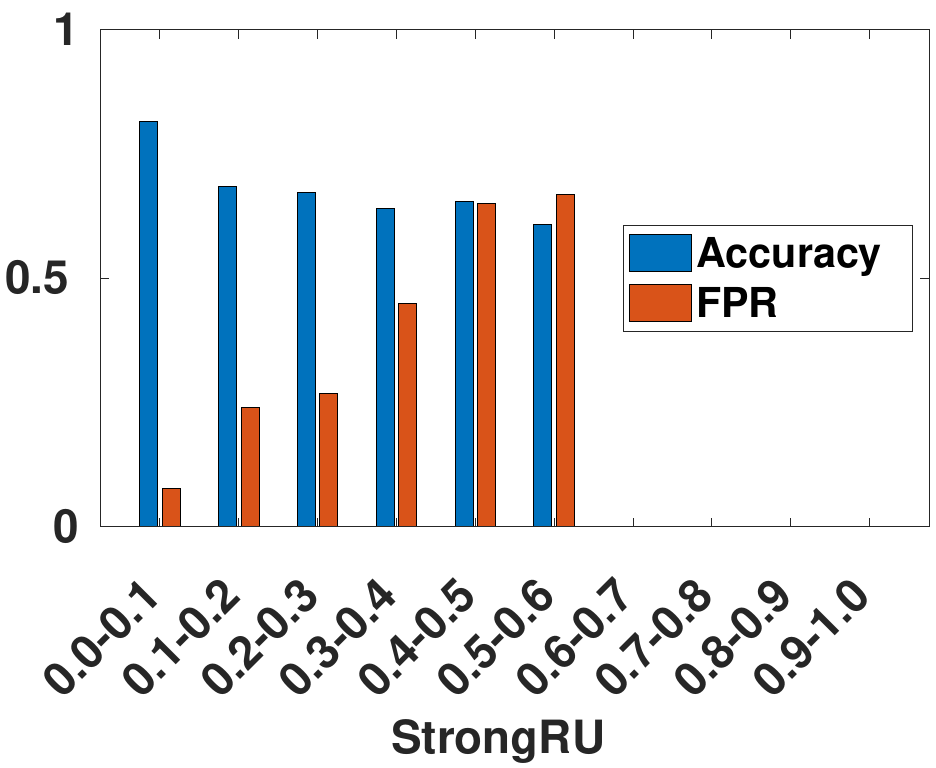}
        \vspace{-5mm}\caption{{\it DCC}, effectiveness of \sru on classification}
        \label{fig:exp-class-bar-real-sdt}
    \end{subfigure}
        \hfill
    \begin{subfigure}[t]{0.23\linewidth}
        \centering
        \includegraphics[width=\textwidth]{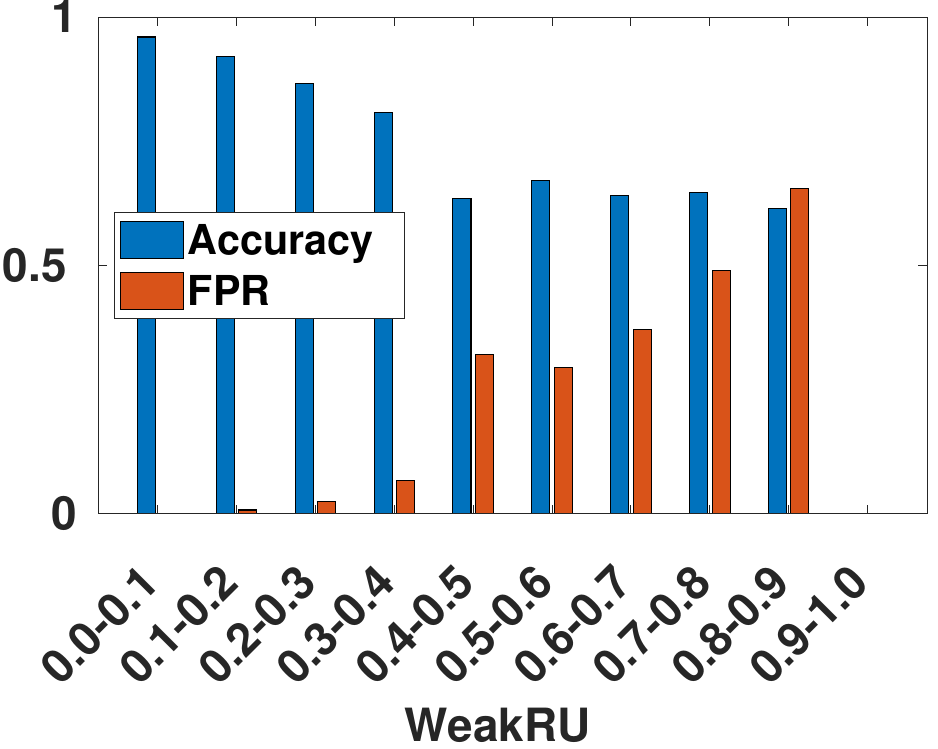}
        \vspace{-5mm}\caption{{\it DCC}, effectiveness of \wru on classification}
        \label{fig:exp-class-bar-real-wdt}
    \end{subfigure}
    \begin{subfigure}[t]{0.23\linewidth}
        \centering
        \includegraphics[width=\textwidth]{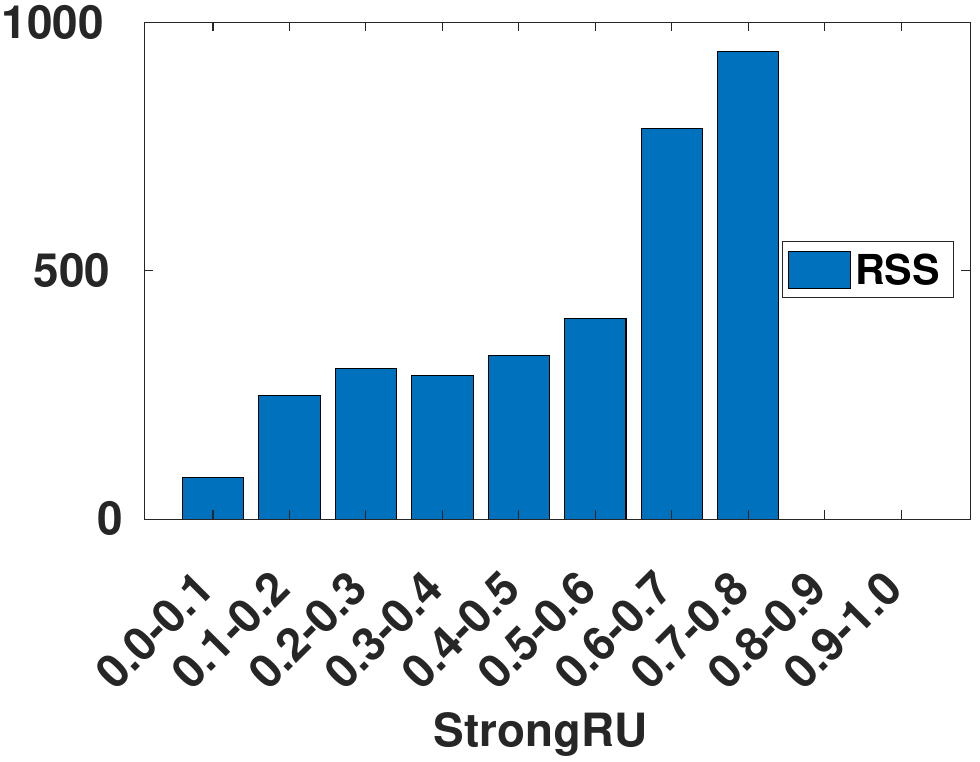}
        \vspace{-5mm}\caption{{\it RN}, effectiveness of \sru on regression}
        \label{fig:exp-reg-bar-sdt}
    \end{subfigure}
    
    \begin{subfigure}[t]{0.23\linewidth}
        \centering
        \includegraphics[width=\textwidth]{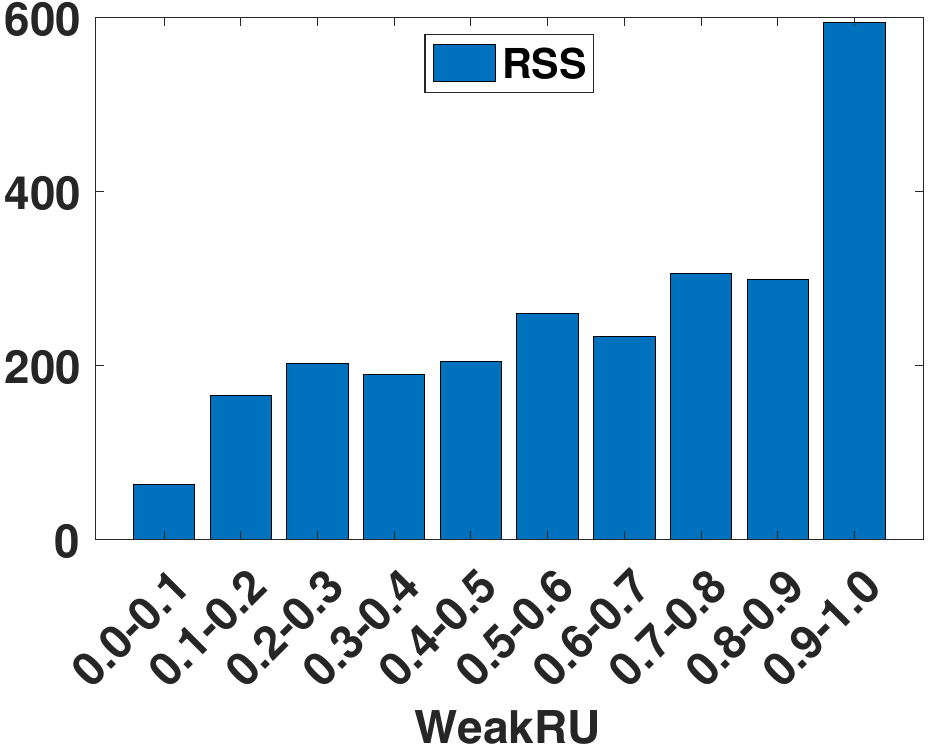}
        \vspace{-5mm}\caption{{\it RN}, effectiveness of \wru  on regression}
        \label{fig:exp-reg-bar-wdt}
    \end{subfigure}
    \hfill
    \begin{subfigure}[t]{0.23\linewidth}
        \centering
        \includegraphics[width=\textwidth]{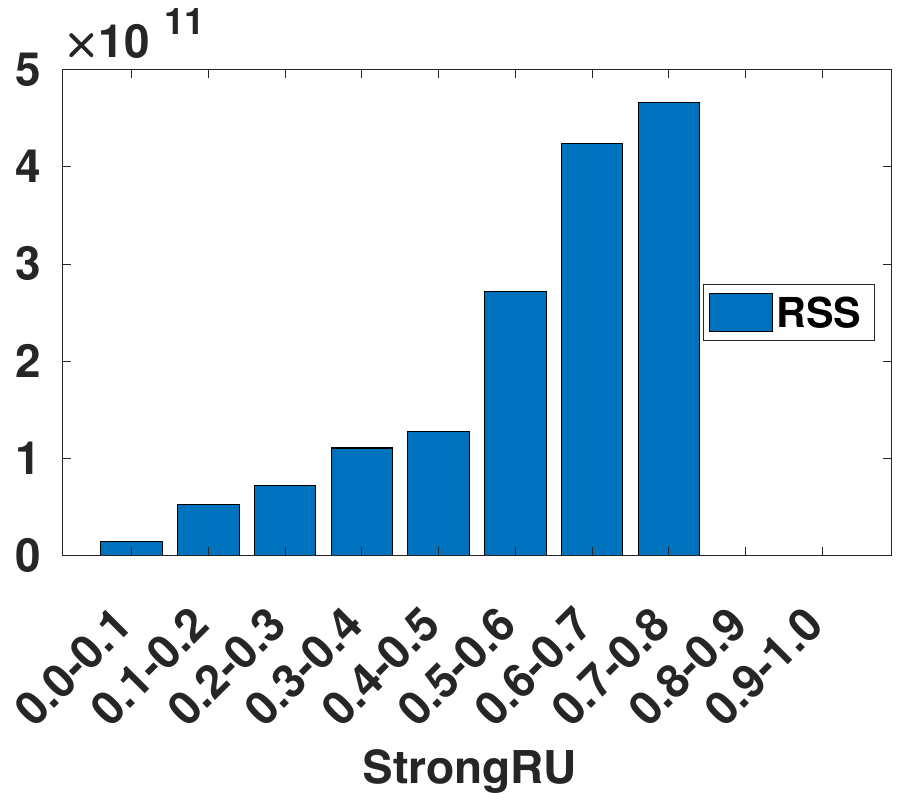}
        \vspace{-5mm}\caption{{\it HS}, effectiveness of \sru on regression}
        \label{fig:exp-housing-bar-sdt}
    \end{subfigure}
    \hfill
    \begin{subfigure}[t]{0.23\linewidth}
        \centering
        \includegraphics[width=\textwidth]{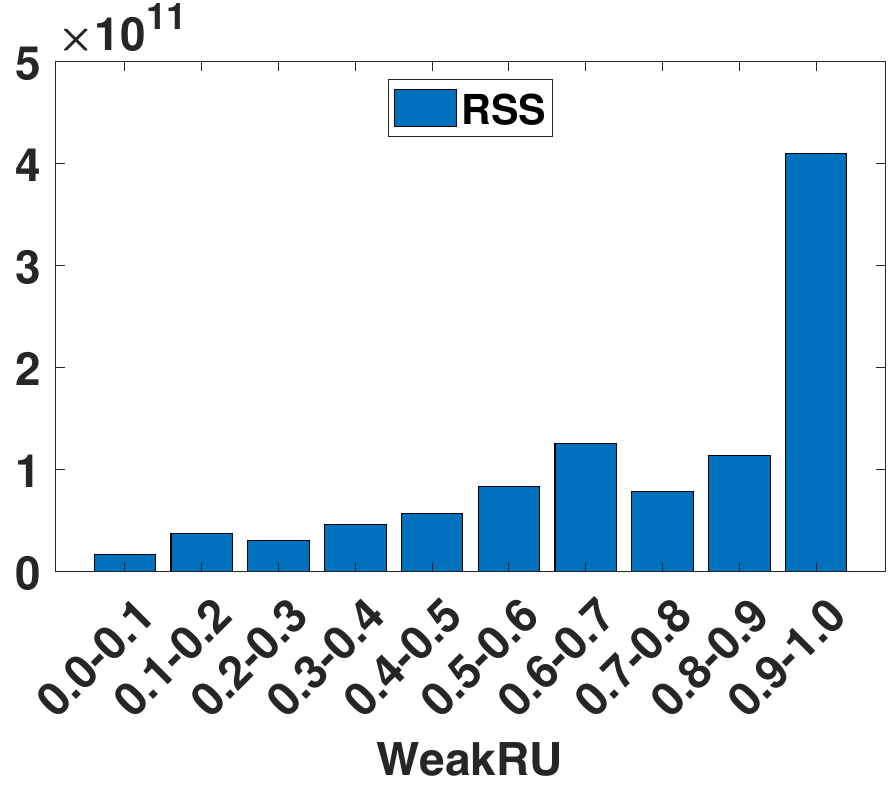}
        \vspace{-5mm}\caption{{\it HS}, effectiveness of \wru on regression}
        \label{fig:exp-housing-bar-wdt}
    \end{subfigure}
\vspace{-2mm}
\caption{proof of concept results: consistent correlation between distrust values and ML performance metrics.}\label{fig:proof:10}
\vspace{-5mm}
\end{figure*}

\vspace{-1mm}
\stitle{Visual validation}\label{exp:example} Consider the 2D data set $\dee$ shown in Figure \ref{fig:exp-cat-train}. $\dee$ is borrowed from {\it SYN} as one of the 60 data sets with the shape of the cat as the ground truth (we obtained similar results for other data sets, as reflected in the aggregate values we shall report next). 
We compute \sru and \wru values for each query point in the uniform sample over the space using the default settings. In Figures \ref{fig:exp-cat-sdt} and \ref{fig:exp-cat-wdt}, the query space is colored by assigning a tone based on the corresponding values of \sru and \wru respectively. As shown in Figure \ref{fig:exp-cat-sdt}, the untrustworthy regions are the set of query points in the space that are both outliers w.r.t. the tuples in $\dee$ {\bf and} also uncertain since the entropy in their $k$-vicinity is high. On the other hand, in Figure \ref{fig:exp-cat-wdt}, the untrustworthy regions are the set of query points that are either outliers {\bf or} uncertain. The closer the color to red, the more untrustworthy the region will be and the opposite goes for green. Next, we train an arbitrary classifier (LR in this case) on data set $\dee$ and evaluate the model's prediction. In this regard, we bucketize the uniform query space in a $10\times10$ grid and we evaluate the model for each cell to see where it falls short ({\it a.k.a.} model predictions become less reliable) then we create a heatmap on top of the grid based on the values for each cell. Results are provided in Figures \ref{fig:exp-cat-heatmap-f1} and \ref{fig:exp-cat-heatmap-fpr}. In a side-by-side comparison between the heatmaps and the colored space based on \sru and \wru (see Figures \ref{fig:exp-cat-sdt}, \ref{fig:exp-cat-wdt}, \ref{fig:exp-cat-heatmap-f1}, and \ref{fig:exp-cat-heatmap-fpr}), it is easy to see that the model failed in the regions where the \ru measures are producing large values. Finally, we generate the bar graphs with a similar procedure as later discussed in \ref{exp:effectiveness} and are shown in Figures \ref{fig:exp-cat-bar-sdt} and \ref{fig:exp-cat-bar-wdt}. As the \ru value increases, the F1 score of the model drops while the FPR rises meaning that, the model fails for the regions that are untrustworthy w.r.t. our \ru measures.
In Figure \ref{fig:exp-cat-bar-sdt}, as \sru increases, the accuracy measures for the model rapidly drop, and for the 0.5-0.6 bucket, F1 is near zero. This confirms that while \wru is a weaker warning, \sru should be viewed as a red flag.

\begin{figure*}[!htb]
    \begin{subfigure}[t]{0.23\linewidth}
        \centering
        \includegraphics[width=\textwidth]{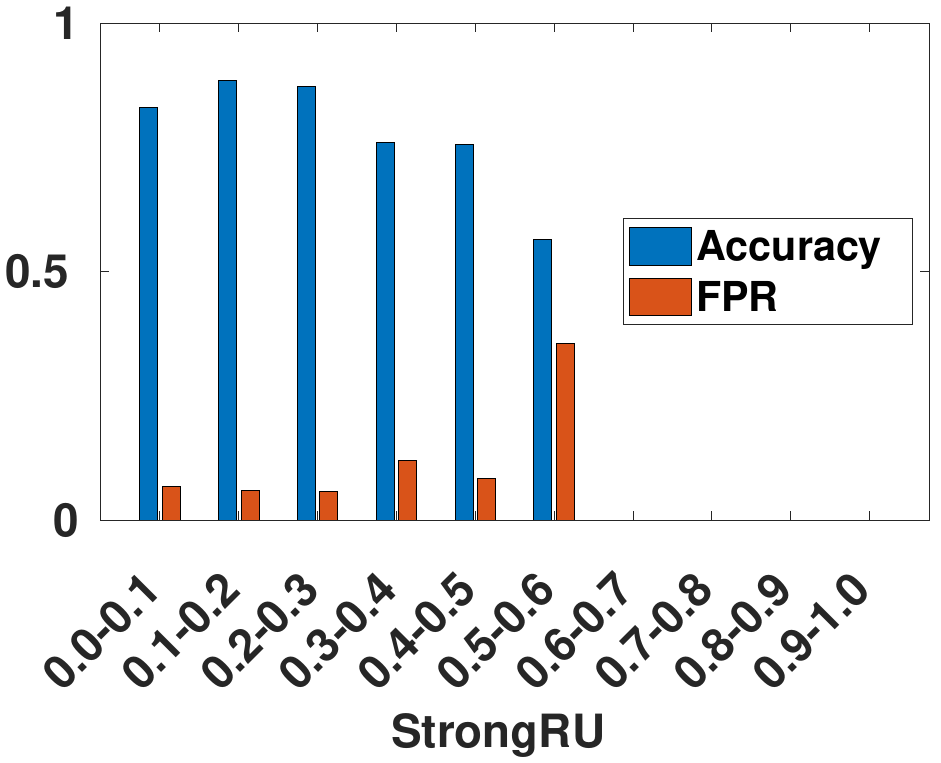}
        \vspace{-5mm}\caption{{\it AD}, effectiveness of \sru  on classification}
        \label{fig:exp-adult-sdt}
    \end{subfigure}\hfill
    \begin{subfigure}[t]{0.23\linewidth}
        \centering
        \includegraphics[width=\textwidth]{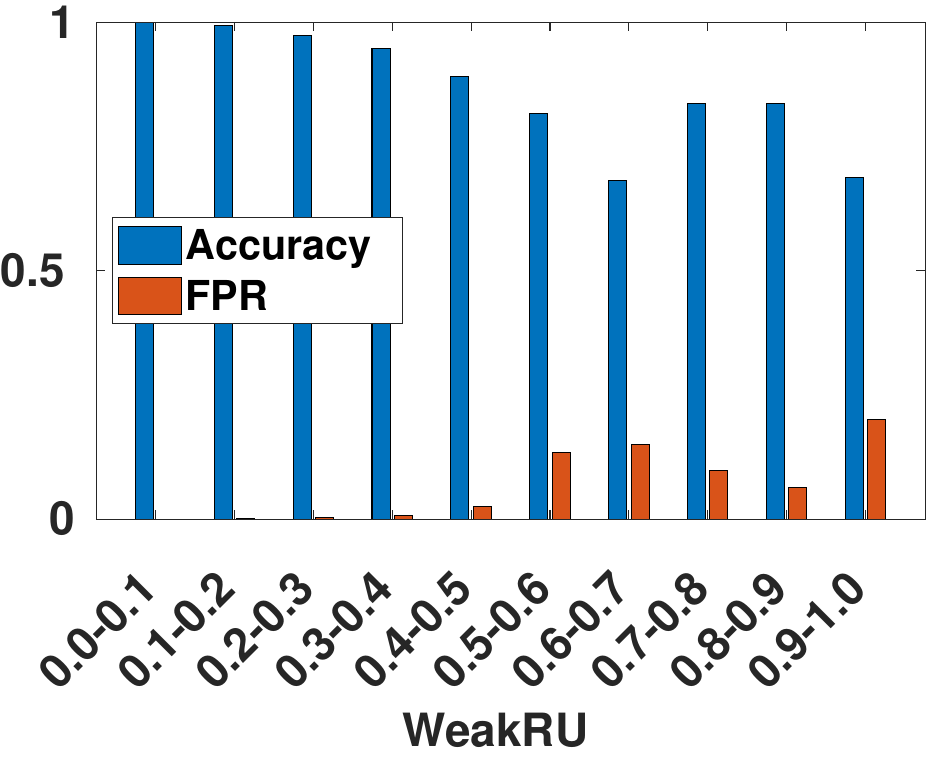}
        \vspace{-5mm}\caption{{\it AD}, effectiveness of \wru  on classification}
        \label{fig:exp-adult-wdt}
    \end{subfigure}\hfill
    \begin{subfigure}[t]{0.23\linewidth}
        \centering
        \includegraphics[width=\textwidth]{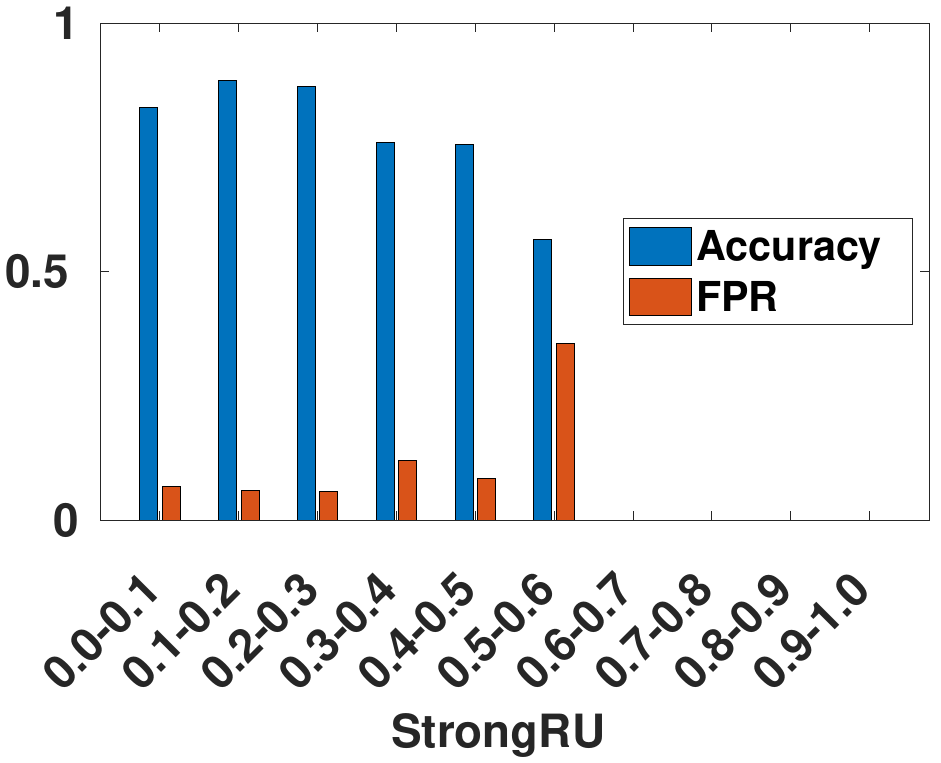}
        \vspace{-5mm}\caption{{\it AD}, effectiveness of \sru  on deep learned classification}
        \label{fig:exp-adult-sdt_dl}
    \end{subfigure}\hfill
    \begin{subfigure}[t]{0.23\linewidth}
        \centering
        \includegraphics[width=\textwidth]{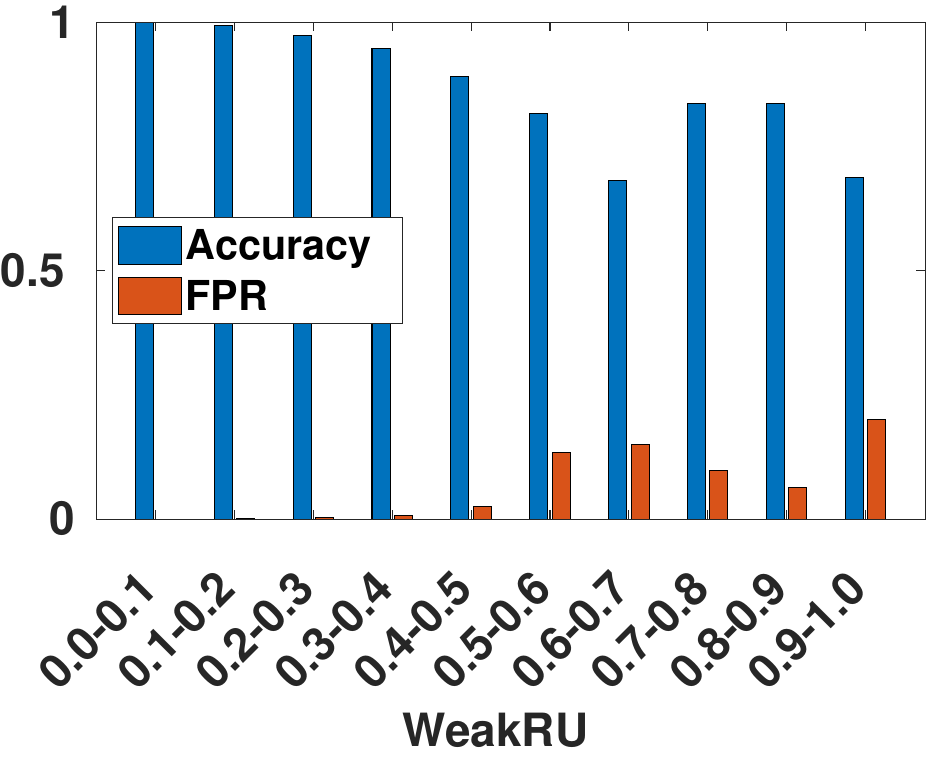}
        \vspace{-5mm}\caption{{\it AD}, \wru on deep learned classification}
        \label{fig:exp-adult-wdt_dl}
    \end{subfigure}

    \begin{subfigure}[t]{0.23\linewidth}
        \centering
        \includegraphics[width=\textwidth]{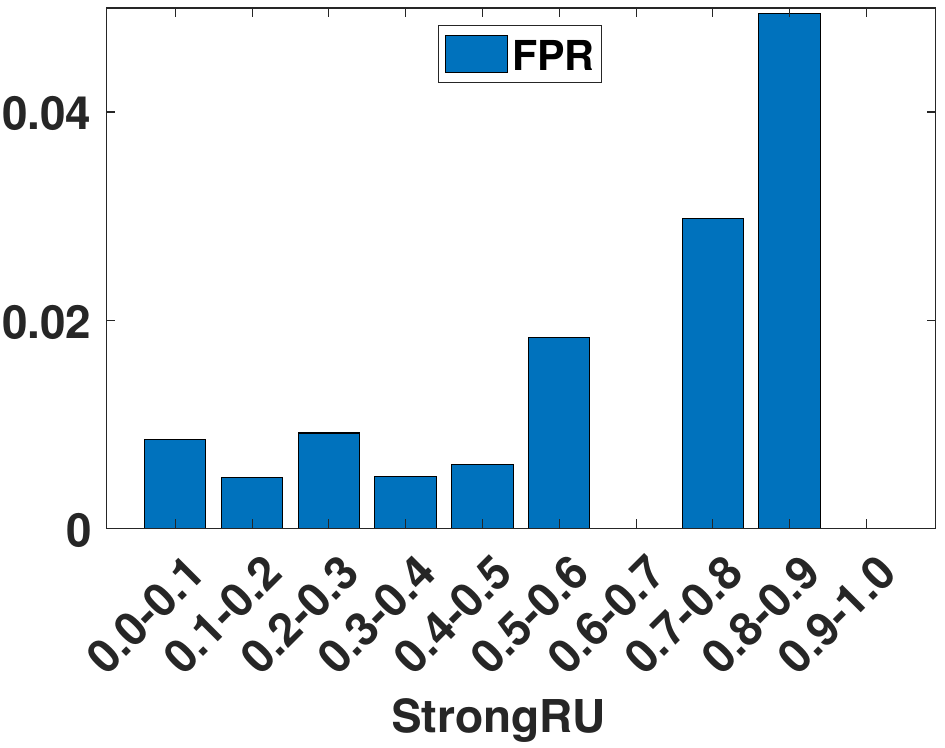}
        \vspace{-5mm}\caption{{\it RS}, effectiveness of \sru  on  deep learned classification}
        \label{fig:exp-reg-real-sim-sdt}
    \end{subfigure}\hfill
    \begin{subfigure}[t]{0.23\linewidth}
        \centering
        \includegraphics[width=\textwidth]{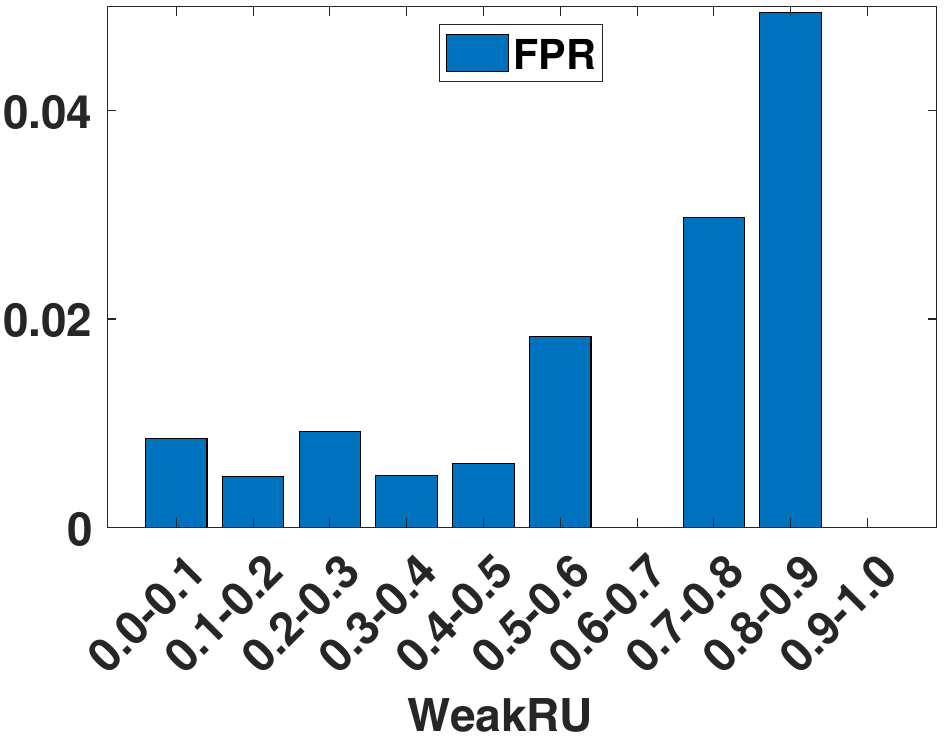}
        \vspace{-5mm}\caption{{\it RS}, effectiveness of \wru  on deep learned classification}
        \label{fig:fig:exp-reg-real-sim-wdt}
    \end{subfigure}\hfill
    \begin{subfigure}[t]{0.23\linewidth}
        \centering
        \includegraphics[width=\textwidth]{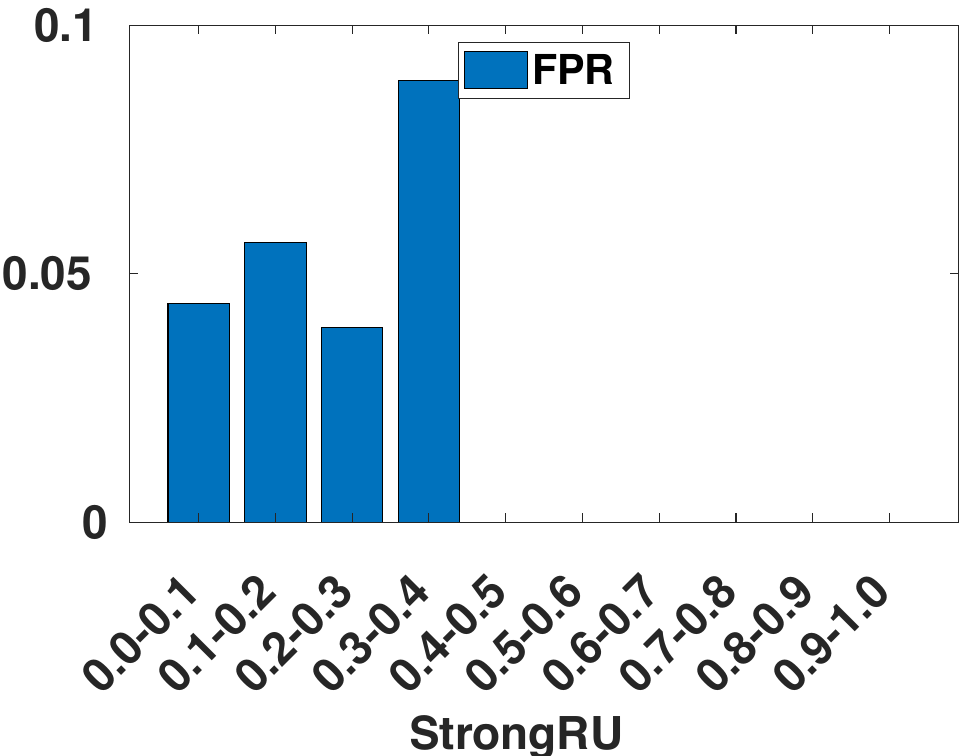}
        \vspace{-5mm}\caption{{\it GS}, effectiveness of \sru on deep learned classification}
        \label{fig:exp-reg-gisette-sdt_dl}
    \end{subfigure}\hfill
    \begin{subfigure}[t]{0.23\linewidth}
        \centering
        \includegraphics[width=\textwidth]{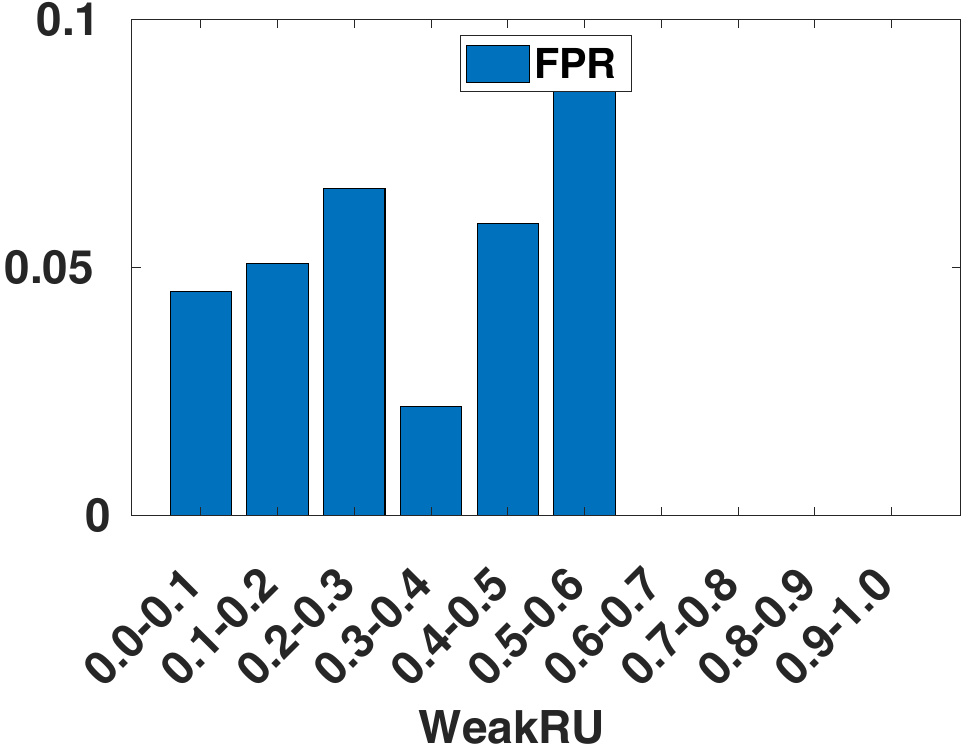}
        \vspace{-5mm}\caption{{\it GS}, \wru  on deep learned classification}
        \label{fig:fig:exp-reg-gisette-wdt_dl}
    \end{subfigure}
    \begin{subfigure}[t]{0.23\linewidth}
        \centering
        \includegraphics[width=\textwidth]{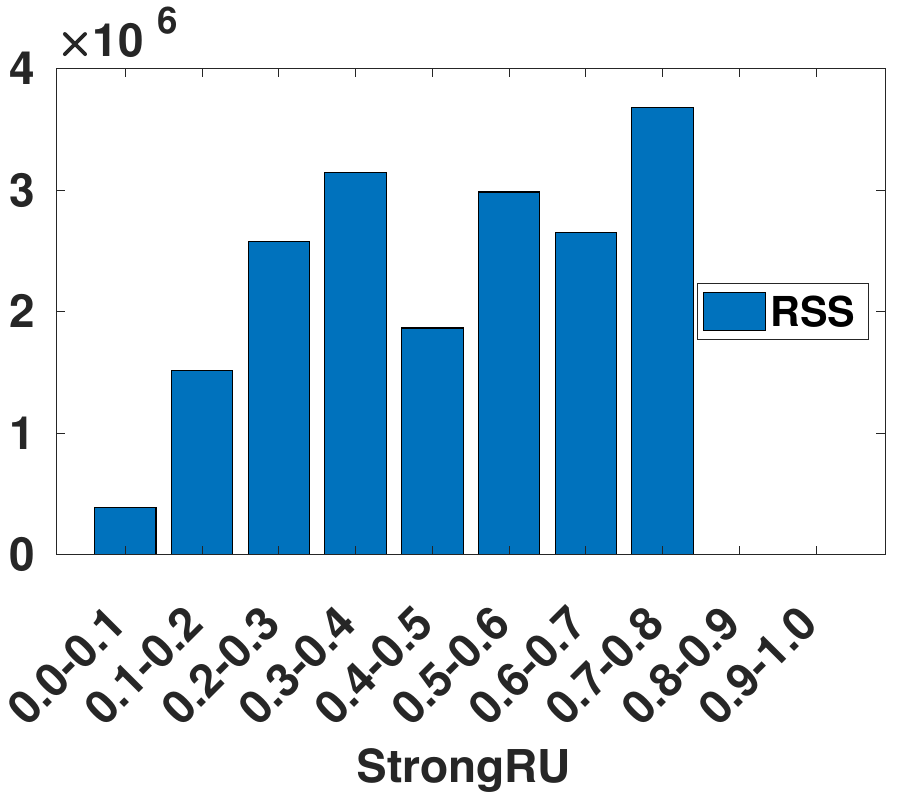}
        \vspace{-5mm}\caption{{\it DI}, effectiveness of \sru  on regression}
        \label{fig:exp-reg-diamond-sdt}
    \end{subfigure}\hfill
    \begin{subfigure}[t]{0.23\linewidth}
        \centering
        \includegraphics[width=\textwidth]{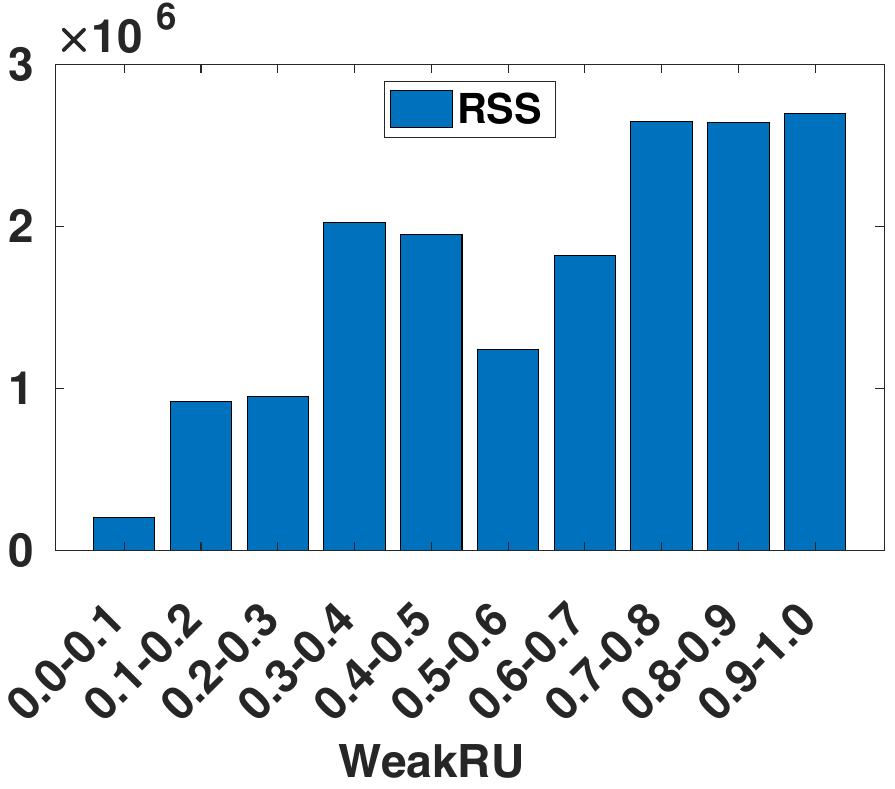}
        \vspace{-5mm}\caption{{\it DI}, effectiveness of \wru  on regression}
        \label{fig:fig:exp-reg-diamond-wdt}
    \end{subfigure}\hfill
    \begin{subfigure}[t]{0.23\linewidth}
        \centering
        \includegraphics[width=\textwidth]{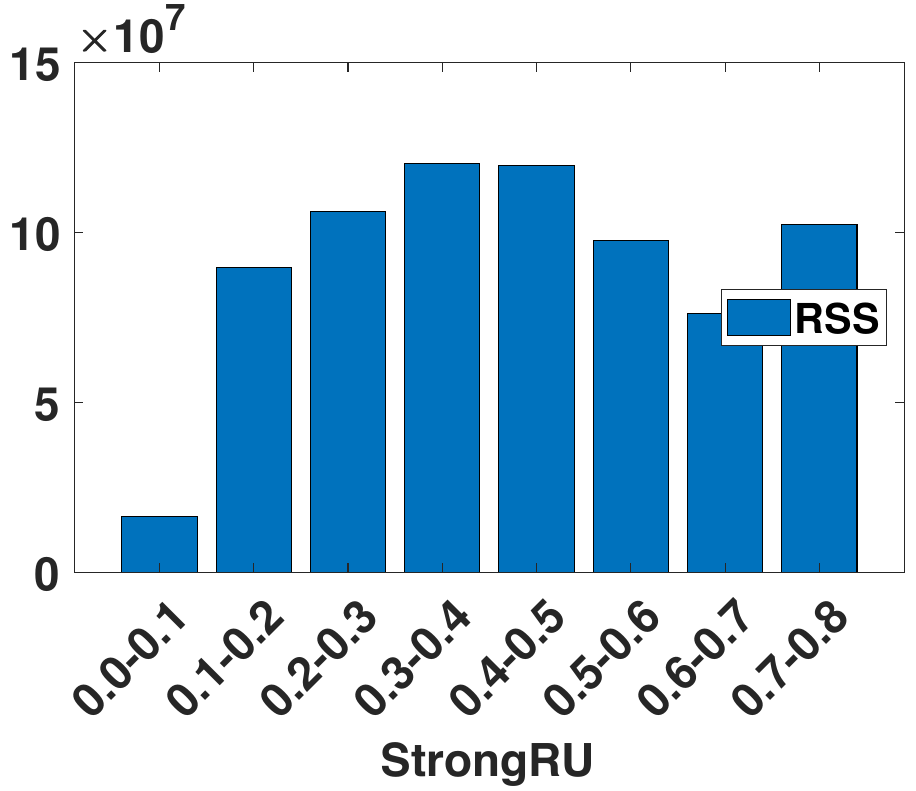}
        \vspace{-5mm}\caption{{\it DI}, effectiveness of \sru on deep learned regression}
        \label{fig:exp-reg-diamond-sdt_dl}
    \end{subfigure}\hfill
    \begin{subfigure}[t]{0.23\linewidth}
        \centering
        \includegraphics[width=\textwidth]{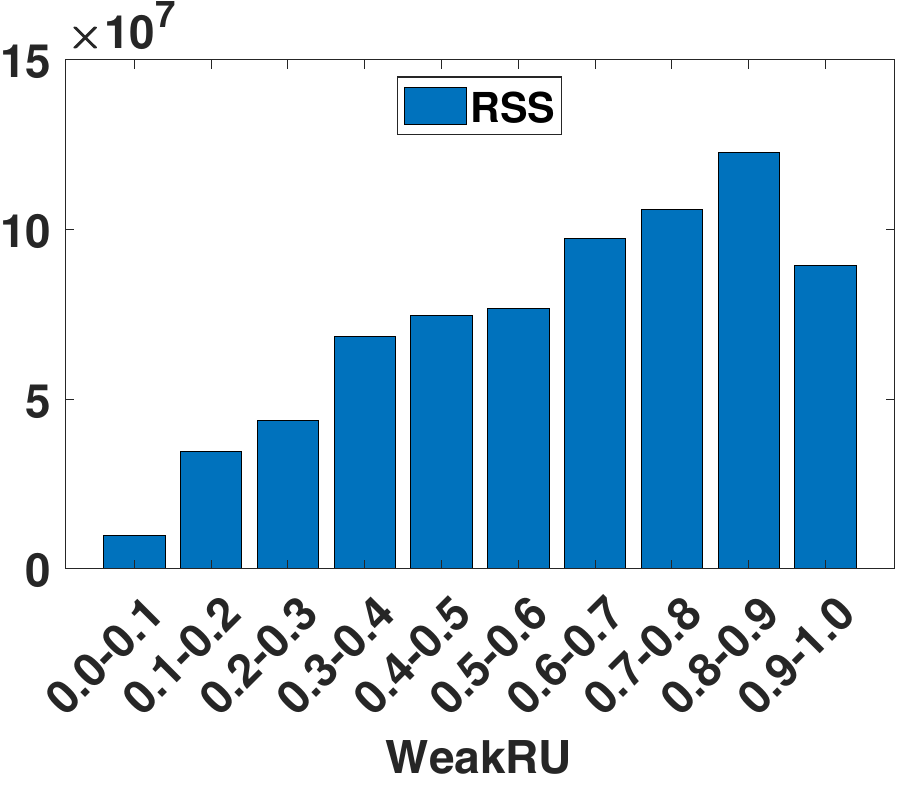}
        \vspace{-5mm}\caption{{\it DI}, \wru  on deep learned regression}
        \label{fig:fig:exp-reg-diamond-wdt_dl}
    \end{subfigure}
    
\vspace{-4mm}
\caption{additional proof of concept results on AD, DI, RS and GS datasets.
}\label{fig:proof:11}
\vspace{-2mm}
\end{figure*}


\begin{figure}[!tbp] 
\centering
    \begin{subfigure}[t]{0.49\linewidth}
        \centering
        \includegraphics[width=\textwidth]{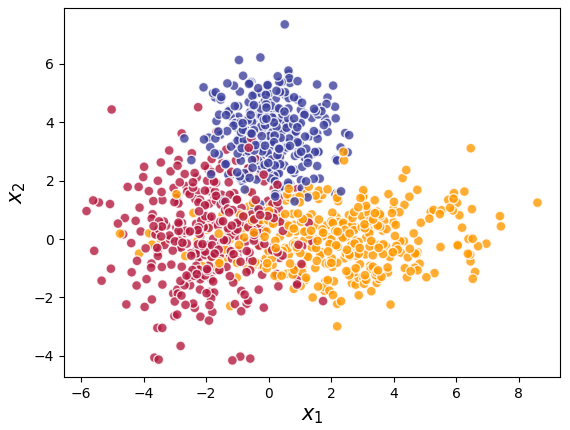}
        \vspace{-6mm}\caption{data set $\dee$ created using 3 Gaussian distributions.}
        \label{fig:ext_1}
    \end{subfigure}
    \hfill
    \begin{subfigure}[t]{0.49\linewidth}
        \centering
        \includegraphics[width=\textwidth]{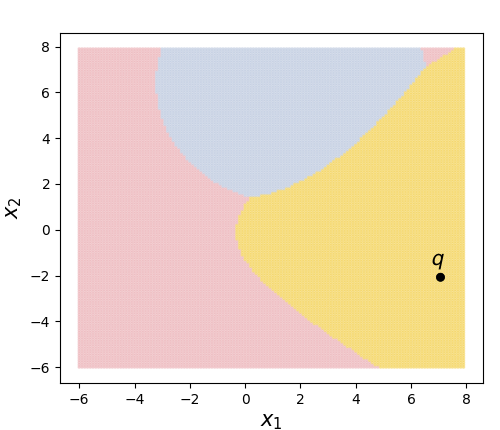}
        \vspace{-6mm}\caption{predicted labels using Gaussian Naive Bayes.}
        \label{fig:ext_2}
    \end{subfigure}
    \hfill
    \begin{subfigure}[t]{0.49\linewidth}
        \centering
        \includegraphics[width=\textwidth]{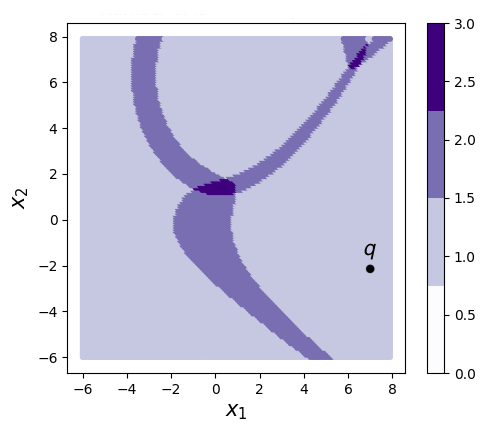}
        \vspace{-6mm}\caption{number of labels for $\alpha=0.05$, softmax conformal score.}
        \label{fig:ext_1_3}
    \end{subfigure}
    \hfill
        \begin{subfigure}[t]{0.49\linewidth}
        \centering
        \includegraphics[width=\textwidth]{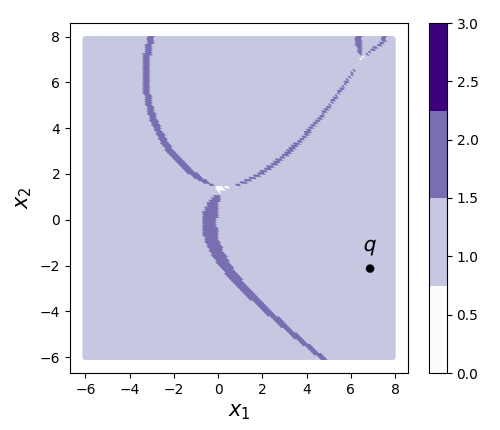}
        \vspace{-6mm}\caption{number of labels for $\alpha=0.1$, softmax conformal score.}
        \label{fig:ext_1_2}
    \end{subfigure}
    \begin{subfigure}[t]{0.49\linewidth}
        \centering
        \includegraphics[width=\textwidth]{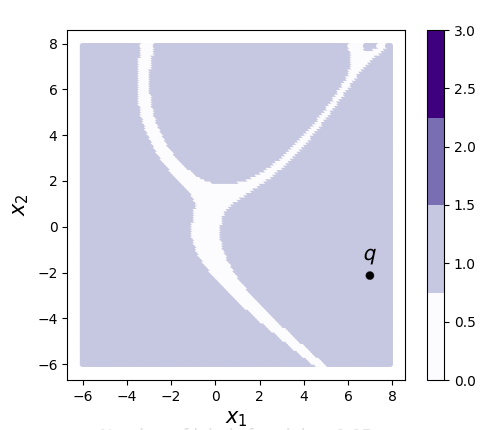}
        \vspace{-6mm}\caption{number of labels for $\alpha=0.2$, softmax conformal score: null pred. sets for uncertain regions.}
        \label{fig:ext_1_1}
    \end{subfigure}
    \hfill
    \begin{subfigure}[t]{0.49\linewidth}
        \centering
        \includegraphics[width=\textwidth]
        {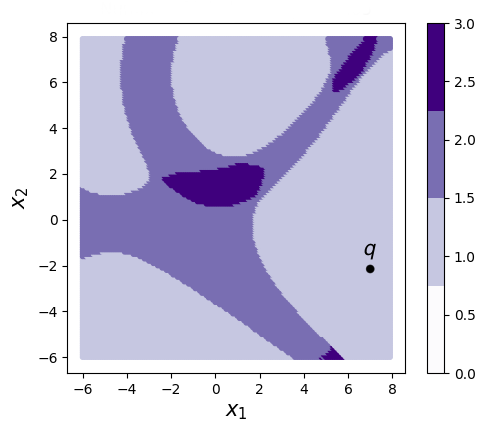}
        \vspace{-6mm}\caption{number of labels for $\alpha=0.05$, cumulative softmax conformal score: wide pred. sets for uncertain regions.}
        \label{fig:ext_2_3}
    \end{subfigure}
    \hfill
    \begin{subfigure}[t]{0.49\linewidth}
        \centering
        \includegraphics[width=\textwidth]{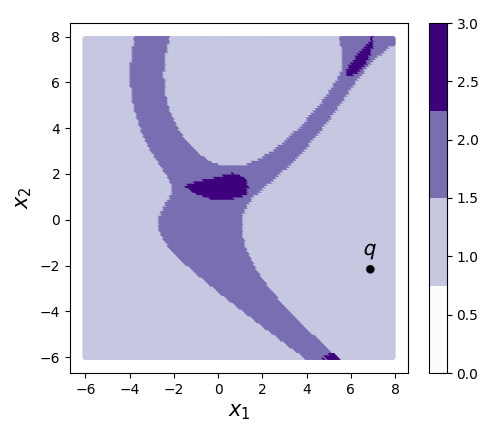}
        \vspace{-6mm}\caption{number of labels for $\alpha=0.1$, cumulative softmax conformal score.}
        \label{fig:ext_2_2}
    \end{subfigure}
    \hfill
    \begin{subfigure}[t]{0.49\linewidth}
        \centering
        \includegraphics[width=\textwidth]{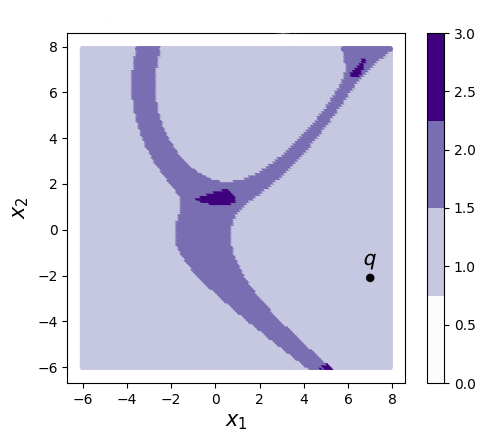}
        \vspace{-6mm}\caption{number of labels for $\alpha=0.2$, cumulative softmax conformal score.}
        \label{fig:ext_2_1}
    \end{subfigure}

    \begin{subfigure}[t]{0.46\linewidth}
        \centering
        \includegraphics[width=\textwidth]{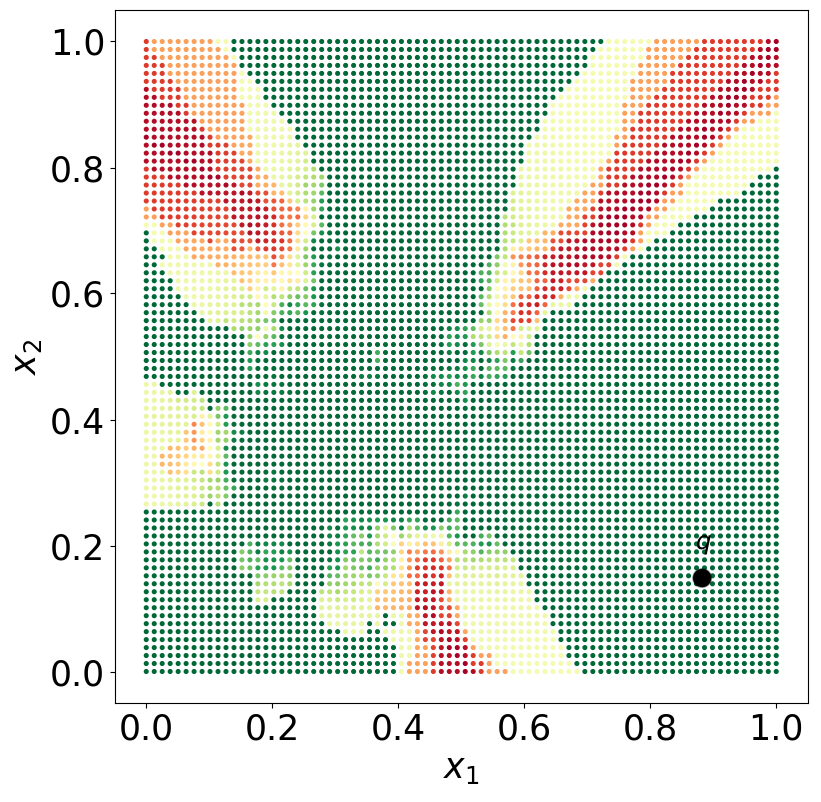}
        \vspace{-6mm}\caption{query space colored based on \sru values with regards to $\dee$.}
        \label{fig:ext_3}
    \end{subfigure}
    \hfill
    \begin{subfigure}[t]{0.46\linewidth}
        \centering
        \includegraphics[width=\textwidth]{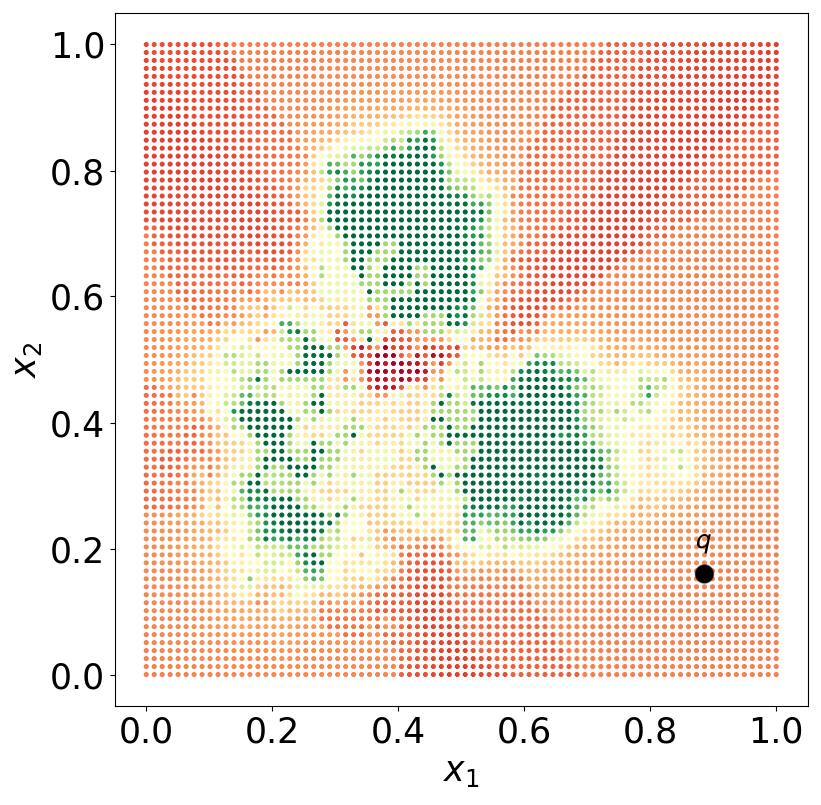}
        \vspace{-6mm}\caption{query space colored based on \wru values with regards to $\dee$.}
        \label{fig:ext_4}
    \end{subfigure} 
    \hfill
\vspace{-2mm}
\caption{conformal prediction fails to detect the prediction unreliability for not well-represented point $\qu$ while \wru correctly captures such unreliability.}
\vspace{-3mm}
\end{figure}

\begin{figure*}[!tbh] 
\begin{minipage}[t]{\linewidth}
\centering
    \begin{subfigure}[t]{0.3\linewidth}
        \centering
        \includegraphics[width=.9\textwidth]{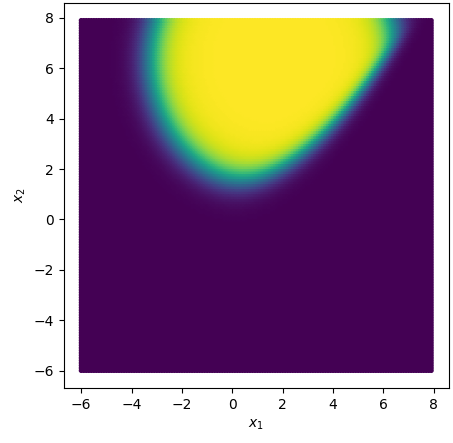}
        \vspace{-3mm}\caption{Prediction probabilities generated by the model for \textit{blue} class.}
        \label{fig:pred_prob_1}
    \end{subfigure}\hfill
    \begin{subfigure}[t]{0.3\linewidth}
        \centering
        \includegraphics[width=.9\textwidth]{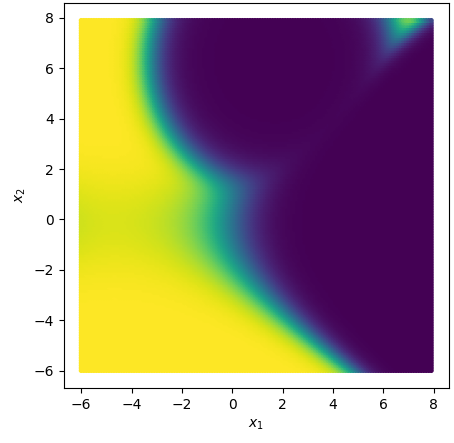}
        \vspace{-3mm}\caption{prediction probabilities generated by the model for \textit{red} class}
        \label{fig:pred_prob_2}
    \end{subfigure}\hfill
    \begin{subfigure}[t]{0.3\linewidth}
        \centering
        \includegraphics[width=.9\textwidth]{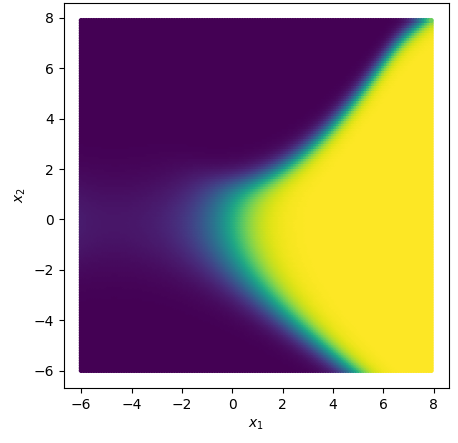}
        \vspace{-3mm}\caption{prediction probabilities generated by the model for \textit{orange} class}
        \label{fig:pred_prob_3}
    \end{subfigure}
\end{minipage}
\vspace{-2mm}
\caption{prediction probabilities of classifiers trained on $\dee$ in Figure \ref{fig:ext_1} fails for query points that are not well-represented.}
\vspace{-4mm}
\end{figure*}

\begin{figure}[!tbh]
\centering
\includegraphics[width=0.33\textwidth]{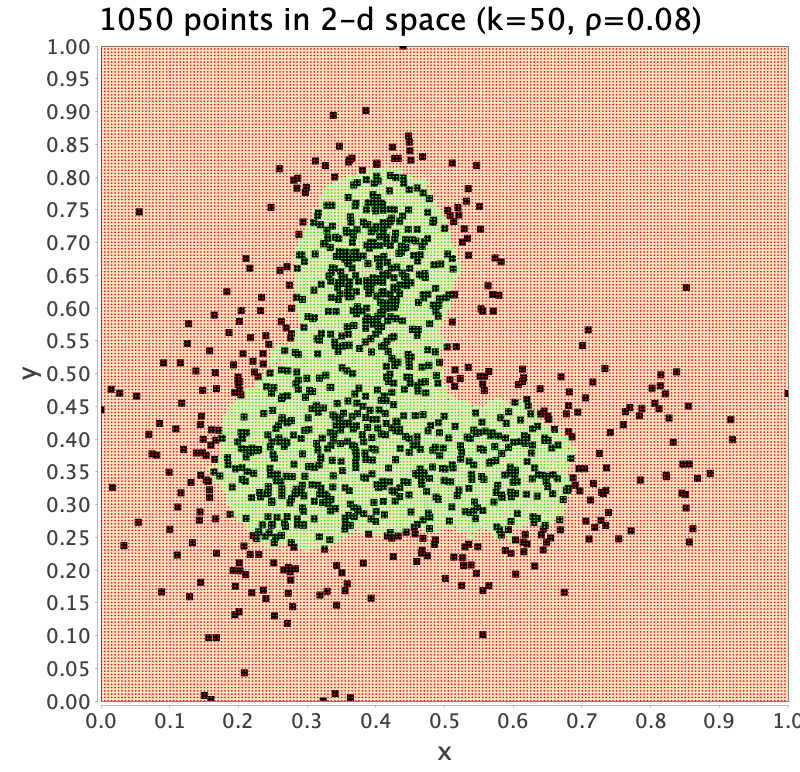}
\vspace{-3mm}\caption{data coverage on data set $\dee$ in Figure \ref{fig:ext_1} fails to capture the unreliability associated with the query points in uncertain regions. The training data ($\dee$) are highlighted as black dots. The regions highlighted in red and green comprise the uncovered and covered regions respectively. Any query point belonging to the green (red) region is considered (un)covered.}

\label{fig:coverage_comparison}
\end{figure}

\begin{figure}[!tbh]
\centering
\vspace{-5mm}\includegraphics[width=0.36\textwidth]{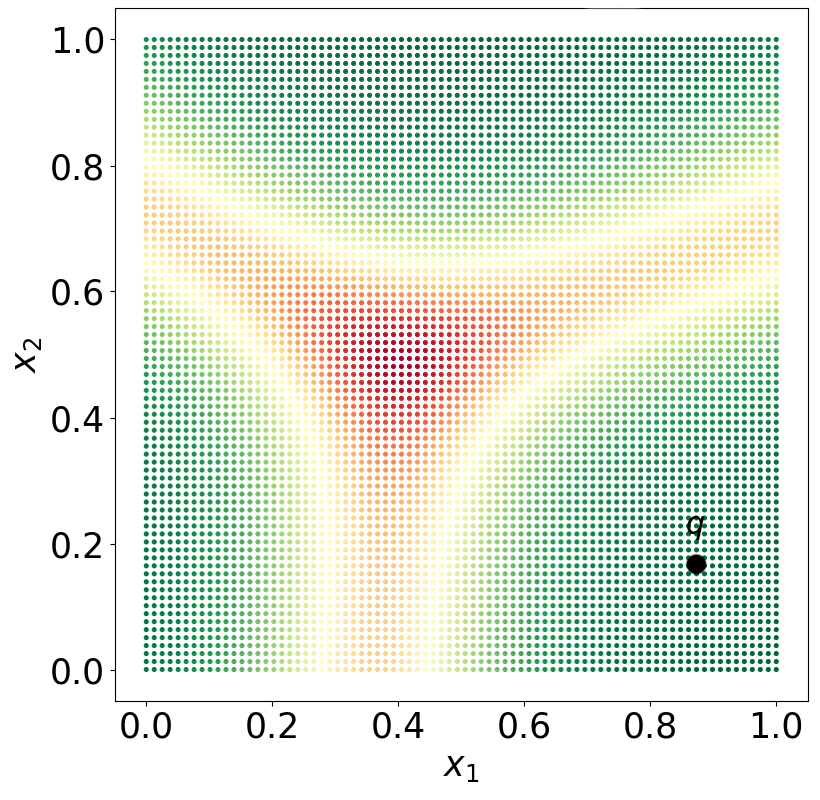}
\vspace{-5mm}\caption{query space colored based on model's Shannon entropy values w.r.t. $\dee$.}
\label{fig:ext_5}
\vspace{-5mm}
\end{figure}

\begin{figure*}[!tbh]
\begin{subfigure}[t]{0.23\linewidth}
        \centering
        \includegraphics[width=\textwidth]{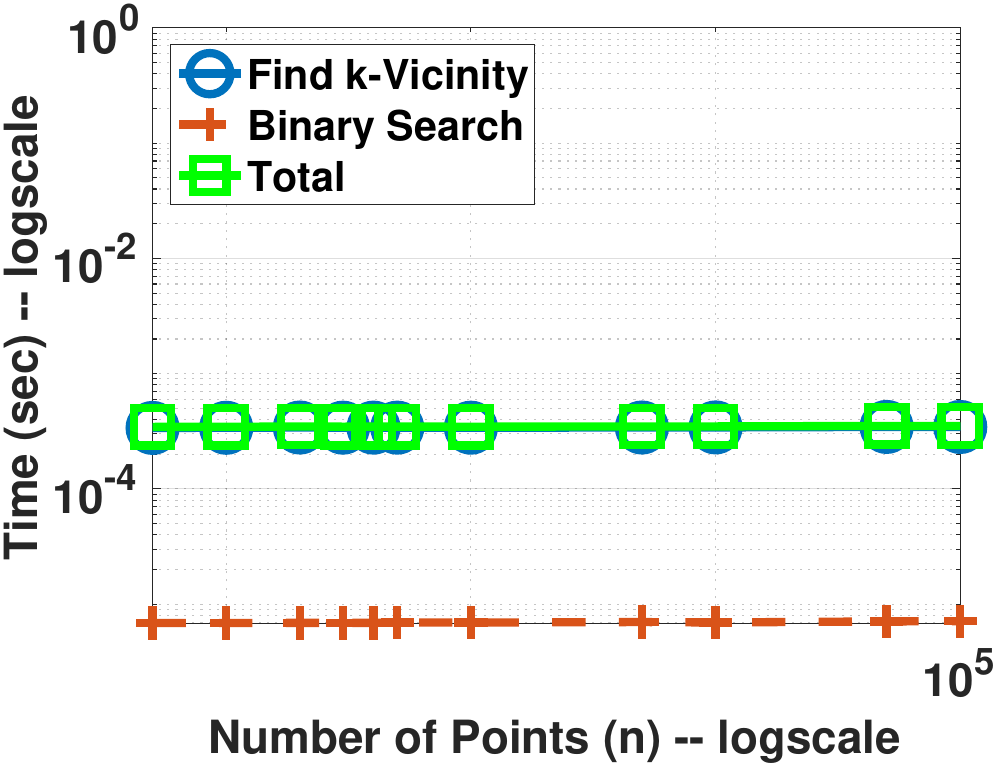}
        \caption{{\it DCC}, impact of $n$ on query time}
        \label{fig:exp-query-time-var-num-point}
    \end{subfigure}
    \hfill
    \begin{subfigure}[t]{0.23\linewidth}
        \centering
        \includegraphics[width=\textwidth]{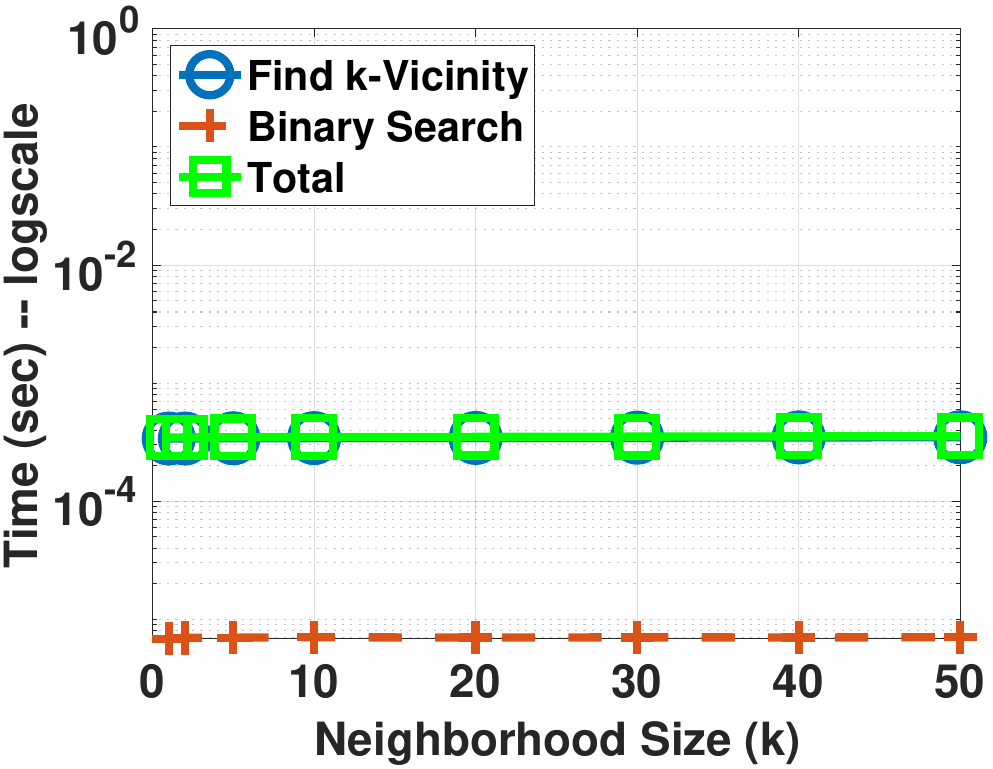}
        \caption{{\it DCC}, impact of $k$ on query time}
        \label{fig:exp-query-time-var-num-neighbors}
    \end{subfigure}
    \hfill
    \begin{subfigure}[t]{0.23\linewidth}
        \centering
        \includegraphics[width=\textwidth]{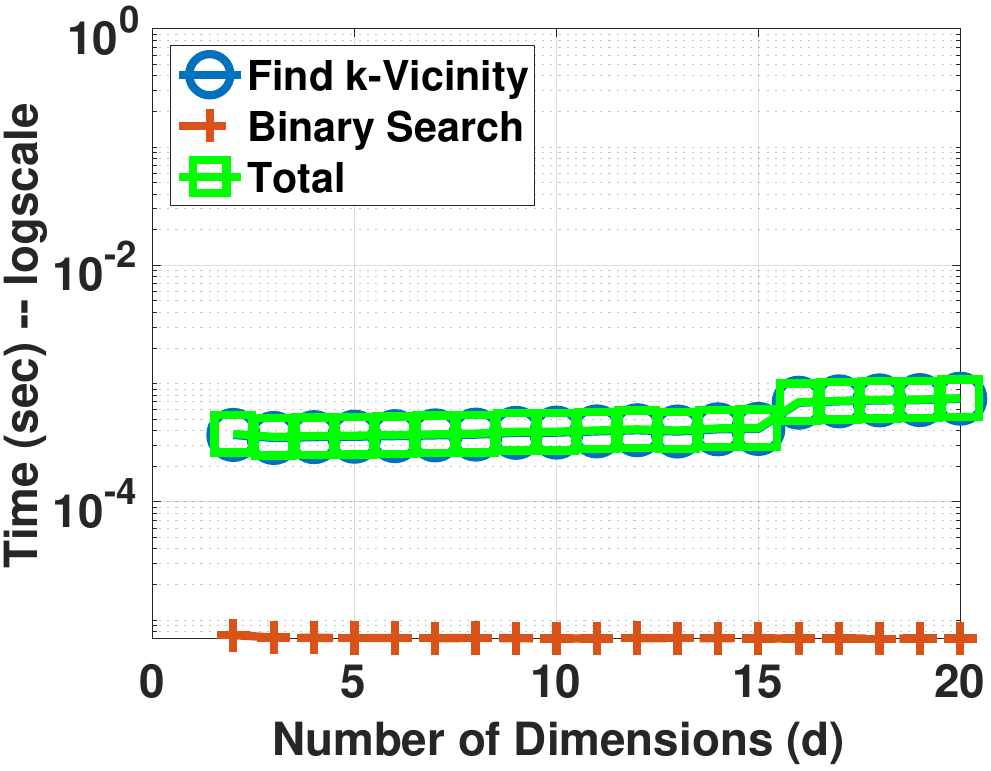}
        \caption{{\it DCC}, impact of $d$ on query time}
        \label{fig:exp-query-time-var-num-dim}
    \end{subfigure}
    \begin{subfigure}[t]{0.23\linewidth}
        \centering
        \includegraphics[width=\textwidth]{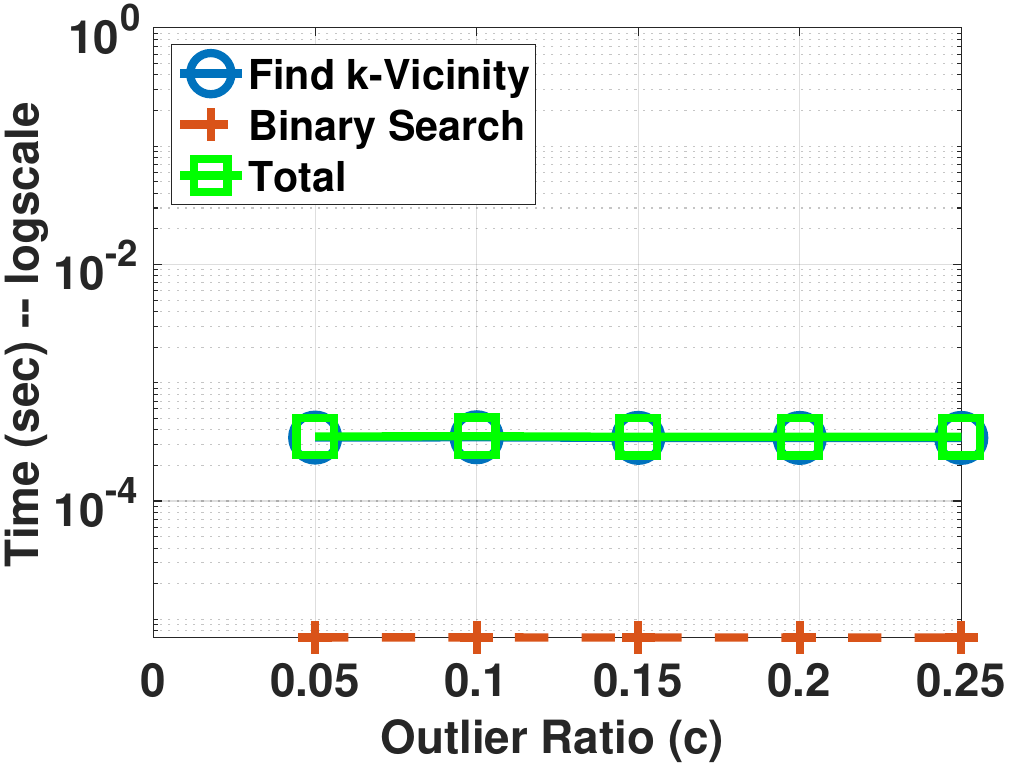}
        \caption{{\it DCC}, impact of $c$ on query time}
        \label{fig:exp-query-time-var-num-outlier_ratio}
    \end{subfigure}
    \hfill
    \begin{subfigure}[t]{0.23\linewidth}
        \centering
        \includegraphics[width=\textwidth]{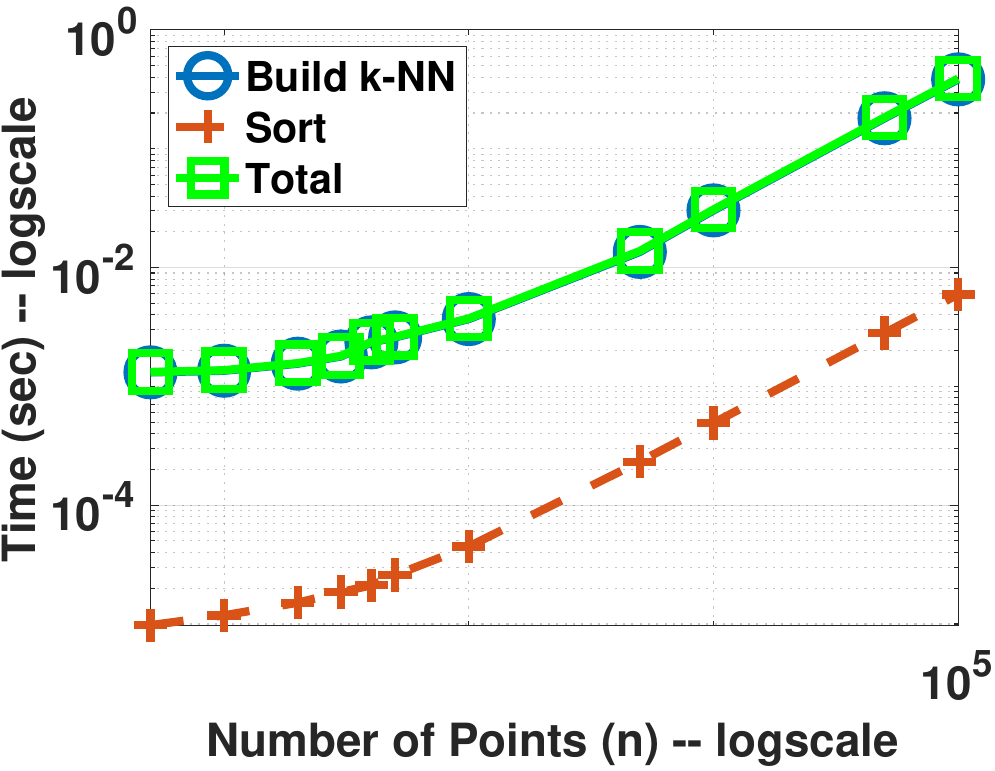}
        \caption{{\it DCC}, impact of $n$ on preprocessing time}
        \label{fig:exp-pre-time-var-num-point}
    \end{subfigure}
    \hfill
    \begin{subfigure}[t]{0.23\linewidth}
        \centering
        \includegraphics[width=\textwidth]{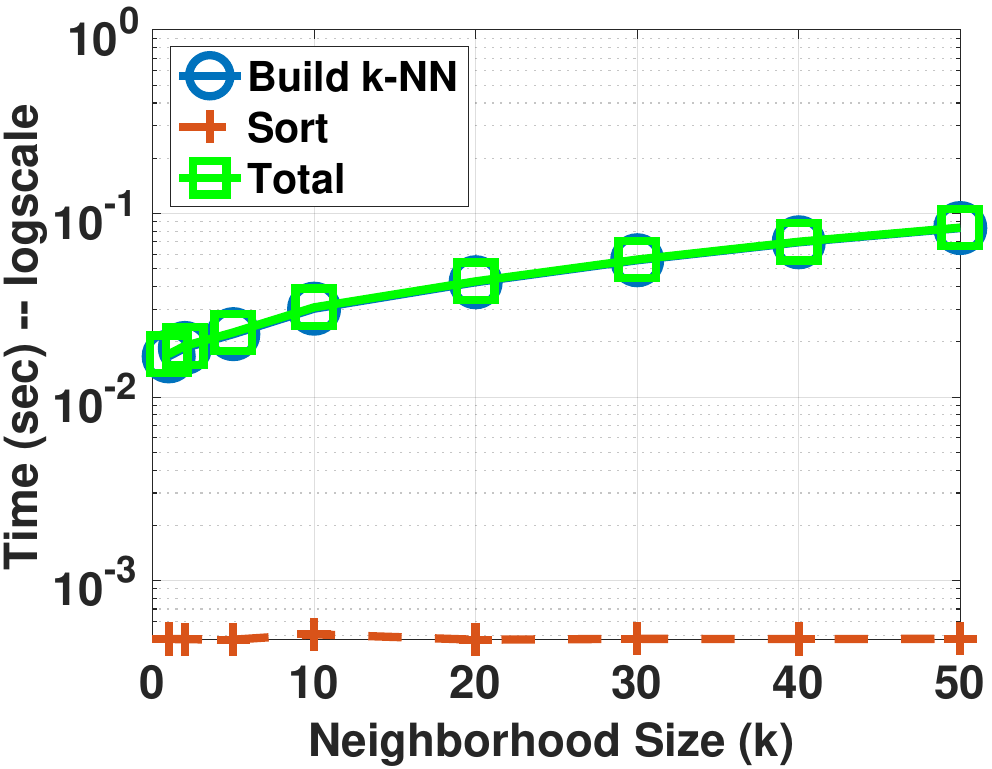}
        \caption{{\it DCC}, impact of $k$ on preprocessing time}
        \label{fig:exp-pre-time-var-num-neighbors}
    \end{subfigure}
    \hfill
    \begin{subfigure}[t]{0.23\linewidth}
        \centering
        \includegraphics[width=\textwidth]{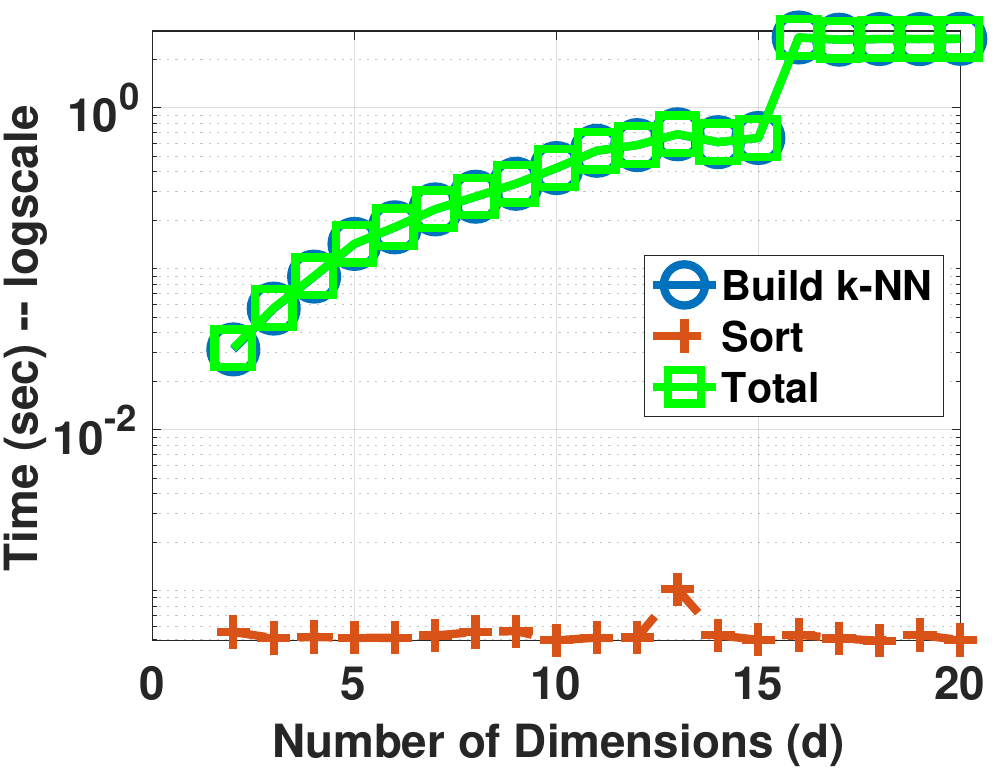}
        \caption{{\it DCC}, impact of $d$ on preprocessing time}
        \label{fig:exp-pre-time-var-num-dim}
    \end{subfigure}
\caption{performance evaluation results}
\end{figure*}

\begin{figure}[!tbh] 
    \begin{subfigure}[t]{0.49\linewidth}
        \centering
        \includegraphics[width=\textwidth]{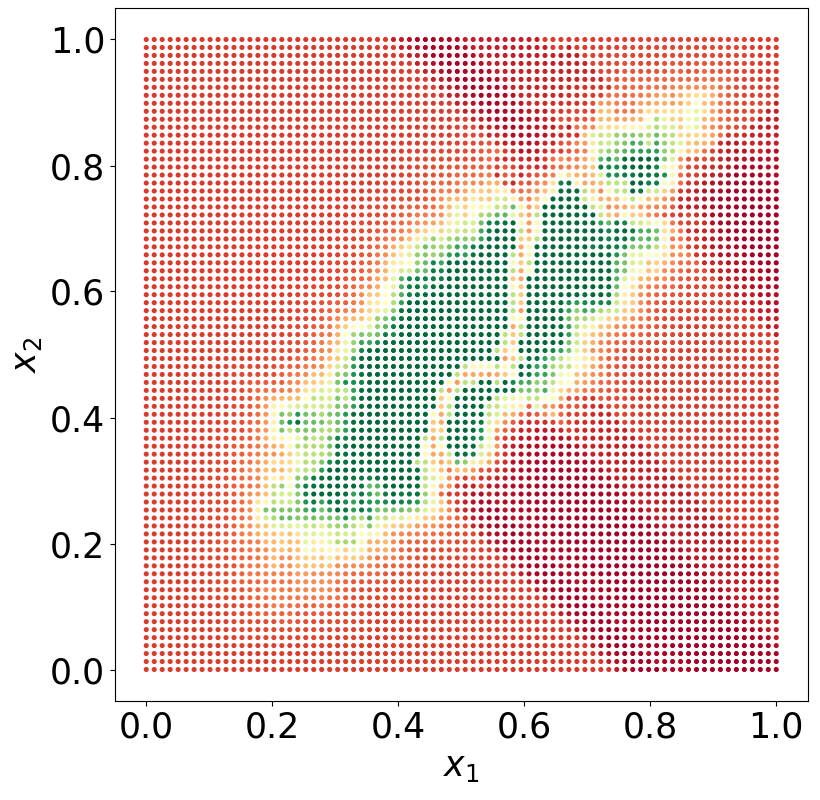}
        \vspace{-5mm}\caption{\wru, chebyshev}
        \label{fig:metric-wdt-3}
    \end{subfigure}
    \hfill
    \begin{subfigure}[t]{0.49\linewidth}
        \centering
        \includegraphics[width=\textwidth]{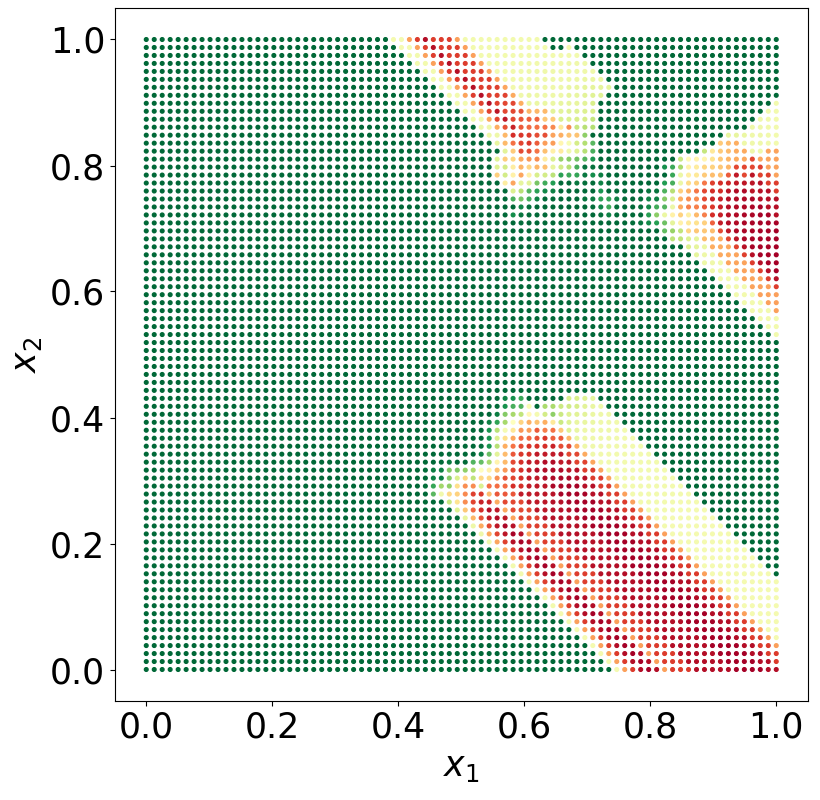}
        \vspace{-5mm}\caption{\sru, chebyshev}
        \label{fig:metric-sdt-3}
    \end{subfigure}
    \hfill
    \begin{subfigure}[t]{0.49\linewidth}
        \centering
        \includegraphics[width=\textwidth]{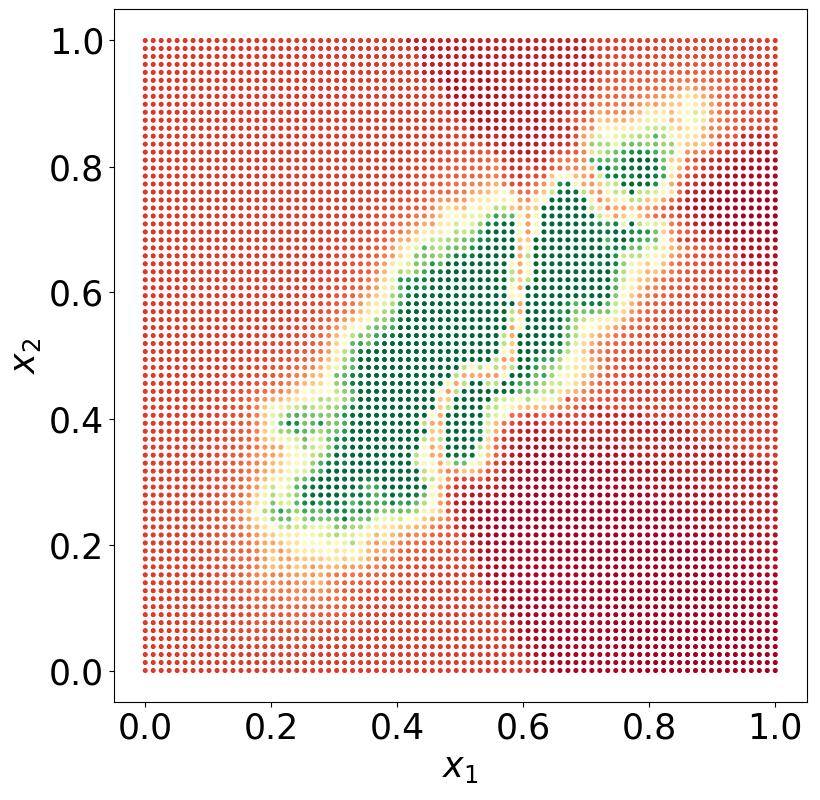}
        \vspace{-5mm}\caption{\wru, euclidean}
        \label{fig:metric-wdt-6}
    \end{subfigure}
    \hfill
    \begin{subfigure}[t]{0.49\linewidth}
        \centering
        \includegraphics[width=\textwidth]{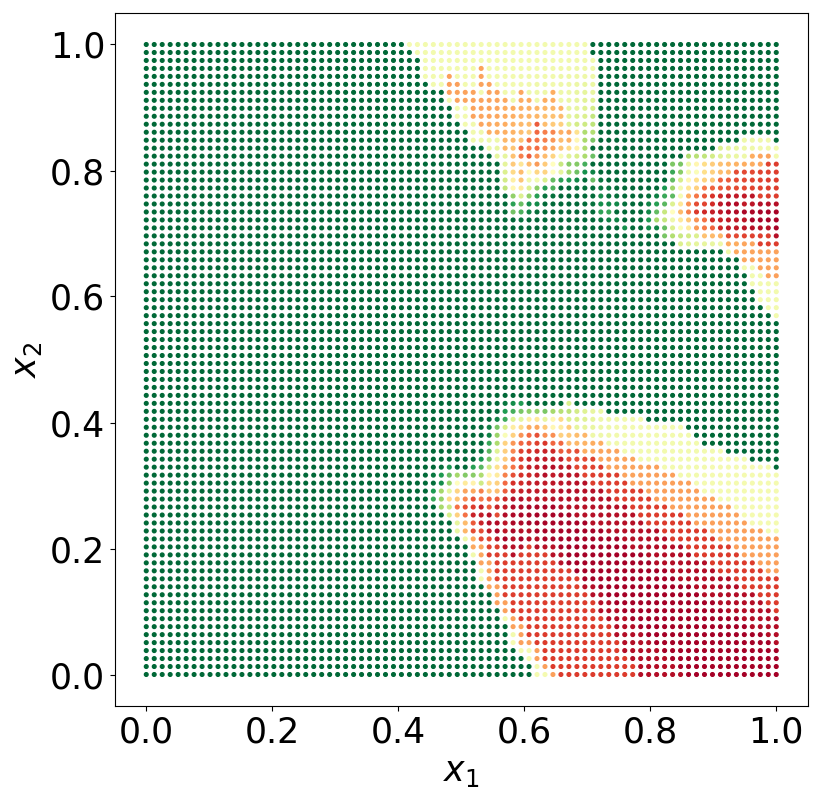}
        \vspace{-5mm}\caption{\sru, euclidean}
        \label{fig:metric-sdt-6}
    \end{subfigure}
    \hfill
    \begin{subfigure}[t]{0.49\linewidth}
        \centering
        \includegraphics[width=\textwidth]{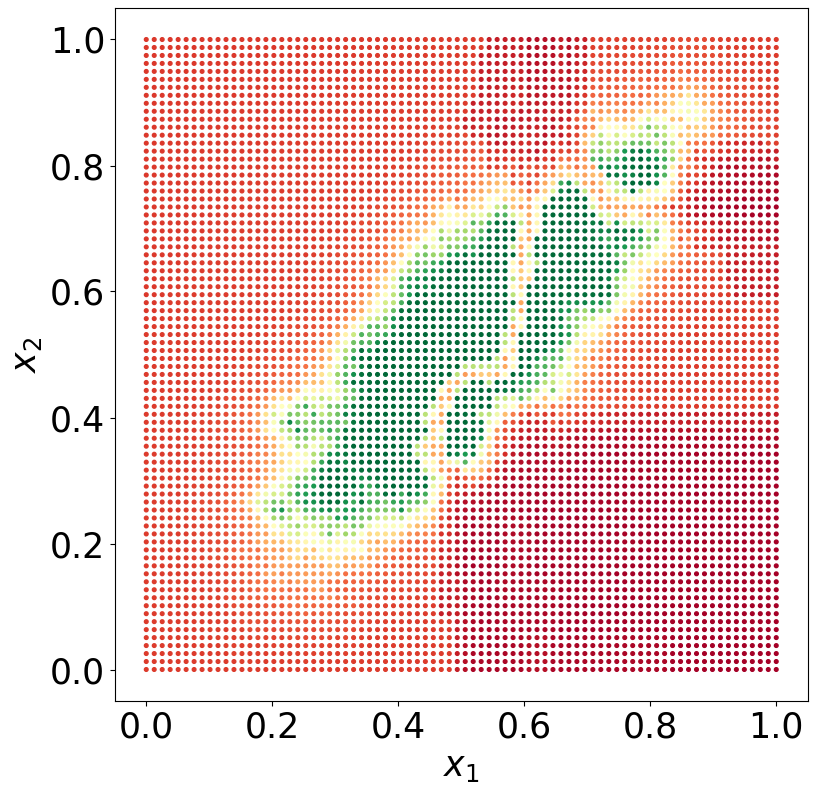}
        \vspace{-5mm}\caption{\wru, manhattan}
        \label{fig:metric-wdt-5}
    \end{subfigure}\hfill
    \begin{subfigure}[t]{0.49\linewidth}
        \centering
        \includegraphics[width=\textwidth]{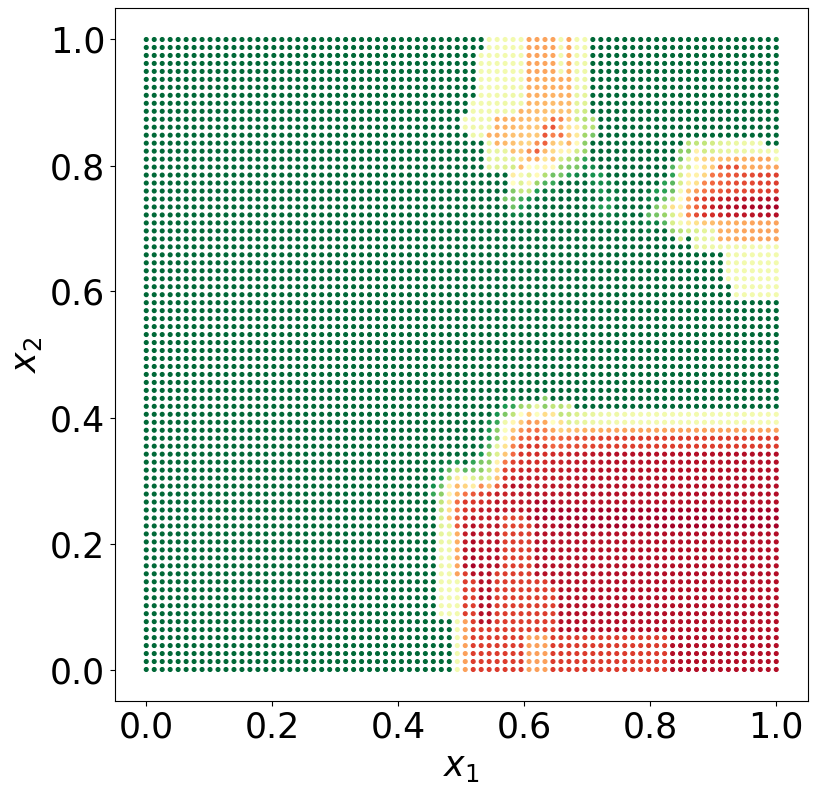}
        \vspace{-5mm}\caption{\sru, manhattan}
        \label{fig:metric-sdt-5}
    \end{subfigure}    
\caption{query space colored based on \wru and \sru values with regards to $\dee$ in Fig.~\ref{fig:exp-cat-train} subject to different distance metrics.}
\label{fig:exp-metrics}
\end{figure}

\vspace{-1mm}
\stitle{Validation on Classification}
Having provided the visual validation results, we next validate our \ru measures on classification tasks. In this regard, using {\it SYN} data set, we first computed the \ru measures for all the query points in the uniform sample and bucketized the points w.r.t. their \ru values in ranges of length 0.1. We repeated this for both \sru and \wru measures. Next, using a classification model that we trained on the (training) data set, we predict the target variable for the points in each range of \ru measure. The values corresponding to the accuracy/error of the classifier over each bucket of \ru values are provided in Figure \ref{fig:exp-class-bar-syn-sdt} and \ref{fig:exp-class-bar-syn-wdt} for \sru and \wru respectively. As the \ru values increase, the accuracy of the model drops while the FNR and FPR rise, and therefore, the model fails to capture the ground truth for the points that fall into untrustworthy regions in the data set.

To also perform the experiments on a real-world data set, we used the {\it DCC} data set with a similar procedure. The results are shown in Figure  \ref{fig:exp-class-bar-real-sdt} and \ref{fig:exp-class-bar-real-wdt} and they follow the same course as the previous experiment. 
We repeated the experiments with five classification algorithms including, Logistic Regression (LR), K-Nearest-Neighbor ($k$-NN), Neural Networks (NN), Random Forest (RF), and SVM. All the classifiers underwent a hyperparameter tuning procedure to achieve optimal prediction correctness. 
We confirm that we obtained similar results in all cases and that all models failed to predict satisfactorily for query points that have a high \ru value. Hence, we provide the results for the NN model, the more advanced and powerful model.   
Depending on the results of the classification, we chose the most appropriate accuracy/error measures that indicated the behavior of the model. For example, if the results were dominated by TPs and TNs, accuracy is the measure that best describes the model, otherwise, F1 might be a better choice. Depending on the number of FPs and FNs and the balance between them, the same rule applies to FPR and FNR measures. 

Finally, to stress test our proposed measures with more complex learning tasks (such as complex classifiers, massive high dimensional data sets based on NLP and vision tasks, sparse data sets, etc.), we extended our experiments to three other classification data sets, {\it AD}, {\it RS}, and {\it GS}. Initially, we repeated the experiments with identical settings as before. For {\it AD}, the results are shown in Figures \ref{fig:exp-adult-sdt} and \ref{fig:exp-adult-wdt} and corroborate our findings. For tuples with high \ru values, models tend to fail more to predict reliably. Next, we repeated the experiments on {\it AD}, using deep learning. We trained a classification model (with tuned parameters) with two hidden layers of size 64 and 32 units respectively. We also constructed and tuned deep-learning models for {\it RS}, and {\it GS} data sets.
The results are illustrated in Figures \ref{fig:exp-adult-sdt_dl}, \ref{fig:exp-adult-wdt_dl}, \ref{fig:exp-reg-real-sim-sdt}, \ref{fig:fig:exp-reg-real-sim-wdt}, \ref{fig:exp-reg-gisette-sdt_dl}, and \ref{fig:fig:exp-reg-gisette-wdt_dl} and are compliant with our previous experiments, showing that even with the choice of more complex models that show promising results in the tasks (~95\% F1 score in the case of {\it RS}), they still are less reliable for query points with high \ru values.  
Models' accuracy in Figures~\ref{fig:exp-reg-real-sim-sdt}--\ref{fig:fig:exp-reg-gisette-wdt_dl} were consistently above 95\% in all cases. Therefore, for visual clarity we only included FPR. \footnotetext{The absence of any records within certain \ru ranges results in a 0 value for error/accuracy (y-axis) in those ranges.}

\vspace{-1mm}
\stitle{Validation on Regression}
In this experiment, we study the effectiveness of our \ru measures in the regression tasks. Accordingly, we used {\it RN} and {\it HS} data sets and computed \sru and \wru values for all the query points in the uniform sample. Thereafter, we repeated the bucketization process as we did in the last experiment, and having trained a regression model over the data set, we evaluated the model's prediction over the tuples from each bucket. The results are presented in Figures \ref{fig:exp-reg-bar-sdt}, \ref{fig:exp-reg-bar-wdt}, \ref{fig:exp-housing-bar-sdt} and \ref{fig:exp-housing-bar-wdt}. As the \ru value increases, the RSS of the regression model follows the same trend denoting that the model fails to perform for tuples with a high \ru value.
We repeated the experiments with 3 different regression algorithms including ElasticNet, DT, and $k$-NN, all three with tuned hyper-parameters. Regardless of the regression model, the outcome was similar and therefore we only report the results for the $k$-NN regressor.

Finally, using the {\it DI} data set, we repeated the experiments with identical settings as before. The results are brought in Figures \ref{fig:exp-reg-diamond-sdt}, \ref{fig:fig:exp-reg-diamond-wdt} verifying our findings. For tuples with high \ru values, models fail more frequently. Next, we repeated the experiments on {\it DI}, using a Deep Learning regression model with tuned parameters.
We constructed a regression model with four hidden layers of size 128, 64, 32, and 16 respectively. The results are shown in Figures \ref{fig:exp-reg-diamond-sdt_dl}, \ref{fig:fig:exp-reg-diamond-wdt_dl} and are consistent with the previous experiments, verifying that even more complex models fail for query points with high \ru values.

The correlations between the \ru values and the model performance metrics are visually clear in Figures~\ref{fig:proof:10} and \ref{fig:proof:11}. 
Still to further verify these correlations we computed the Pearson correlation between the \ru value bins and model performance metrics. The results are provided in Table~\ref{tab:cor}, which confirms the high correlation values for the datasets AD, RN, HS, and DI.
Furthermore, to confirm that these values are not sensitive to the binning choices, we perturbed the bin boundaries. However, the results did not meaningfully change, which confirms their robustness.
\begin{table}[h]
    \centering
    \begin{tabular}{c|c|c|c}
         dataset &pref. metric &dist. metric &correlation  \\ \hline
         AD &accuracy &\sru & -0.88  \\
          &FPR &\sru & 0.84  \\ 
          &accuracy &\wru & -0.88  \\ 
          &FPR &\wru & 0.78  \\ \hline
         RN &MSE &\sru & 0.73  \\
          &MSE &\wru & 0.78  \\ \hline
         HS &MSE &\sru & 0.91  \\
          &MSE &\wru & 0.70  \\ \hline
         DI &MSE &\sru & 0.78  \\
          &MSE &\wru & 0.88  \\
    \end{tabular}
    \caption{Pearson correlations between the \ru values and model performance metrics for various datasets.}
    \label{tab:cor}
    \vspace{-5mm}
\end{table}

\noindent \underline{\bf Summary of Proof of Concept:}
In short, experiments consistently demonstrate that as the \ru values grow, the ML models become less reliable in capturing the truth for the corresponding regions. \rev{Consequently, when \ru values for a query point in a data set are high, one should \underline{discard} or at least \underline{not rely on} the outcome of the model constructed on it for critical decisions.} 

\vspace{-1em}
\subsection{Comparison with the Existing Work}\label{exp:comparison}

In this section, we thoroughly evaluate the \ru measures in the context of the existing approaches discussed in \S \ref{sec:related}, demonstrate why the existing approaches fail, and how \ru measures are superior in capturing the unreliability of individual predictions.

Consider data set $\dee$ as shown in Figure \ref{fig:ext_1} created with three Gaussian distributions representing classes \textit{red, blue, orange}. An arbitrary classification model (e.g. Gaussian Naive Bayes classifier) as the base classifier is trained on $\dee$ and the predicted labels are depicted in Figure \ref{fig:ext_2}. Finally, Figures~\ref{fig:ext_3} and ~\ref{fig:ext_4} show the corresponding \sru and \wru values for data set $\dee$. 

Through the rest of this section, we use the data set $\dee$ and the classifier described above to evaluate existing methods for the reliability of individual predictions.

\subsubsection{Conformal Prediction} We start by employing the conformal prediction (CP) framework \footnote{We used MAPIE (Model Agnostic Prediction Interval Estimator), which is an implementation of Conformal Prediction works such as \cite{agarwal2017efficient,sadinle2019least}. See more at: https://mapie.readthedocs.io/} with confidence level $\alpha$ of 0.2, 0.1, and 0.05 and softmax score output of the base classifier as the conformity score. Results are shown in Figures \ref{fig:ext_1_1}, \ref{fig:ext_1_2}, \ref{fig:ext_1_3}. As can be seen in Figure \ref{fig:ext_1_1}, CP is creating empty prediction sets for $\alpha=0.2$ for query points around the uncertain areas which are faulty and show that CP is highly dependent on the choice of $\alpha$. 
The null region disappears for smaller $\alpha$ values but ambiguous classification regions arise with several labels included in the prediction sets highlighting the uncertain behavior of the base classifier.
By choosing the cumulative softmax conformal score, the empty prediction set problem is resolved however, uncertain regions are emphasized by wider boundaries. Now consider the query point $q$; according to the model prediction, $q$ belongs to the \textit{orange} class and regardless of the chosen $\alpha$, CP confirms that. However, this is only true if the true decision boundary is identical to the one estimated by the base classifier~\cite{shafer2008tutorial}; still, as previously discussed in \S~\ref{sec:measure:toy}, this may not always be the case. Therefore, although $q$ is in an uncertain region, CP fails to capture it as it always returns a prediction set of size 1. Conversely, as can be seen in Figure \ref{fig:ext_4}, the \wru measure can successfully capture the \ru associated with $q$ as it is an outlier, yet does not belong to an uncertain region.

\subsubsection{Prediction Probabilities} In the next experiment, we evaluate the prediction probabilities generated by probabilistic classification models and demonstrate their failure for query points that are not represented by data. To do so, we employ data set $\dee$ and train an arbitrary probabilistic classifier such as Gaussian Naive Bayes on it (remember that we can use any classifier, however, if the model is not intrinsically probabilistic, we need to make sure that the probabilistic outcomes are calibrated). Figures \ref{fig:pred_prob_1}, \ref{fig:pred_prob_2}, and \ref{fig:pred_prob_3} show the prediction probabilities assigned to either of the classes \textit{red, blue} and \textit{orange}. As observed, prediction probabilities fail to capture query points that belong to unrepresented regions and assign a negligible chance of belonging to any other class but the one determined by the decision boundary, however, this is only true if the true decision boundary is identical to the one estimated by the base classifier and as previously discussed in \S \ref{sec:measure:toy}, this may not always be the case if the distribution between training and production data vary.

\subsubsection{Data Coverage} Finally, we conduct an experiment to assess the capacity of data coverage techniques to create proper warnings and demonstrate their failure for query points that are in uncertain regions. To this end, using the continuous notion of data coverage \cite{asudeh2021identifying} and tuned parameters of $k=50$ and $\rho=0.08$, we identify the uncovered region on data set $\dee$ as illustrated in Figure \ref{fig:coverage_comparison}. The training data points ($\dee$) are highlighted as black dots. The regions highlighted in red and green comprise the uncovered and covered regions respectively. Any query point belonging to the green region is covered and all the query points in the red region are uncovered.
While the uncovered region can raise warning signals for the unreliability of underrepresented query points, it fails to capture the unreliability associated with the uncertain regions (regions close to the decision boundary in the case of data set $\dee$). Furthermore, even for the query points in underrepresented regions, data coverage creates a binary value that is sensitive to the choice of parameters such as the radius $\rho$. This issue is further highlighted considering the sharp transition from uncovered to covered, specified by the uncovered region's decision boundary. As a result, while two points close to each other, where one is inside and the other outside of the decision boundary, are almost equally miss-presented, for one the output signal is covered (no warning at all) while the other is uncovered (maximum warning).

\subsubsection{Uncertainty Sampling} Uncertainty sampling is yet another model-centric notion of uncertainty quantification, which is the dominant approach for active learning. 
In an active learning setting, the model is given a pool of unlabeled samples and should decide which sample to label next. In uncertainty sampling, the idea is to label the sample that the model is most uncertain about. This uncertainty is evaluated based on the model's (probabilistic) confidence for each training point using Shannon entropy. 
In this experiment, we used the class probabilities assigned by the base classifier to calculate the Shannon entropy for each query point.
Figure \ref{fig:ext_5} shows the query space colored based on the model's confidence (using Shannon entropy). As expected, besides being dependent on the model itself, this approach fails for query points that lack sufficient representation.

\vspace{-5mm}
\subsection {Performance Evaluations}
After demonstrating the effectiveness of the \ru measures, we now focus on the performance of our algorithms. 
In this section, we use the {\it DCC} and {\it RN} data sets to evaluate the time efficiency of algorithms. We obtained similar results with almost identical plots for both classification and regression tasks. In the following, we present the results for classification tasks using {\it DCC} data set with different settings.

\vspace{-2mm}
\subsubsection{Query Time}
The query time consists of (i) the time to find the $k$-vicinity of the query point $q$ and identifying the tuple in $k$-vicinity that has the maximum distance from $q$, and (ii) the time to apply binary search on the sorted multi-sets.

\vspace{-1mm}
\stitle{Varying $n$}
To study the impact of the number of tuples $n$ on the performance of the query time, we gradually increase the size of the data set from 50 to 100K. The results are provided in Figure \ref{fig:exp-query-time-var-num-point}. 
The total query time is dominated by the first bottleneck and the time to binary search the lists is negligible compared to it. In our experiments, the query time did not (meaningfully) change as the data set size increased, showing the scalability of our algorithm to the very large settings.

\vspace{-1mm}
\stitle{Varying $k$}
Next, we vary the neighborhood size $k$ from 1 to 50. The results in Figure \ref{fig:exp-query-time-var-num-neighbors} suggest that the query time is (almost) independent from the $k$.

\vspace{-1mm}
\stitle{Varying $d$}
We next study the impact of the number of attributes $d$ by varying it from 2 to 20. The results in Figure \ref{fig:exp-query-time-var-num-dim} verify the scalability of our algorithms concerning the number of dimensions.

\vspace{-1mm}
\stitle{Varying $c$}
In our final experiment, we change $c$ from 0.05 to 0.25. The results are brought in Figure \ref{fig:exp-query-time-var-num-outlier_ratio}. Results verify that the query time is independent of $c$.

\subsubsection{Preprocessing Time}
Our preprocessing time consists of two parts. The first is the time to build the $k$-NN data structure, identifying the $k$-vicinity radius (in Algorithm \ref{alg:po}) and computing uncertainty (in Algorithm \ref{alg:pe}) for each tuple in the data set. The second one is the time to construct the sorted multi-sets of $k$-vicinity radii and uncertainty values. 
We use the exact $k$-vicinity radii and entropy values in this experiment.

\stitle{Varying $n$} In this experiment, we study the impact of the number of tuples $n$ in the data set on the preprocessing time by gradually increasing the size of the data set from 50 to 100K. We then measure the time to build the $k$-NN data structure and construct the sorted multi-sets. The results are provided in Figure \ref{fig:exp-pre-time-var-num-point}. The total preprocessing time is dominated by the time to build the $k$-NN data structure and the time to build the multi-sets is almost negligible compared to it. 
Nevertheless, the cumulative preprocessing time was small enough that the algorithm could scale to larger settings, finishing in less than a second for $n$=100K.   

\stitle{Varying $k$} To study the impact of neighborhood size $k$ on the preprocessing time, we vary $k$ from 1 to 50. The results can be seen in Figure \ref{fig:exp-pre-time-var-num-neighbors}. Similarly, the total preprocessing time in this experiment is also dominated by the time to build the $k$-NN data structure and the algorithm was efficient in all settings, finishing in less than a fraction of a second for $k$=50.

\stitle{Varying $d$} To study the impact of the number of attributes $d$ of the data set on the preprocessing time, we gradually change $d$ from 2 to 20. The results are brought in Figure \ref{fig:exp-pre-time-var-num-dim}. Like the previous settings, the total preprocessing time is dominated by the time to build the $k$-NN data structure and the algorithm linearly scales to larger settings, finishing in less than 3 seconds for $d$=20.

\vspace{-5mm}
\subsection{Impact of Distance Measures on \ru Values}
\vspace{-2mm}
In this experiment, we study the effect of the distance metric chosen to determine the neighborhood of a data point (in $k$-NN component) on the \ru values. To do so, we employ data set $\dee$ (Figure \ref{fig:exp-cat-train}) and calculate the \ru values for the entire query space w.r.t. 3 distance measures of \textit{chebyshev, manhattan} and \textit{euclidean}. Although the distance metric is expected as an input in our implementation, however, the results show general consistency across both measures (Figures \ref{fig:exp-metrics}).

\vspace{-2mm}
\subsubsection{Training Regression Models for No Data Access}
\begin{figure}[!tbh]
    \centering
    \includegraphics[width=.40\textwidth]{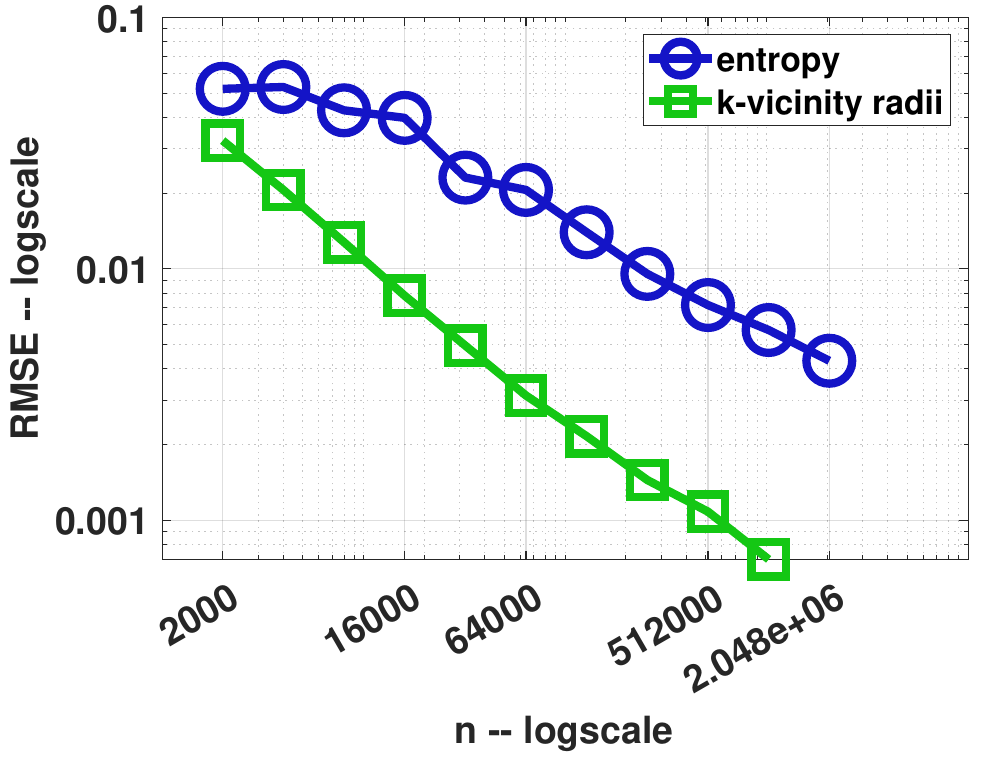}
    \vspace{-4mm}\caption{effectiveness of exponential search in reducing the error of learning distrust parameters}
    \label{fig:sampling}
    \vspace{-8mm}
\end{figure}

\vspace{-2mm}
As our final experiment, we study the efficiency of the training process for building the regression models to estimate the values of $k$-vicinity radius and entropy.
As explained in \S~\ref{sec:sampling}, we apply exponential sampling to generate the right amount of data such that the trained models satisfy the user-specified error.
As a heuristic, we generate a fraction of samples uniformly and the others from the underlying distribution of the training data using GAN methods or Gaussian copula distribution functions \cite{7796926}.
Figure \ref{fig:sampling} illustrates the monotonic drop in the error of both regression models trained to predict the entropy and $k$-vicinity radii of query points (as discussed in \S~\ref{sec:sampling}), as the number of synthetic i.i.d samples increases. Since both entropy and $k$-vicinity radii values are in the range of [0,1], selecting a sufficiently small (RMSE) error threshold  (e.g. $10^{-3}$ or $0.01\%$ average difference between the actual and predicted values) guarantees not going overboard with generating too many samples (affecting preprocessing time) while achieving good prediction accuracy.

\vspace{-5mm}
\subsection{Sensitivity Analyses}
\vspace{-3mm}
In the following, we provide the complimentary experiments by choosing alternative settings compared to the previous experiments.
In short, in the complimentary experiments, we observed consistency with our previous results, further validating and verifying our proposal.
\vspace{-2mm}
\subsubsection{Proof of Concept Experiments with Various Underlying Distributions}
\begin{figure*}[!tbh] 
\begin{minipage}[t]{\linewidth}
    \begin{subfigure}[t]{0.23\linewidth}
        \centering
        \includegraphics[width=\textwidth]{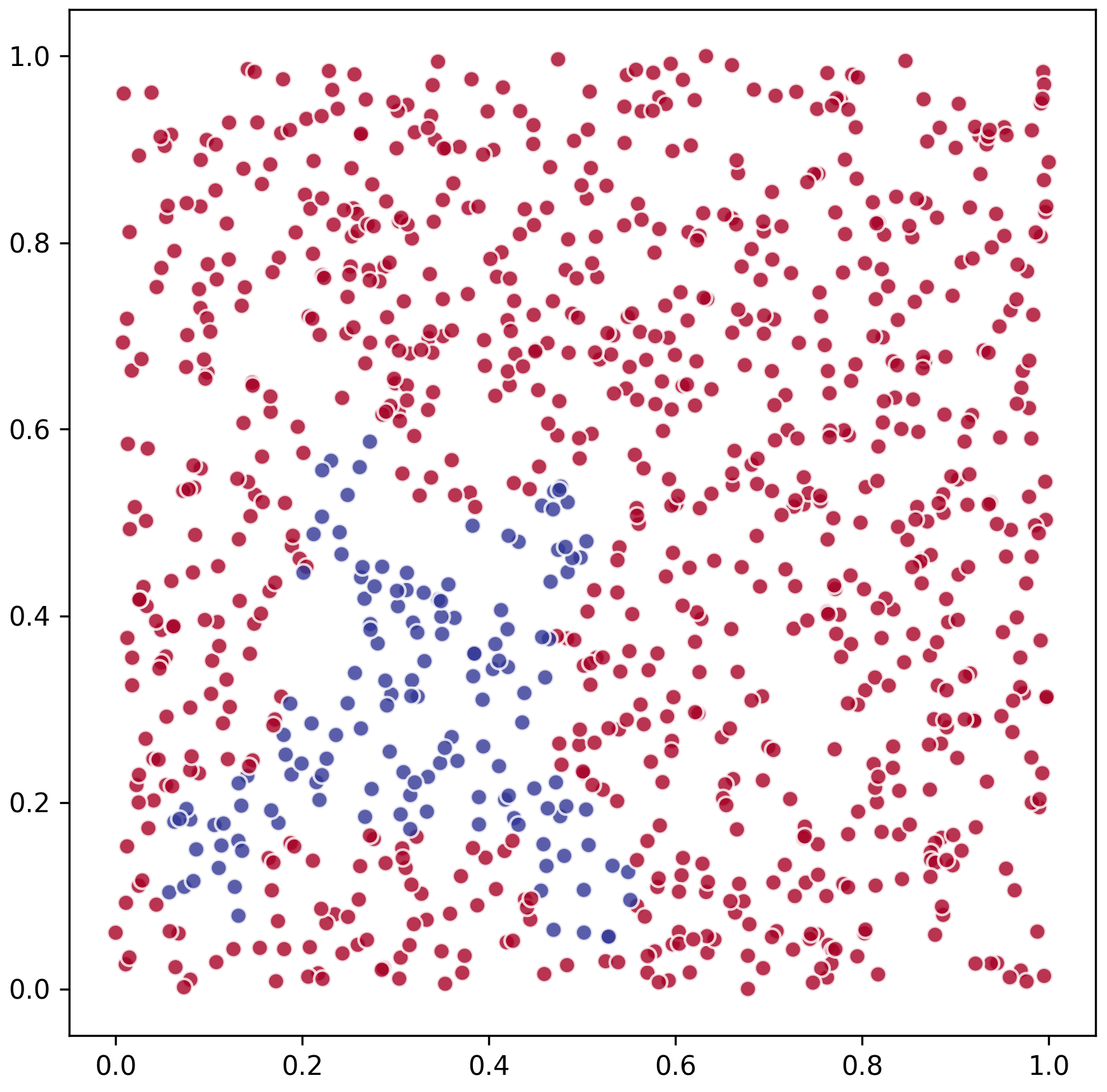}
        \vspace{-5mm}\caption{a data set $\dee_u$ in SYN following a uniform distribution}
        \label{fig:dataset_uniform}
    \end{subfigure}\hfill
    \begin{subfigure}[t]{0.23\linewidth}
        \centering
        \includegraphics[width=\textwidth]{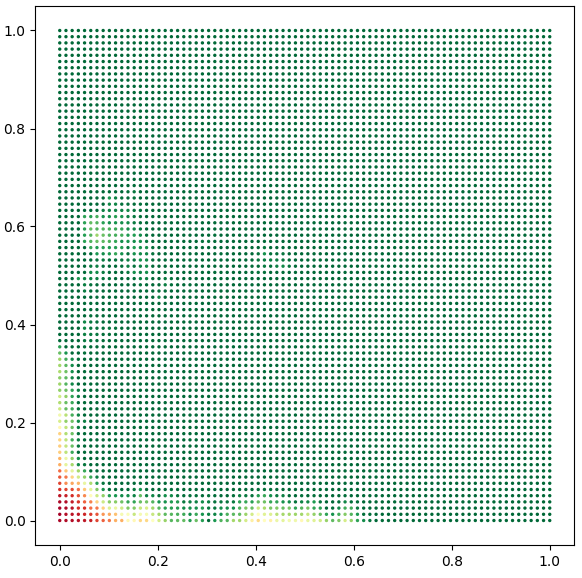}
        \vspace{-5mm}\caption{query space colored based on \sru values w.r.t. $\dee_u$ in Fig.~\ref{fig:dataset_uniform}}
        \label{fig:uniform_sdt}
    \end{subfigure}\hfill
    \begin{subfigure}[t]{0.23\linewidth}
        \centering
        \includegraphics[width=\textwidth]{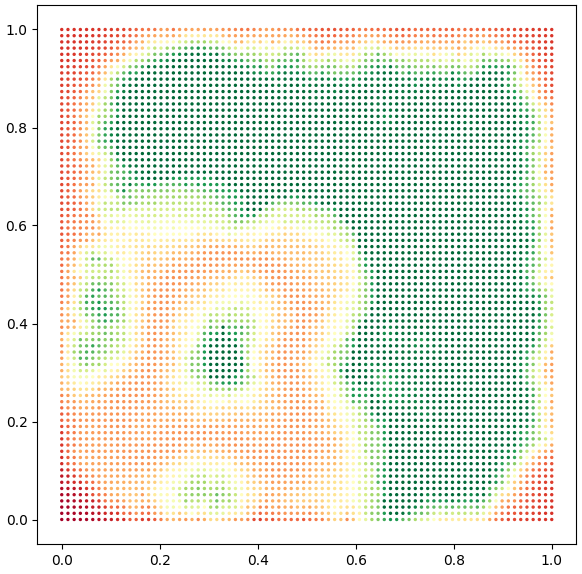}
        \vspace{-5mm}\caption{query space colored based on \wru values w.r.t. $\dee_u$ in Fig.~\ref{fig:dataset_uniform}}
        \label{fig:uniform_wdt}
    \end{subfigure}\hfill
    \begin{subfigure}[t]{0.25\linewidth}
        \centering
        \includegraphics[width=\textwidth]{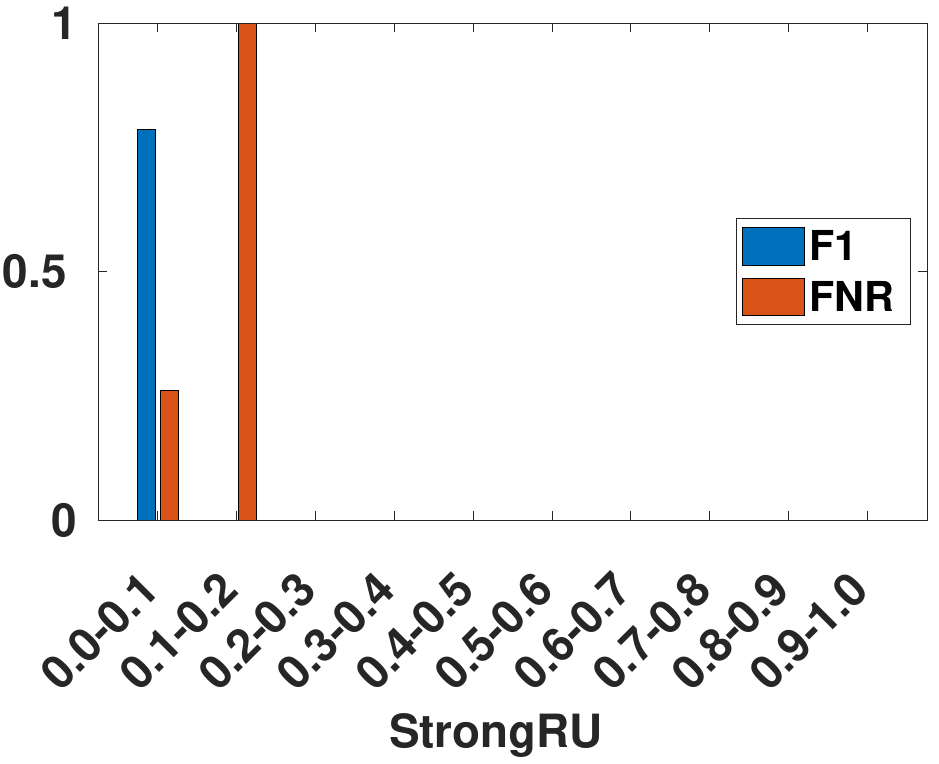}
        \vspace{-5mm}\caption{effectiveness of \sru over $\dee_u$ in Fig.~\ref{fig:dataset_uniform}}
        \label{fig:uniform_sdt_effectiveness}
    \end{subfigure}
\end{minipage}
\begin{minipage}[t]{\linewidth}
    \begin{subfigure}[t]{0.25\linewidth}
        \centering
        \includegraphics[width=\textwidth]{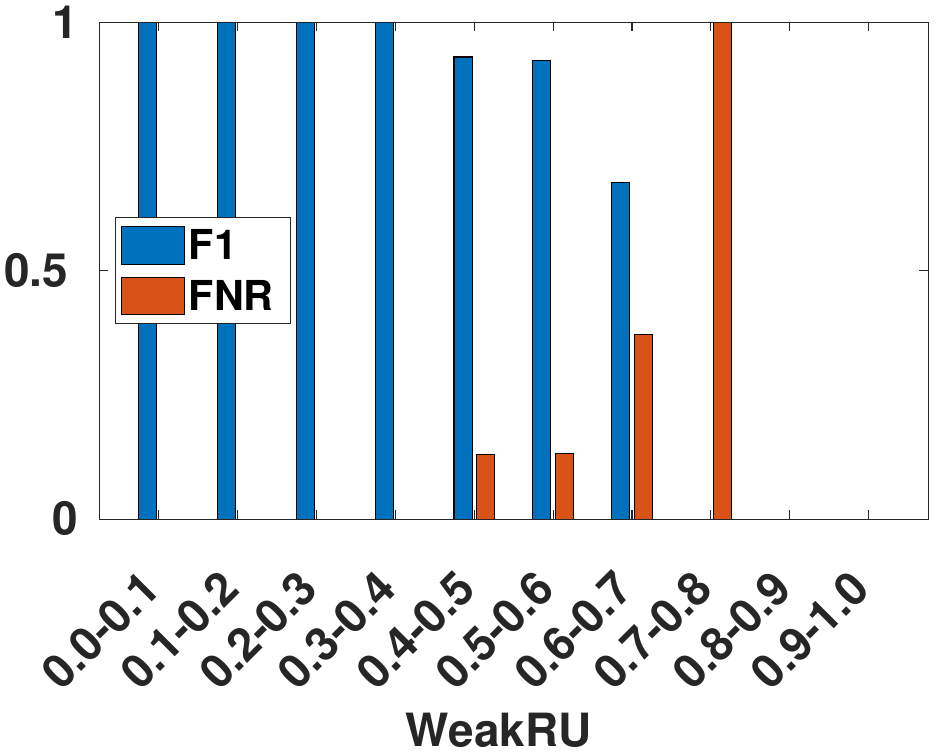}
        \vspace{-5mm}\caption{effectiveness of \wru over $\dee_u$ in Fig.~\ref{fig:dataset_uniform}}
        \label{fig:uniform_wdt_effectiveness}
    \end{subfigure}\hfill
    \begin{subfigure}[t]{0.23\linewidth}
        \centering
        \includegraphics[width=\textwidth]{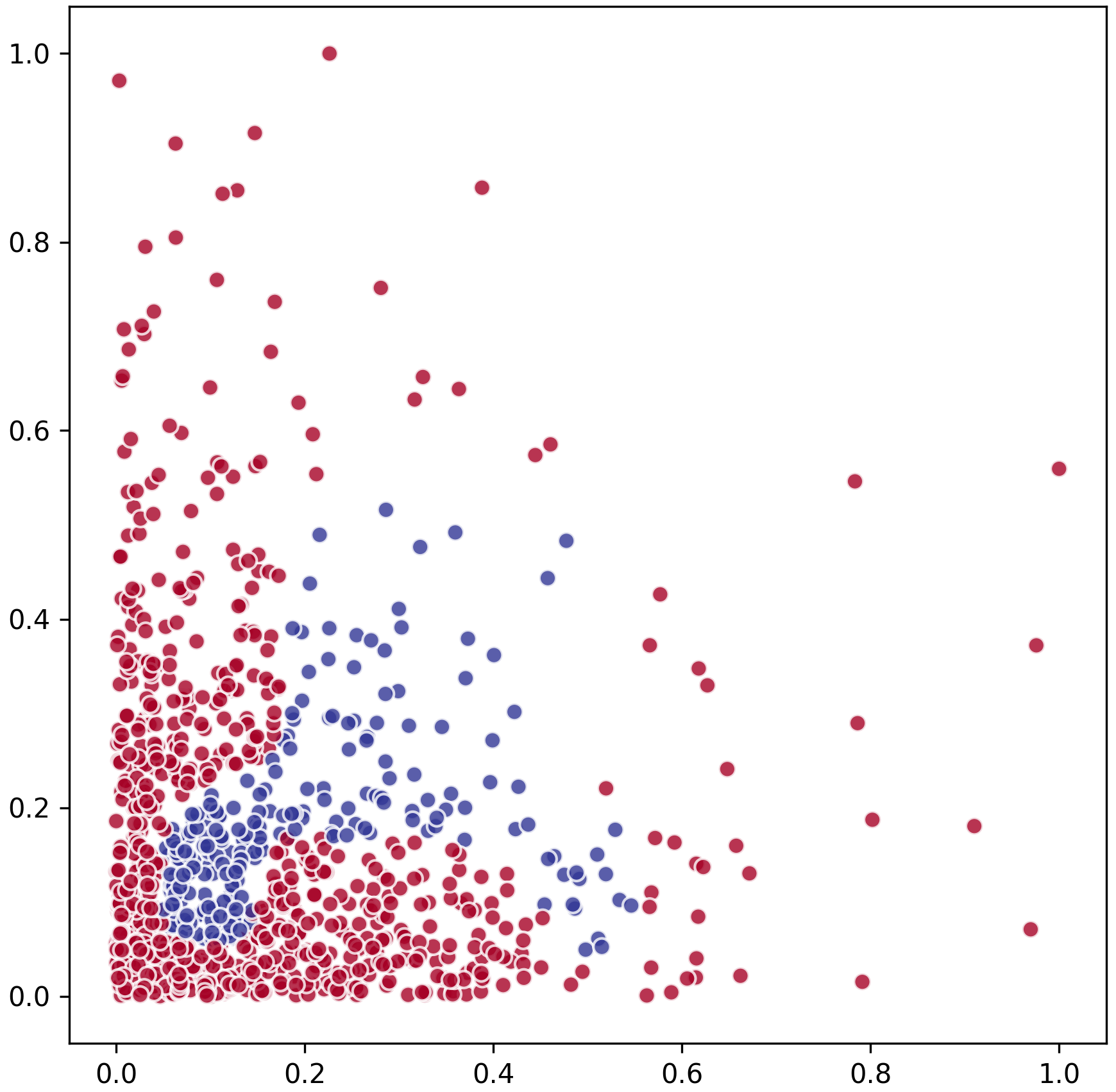}
        \vspace{-5mm}\caption{a data set $\dee_e$ in SYN following a exponential distribution}
        \label{fig:dataset_exponential}
    \end{subfigure}\hfill
    \begin{subfigure}[t]{0.23\linewidth}
        \centering
        \includegraphics[width=\textwidth]{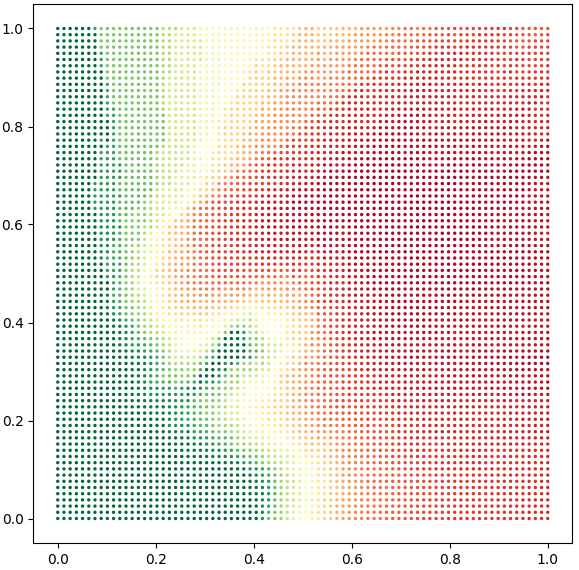}
        \vspace{-5mm}\caption{query space colored based on \sru values w.r.t. $\dee_e$ in Fig.~\ref{fig:dataset_exponential}}
        \label{fig:exponential_sdt}
    \end{subfigure}\hfill
    \begin{subfigure}[t]{0.23\linewidth}
        \centering
        \includegraphics[width=\textwidth]{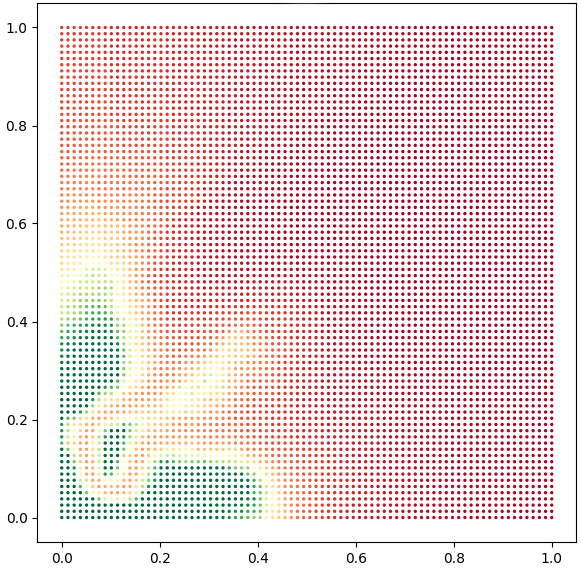}
        \vspace{-5mm}\caption{query space colored based on \wru values w.r.t. $\dee_e$ in Fig.~\ref{fig:dataset_exponential}}
        \label{fig:exponential_wdt}
    \end{subfigure}
\end{minipage}

\begin{minipage}[t]{\linewidth}
    \begin{subfigure}[t]{0.25\linewidth}
        \centering
        \includegraphics[width=\textwidth]{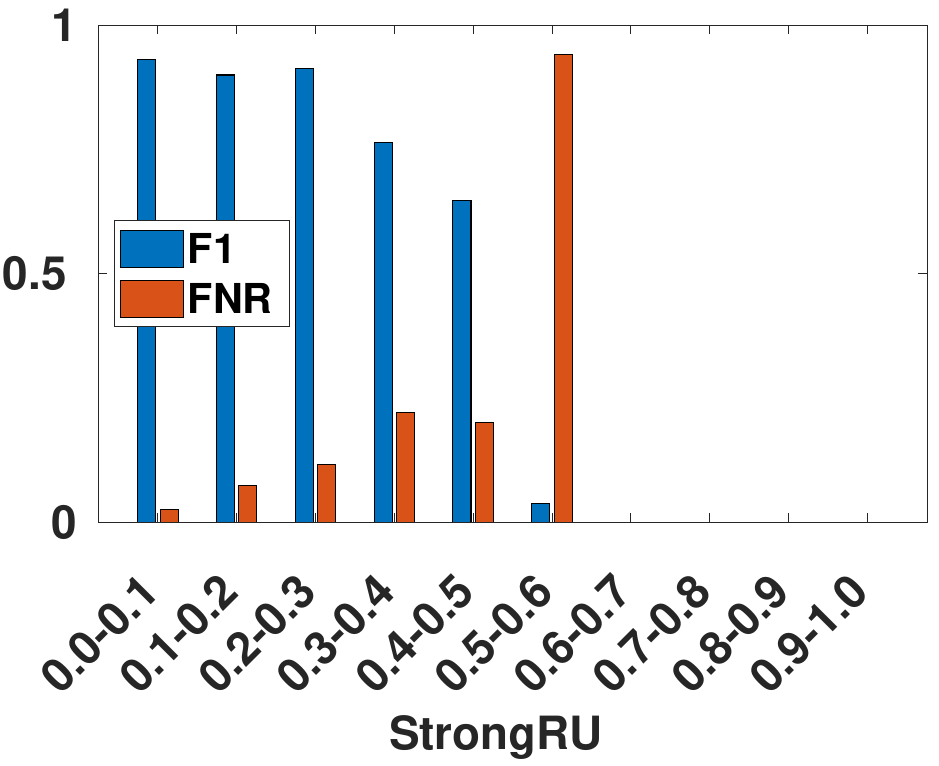}
        \vspace{-5mm}\caption{effectiveness of \sru over $\dee_e$ in Fig.~\ref{fig:dataset_exponential}}
        \label{fig:exponential_sdt_effectiveness}
    \end{subfigure}\hfill
    \begin{subfigure}[t]{0.25\linewidth}
        \centering
        \includegraphics[width=\textwidth]{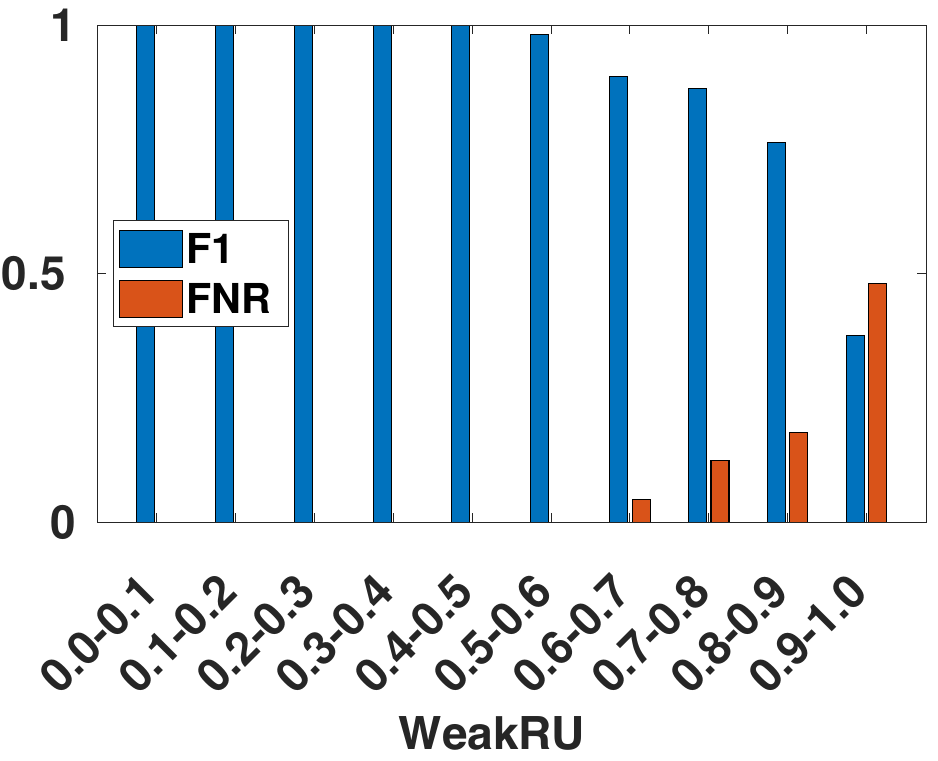}
        \vspace{-5mm}\caption{effectiveness of \wru over $\dee_e$ in Fig.~\ref{fig:dataset_exponential}}
        \label{fig:exponential_wdt_effectiveness}
    \end{subfigure}\hfill
    \begin{subfigure}[t]{0.23\linewidth}
        \centering
        \includegraphics[width=\textwidth]{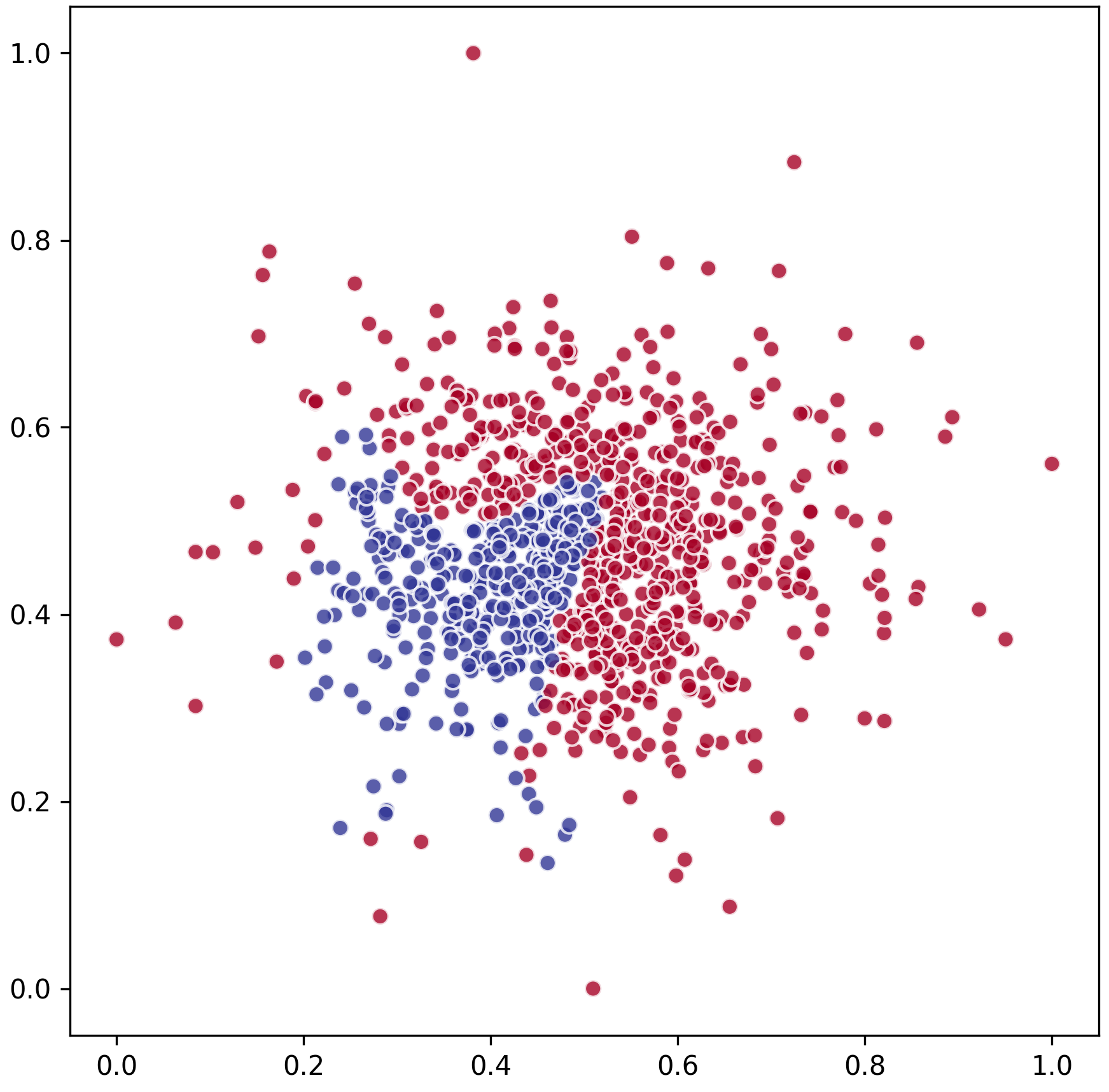}
        \vspace{-5mm}\caption{a data set $\dee_l$ in SYN following a logistic distribution}
        \label{fig:dataset_logistic}
    \end{subfigure}\hfill
    \begin{subfigure}[t]{0.23\linewidth}
        \centering
        \includegraphics[width=\textwidth]{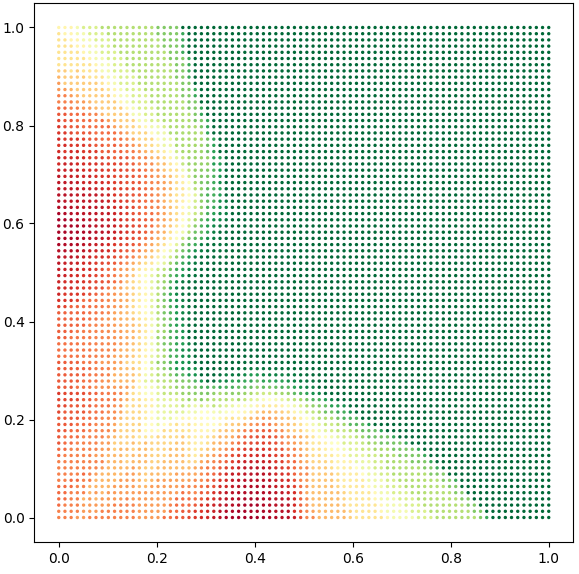}
        \vspace{-5mm}\caption{query space colored based on \sru values w.r.t. $\dee_l$ in Fig.~\ref{fig:dataset_logistic}}
        \label{fig:logistic_sdt}
    \end{subfigure}\hfill
\end{minipage}
\begin{minipage}[t]{\linewidth}
    \begin{subfigure}[t]{0.23\linewidth}
        \centering
        \includegraphics[width=\textwidth]{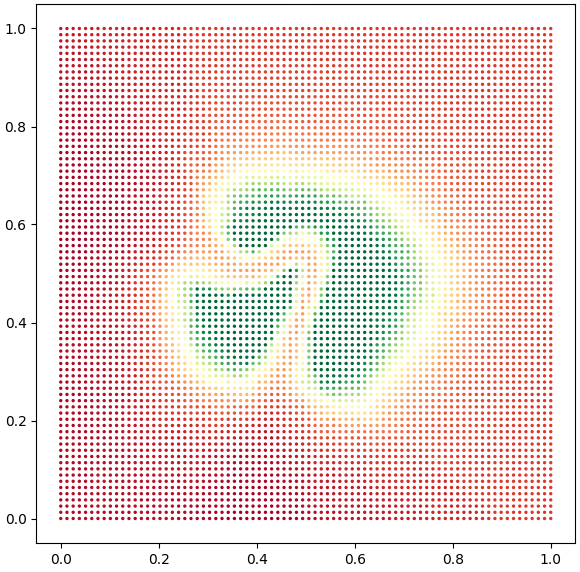}
        \vspace{-5mm}\caption{query space colored based on \wru values w.r.t. $\dee_l$ in Fig.~\ref{fig:dataset_logistic}}
        \label{fig:logistic_wdt}
    \end{subfigure}\hfill
    \begin{subfigure}[t]{0.25\linewidth}
        \centering
        \includegraphics[width=\textwidth]{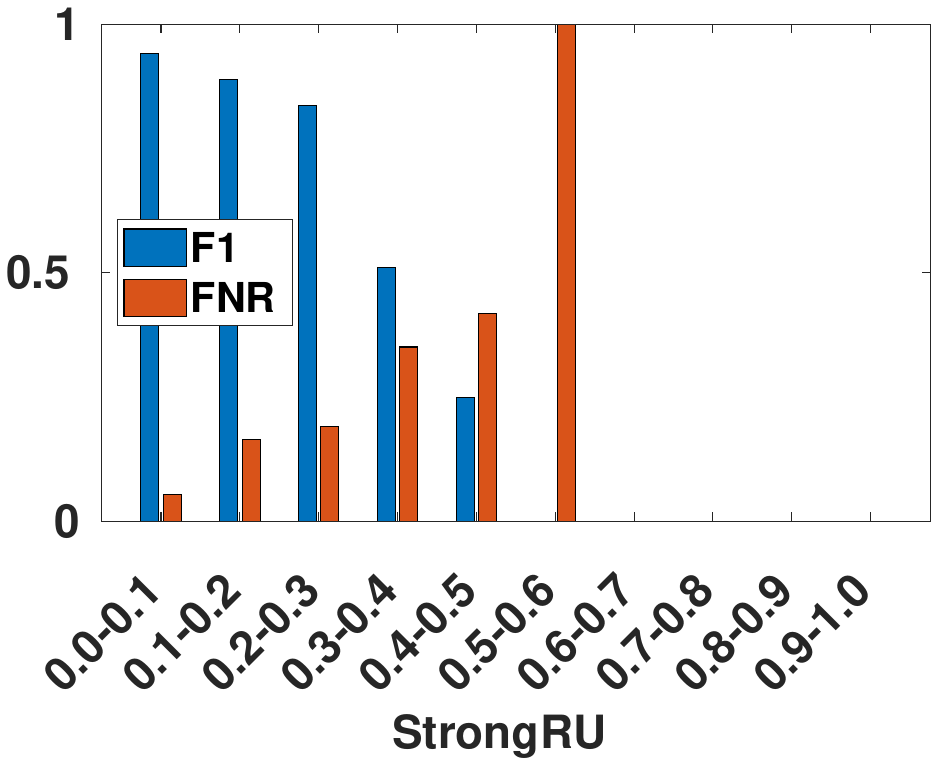}
        \vspace{-5mm}\caption{effectiveness of \sru over $\dee_l$ in Fig.~\ref{fig:dataset_logistic}}
        \label{fig:logistic_sdt_effectiveness}
    \end{subfigure}\hfill
    \begin{subfigure}[t]{0.25\linewidth}
        \centering
        \includegraphics[width=\textwidth]{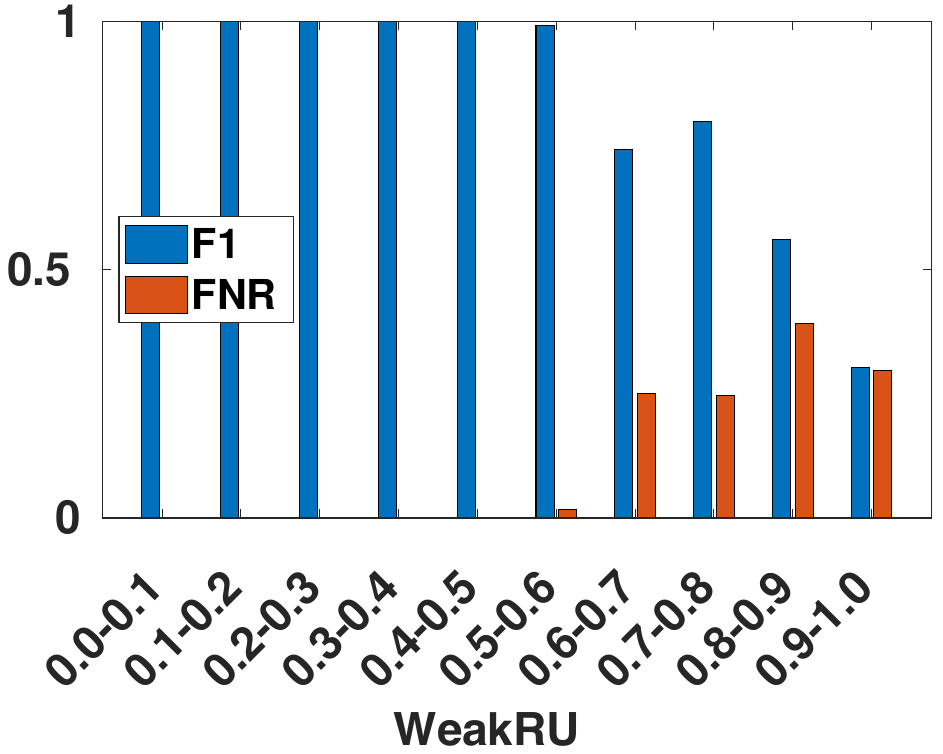}
        \vspace{-5mm}\caption{effectiveness of \wru over $\dee_l$ in Fig.~\ref{fig:dataset_logistic}}
        \label{fig:logistic_wdt_effectiveness}
    \end{subfigure}
\end{minipage}
\caption{{effect of sampling following different distributions while generating SYN data sets on distrust values}}
\label{fig:varying_distributions}
\vspace{-5mm}
\end{figure*}

\vspace{-2mm}
We used Gaussian as the underlying distribution of our synthetic datasets. In this experiment, we study whether the underlying distribution of the data would affect the capacity of the \ru measures in revealing unreliability. To do so, we follow the same procedure outlined in the construction of synthetic datasets in \S~\ref{sec:datasets}, however, instead of generating a sample following a Gaussian distribution, we opt to utilize alternative distributions such as {\em Uniform, Exponential, and Logistic}. The results are illustrated in Figure \ref{fig:varying_distributions}. In summary, aligned with our previous results, there is a consistent correlation between \ru values and a model's potential to predict correctly, with models exhibiting more error as \ru values increase. 
For the Uniform distribution (Figures \ref{fig:dataset_uniform}-\ref{fig:uniform_wdt_effectiveness}), we expected the entire query space should be equally represented by the training data and hence, the lack of representation scores to be universally low. As a result, the \sru scores remain low. \wru in this case, mostly reflects the impact of uncertainty. Consistent with our previous results, one can see the high correlation between \wru and model performance in Figure~\ref{fig:uniform_wdt_effectiveness}. 
\vspace{-2mm}
\subsubsection{Effect of Outlier Detection Metric}
\begin{figure*}[!tbh] 
\begin{minipage}[t]{\linewidth}
    \begin{subfigure}[t]{0.24\linewidth}
        \centering
        \includegraphics[width=\textwidth]{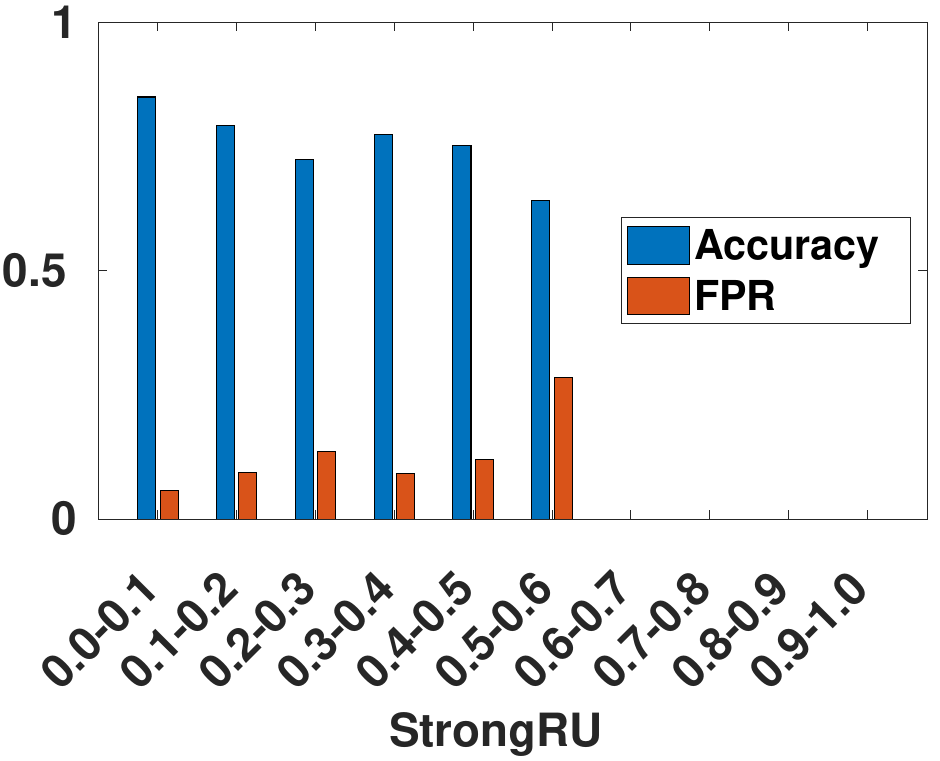}
        \vspace{-5mm}\caption{LOF, effectiveness of \sru  on classification in {\it AD}}
        \label{fig:sdt_adult_lof}
    \end{subfigure}\hfill
    \begin{subfigure}[t]{0.24\linewidth}
        \centering
        \includegraphics[width=\textwidth]{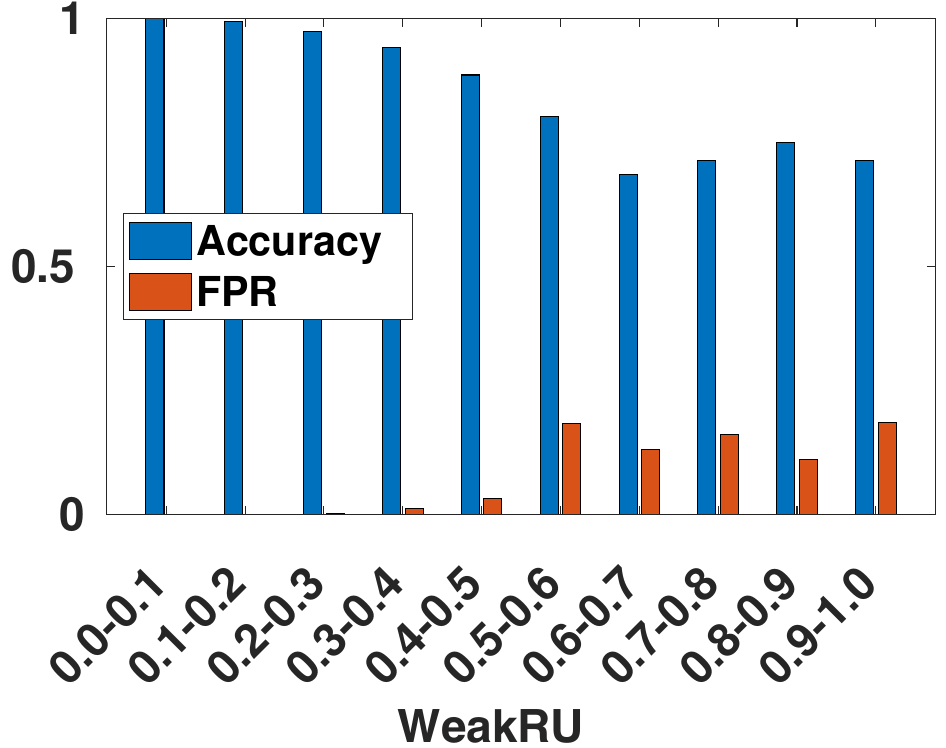}
        \vspace{-5mm}\caption{LOF, effectiveness of \wru on classification in {\it AD}}
        \label{fig:wdt_adult_lof}
    \end{subfigure}\hfill
    \begin{subfigure}[t]{0.24\linewidth}
        \centering
        \includegraphics[width=\textwidth]{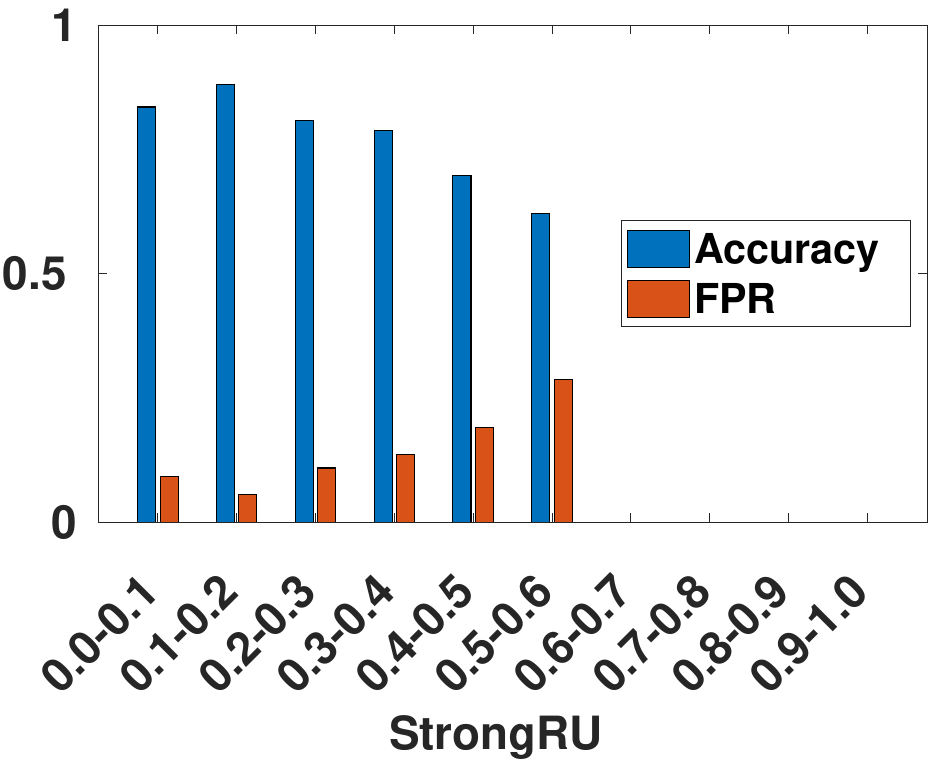}
        \vspace{-5mm}\caption{ECOD, effectiveness of \sru on classification in {\it AD}}
        \label{fig:sdt_adult_ecod}
    \end{subfigure}\hfill
    \begin{subfigure}[t]{0.24\linewidth}
        \centering
        \includegraphics[width=\textwidth]{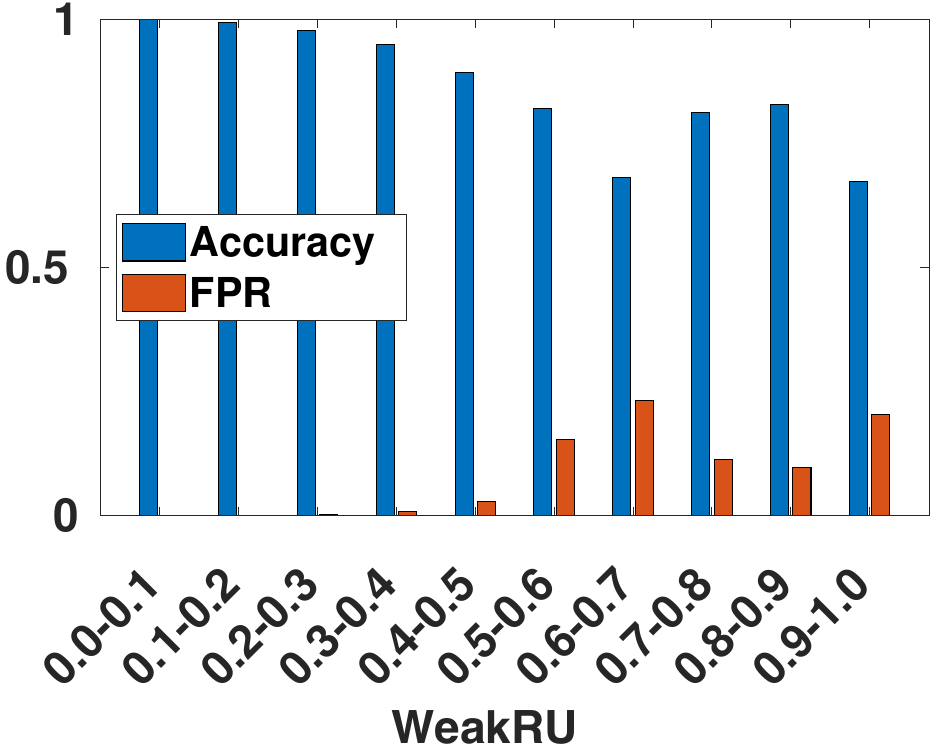}
        \vspace{-5mm}\caption{ECOD, effectiveness of \sru on classification in {\it AD}}
        \label{fig:wdt_adult_ecod}
    \end{subfigure}
\end{minipage}
\begin{minipage}[t]{\linewidth}
    \begin{subfigure}[t]{0.24\linewidth}
        \centering
        \includegraphics[width=\textwidth]{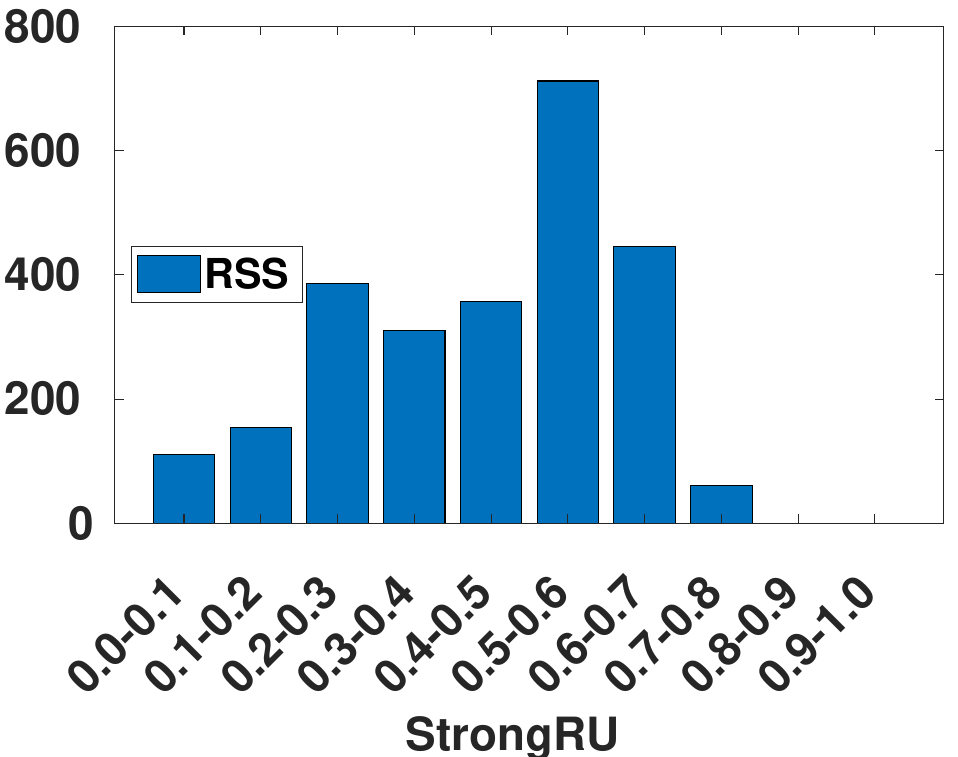}
        \vspace{-5mm}\caption{LOF, effectiveness of \sru on regression in {\it RN}}
        \label{fig:sdt_rn_lof}
    \end{subfigure}\hfill
    \begin{subfigure}[t]{0.24\linewidth}
        \centering
        \includegraphics[width=\textwidth]{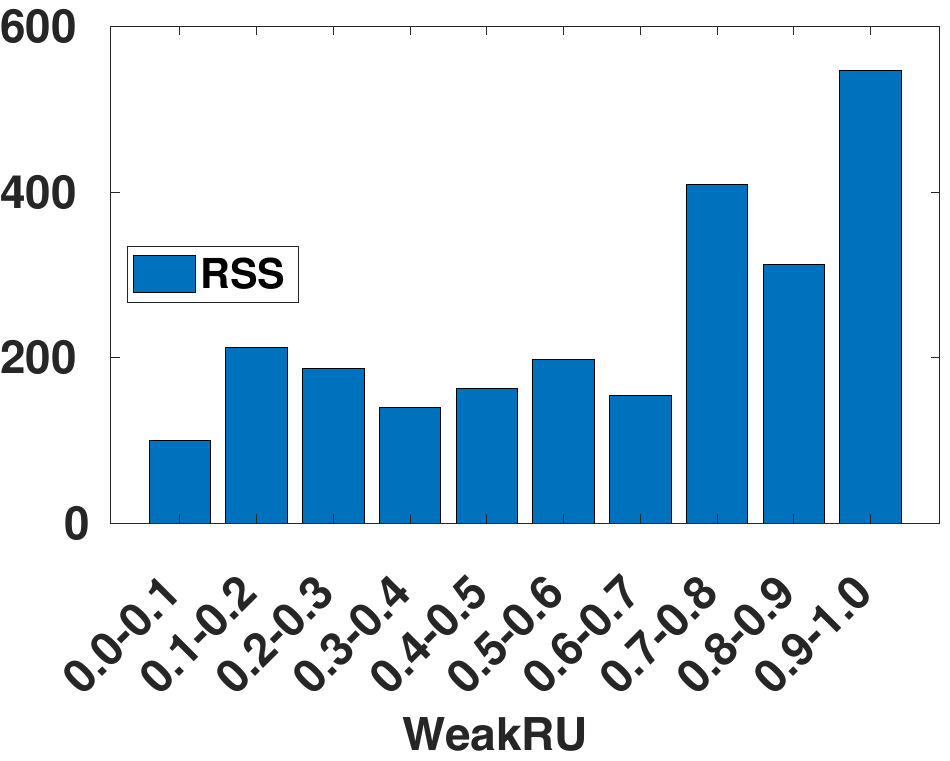}
        \vspace{-5mm}\caption{LOF, effectiveness of \wru on regression in {\it RN}}
        \label{fig:wdt_rn_lof}
    \end{subfigure}\hfill
    \begin{subfigure}[t]{0.24\linewidth}
        \centering
        \includegraphics[width=\textwidth]{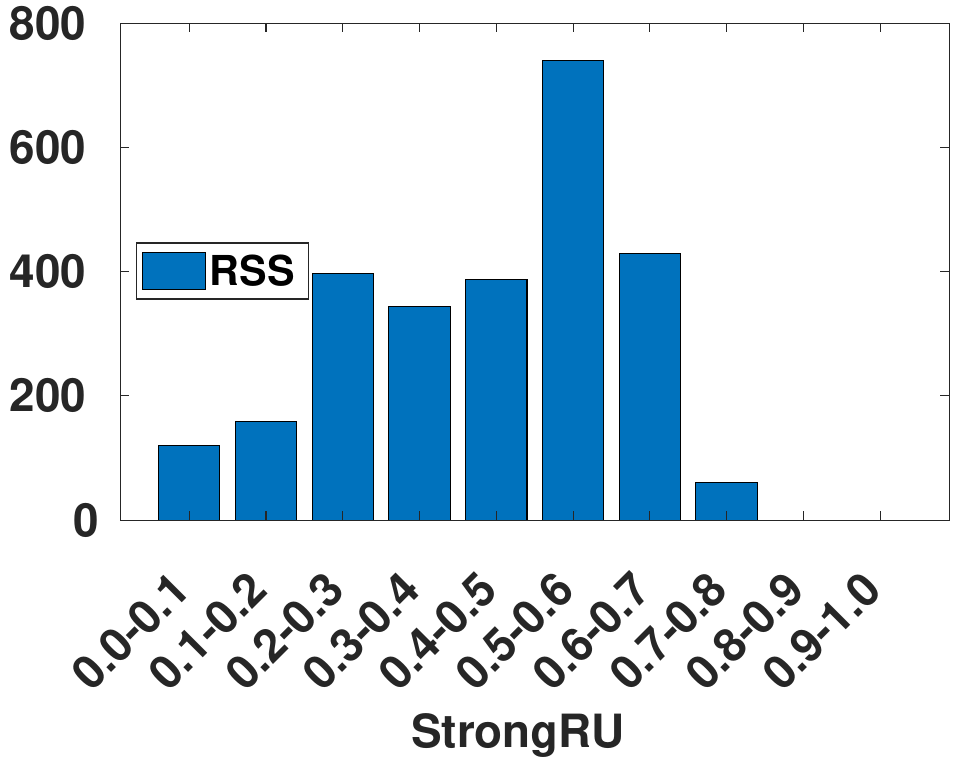}
        \vspace{-5mm}\caption{ECOD, effectiveness of \sru on regression in {\it RN}}
        \label{fig:sdt_rn_ecod}
    \end{subfigure}\hfill
    \begin{subfigure}[t]{0.24\linewidth}
        \centering
        \includegraphics[width=\textwidth]{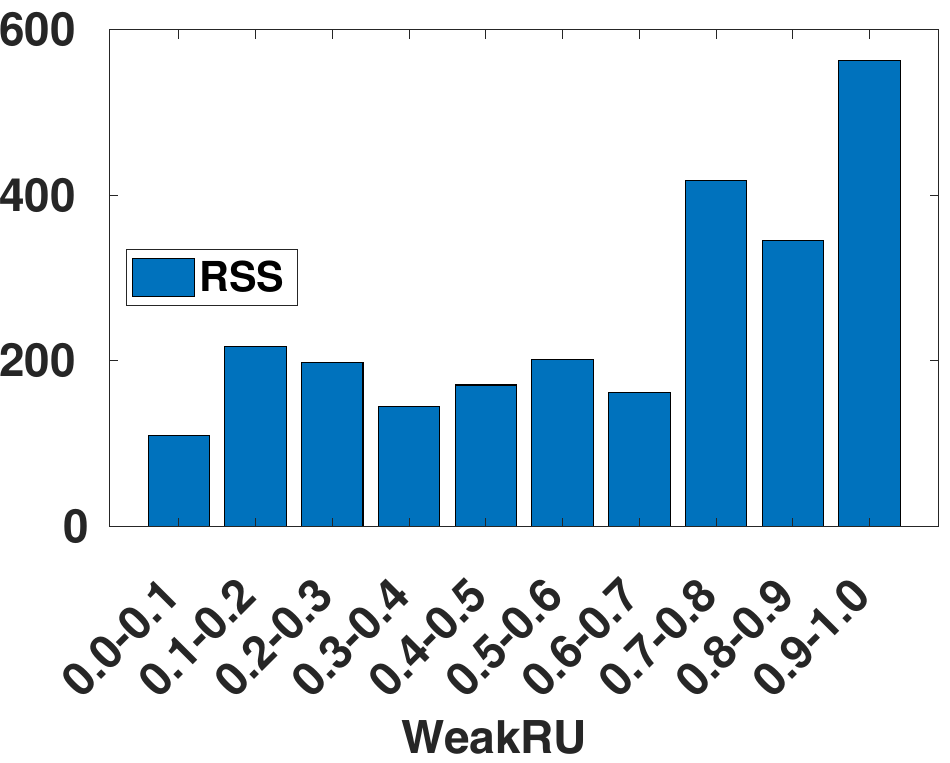}
        \vspace{-5mm}\caption{ECOD, effectiveness of \wru on regression in {\it RN}}
        \label{fig:wdt_rn_ecod}
    \end{subfigure}
\end{minipage}
\vspace{-1em}
\caption{{effect of varying the lack of representation component method of choice on the distrust values}}
\label{fig:varying_outlier_detection_metric}
\end{figure*}

\vspace{-2mm}
As previously noted in \S~\ref{sec:dev}, the \ru measures are agnostic to the choice of the outlier detection technique. While we used the distance of $k-$th nearest tuple to estimate $\pe_o$, in this experiment, we investigate the impact of using alternative outlier detection approaches in computing \ru values. More specifically, we integrate the following two outlier detection metrics in the lack of representation oracle:
\begin{itemize}
    \item Local Outlier Factor (LOF)~\cite{breunig2000lof}: Widely used proximity based outlier detection technique.
    \item Empirical Cumulative Distribution-based Outlier Detection (ECOD)~\cite{li2022ecod}: The SOTA probabilistic outlier detection technique recognized by the PyOD toolbox~\cite{zhao2019pyod}.
\end{itemize}
We conduct this experiment on one classification (using {\it AD} data set) and one regression task (using {\it RN} data set).
The results are shown in Figure \ref{fig:varying_outlier_detection_metric}. In short, by comparing the results corresponding to the same task and data set (e.g. effectiveness of \sru results for {\it AD} as shown in Figures \ref{fig:sdt_adult_lof}, \ref{fig:sdt_adult_ecod} and previously in \ref{fig:exp-adult-sdt}, \ref{fig:exp-adult-sdt_dl}), we observe almost identical results per \ru value ranges. This pattern is repeated for the effectiveness of \wru results for {\it AD} (Figures \ref{fig:wdt_adult_lof}, \ref{fig:wdt_adult_ecod} and previously in \ref{fig:exp-adult-wdt}, \ref{fig:exp-adult-wdt_dl}) and the effectiveness of \sru and \wru results for {\it RN} (Figures \ref{fig:sdt_rn_lof},\ref{fig:sdt_rn_ecod},\ref{fig:wdt_rn_lof},\ref{fig:wdt_rn_ecod} and previously \ref{fig:exp-reg-bar-sdt} and \ref{fig:exp-reg-bar-wdt}), demonstrating the flexibility of \ru measures to the choice of the outlier detection technique when computing $\pe_o$.

\begin{figure*}[!tbh] 
\begin{minipage}[t]{\linewidth}
    \begin{subfigure}[t]{0.23\linewidth}
        \centering
        \includegraphics[width=\textwidth]{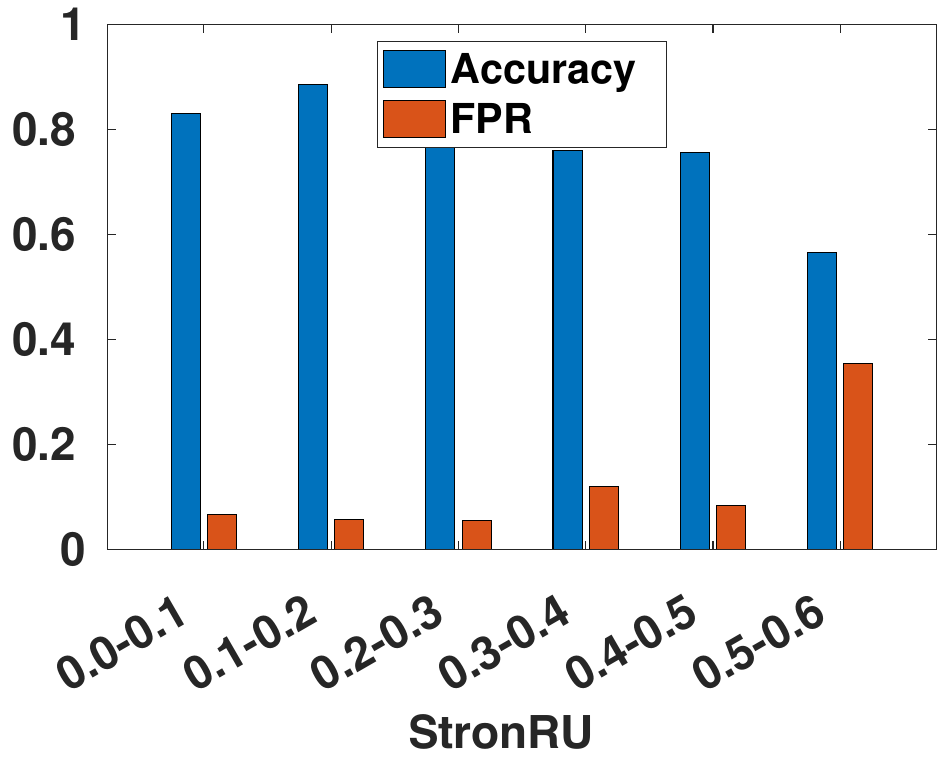}
        \vspace{-5mm}\caption{{\it AD}, effectiveness of \sru on classification}
        \label{fig:sdt_adult_uniform}
    \end{subfigure}\hfill
    \begin{subfigure}[t]{0.23\linewidth}
        \centering
        \includegraphics[width=\textwidth]{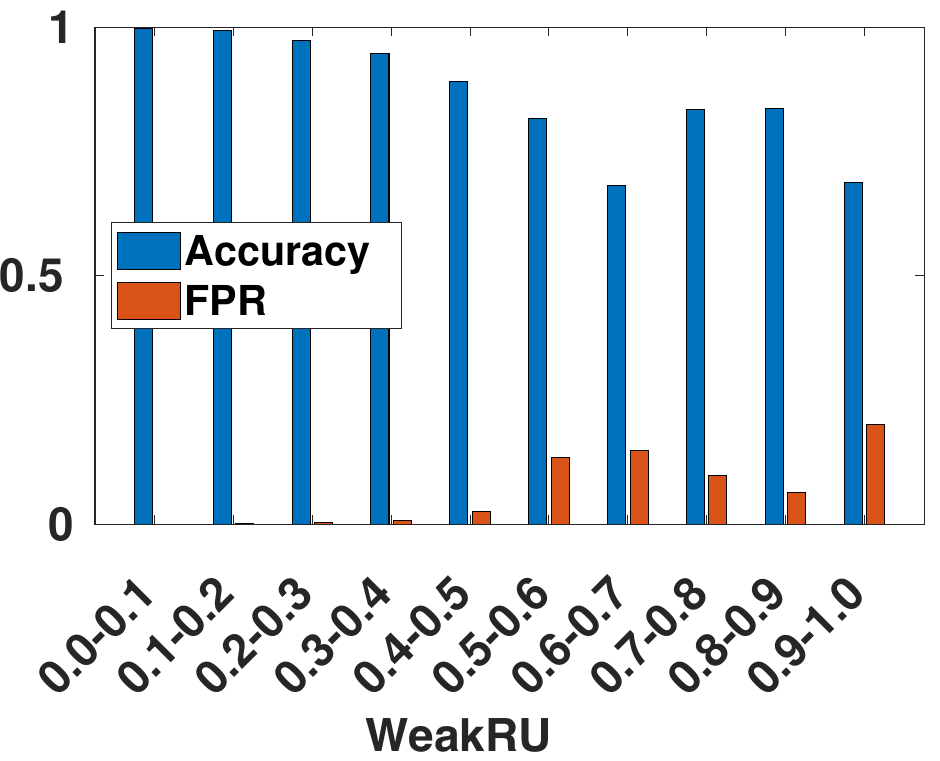}
        \vspace{-5mm}\caption{{\it AD}, effectiveness of \wru on classification}
        \label{fig:wdt_adult_uniform}
    \end{subfigure}\hfill
    \begin{subfigure}[t]{0.23\linewidth}
        \centering
        \includegraphics[width=\textwidth]{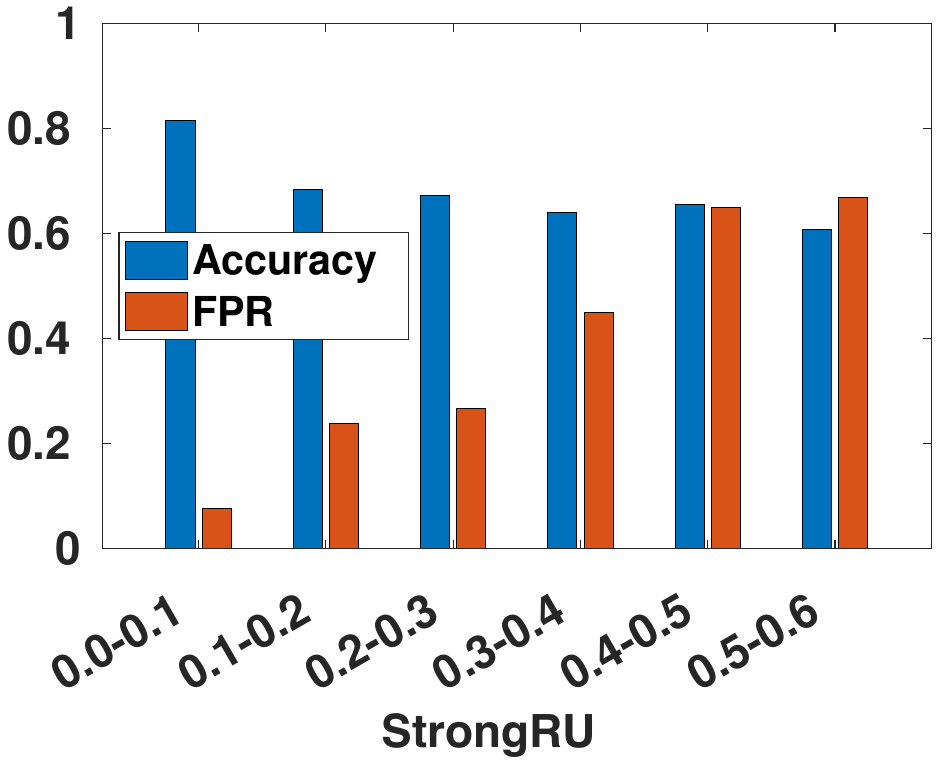}
        \vspace{-5mm}\caption{{{\it DCC}, effectiveness of \sru on classification}}
        \label{fig:sdt_dcc_uniform}
    \end{subfigure}\hfill
    \begin{subfigure}[t]{0.23\linewidth}
        \centering
        \includegraphics[width=\textwidth]{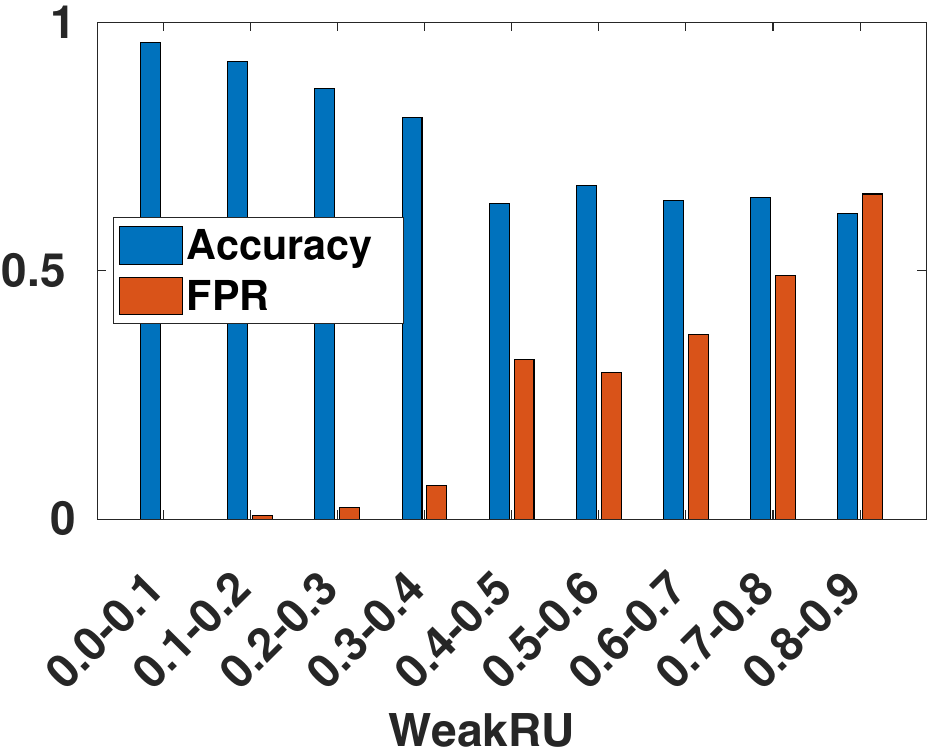}
        \vspace{-5mm}\caption{{{\it DCC}, effectiveness of \wru on classification}}
        \label{fig:wdt_dcc_uniform}
    \end{subfigure}
\end{minipage}
\begin{minipage}[t]{\linewidth}
    \begin{subfigure}[t]{0.23\linewidth}
        \centering
        \includegraphics[width=\textwidth]{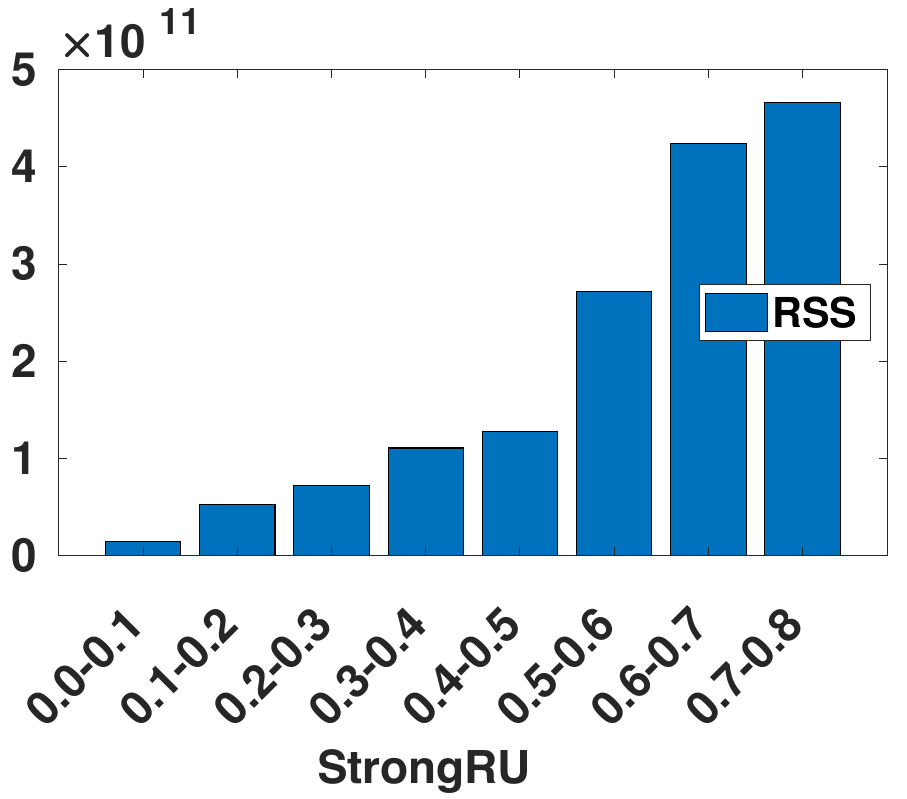}
        \vspace{-5mm}\caption{{{\it HS}, effectiveness of \sru on regression}}
        \label{fig:sdt_hs_uniform}
    \end{subfigure}\hfill
    \begin{subfigure}[t]{0.23\linewidth}
        \centering
        \includegraphics[width=\textwidth]{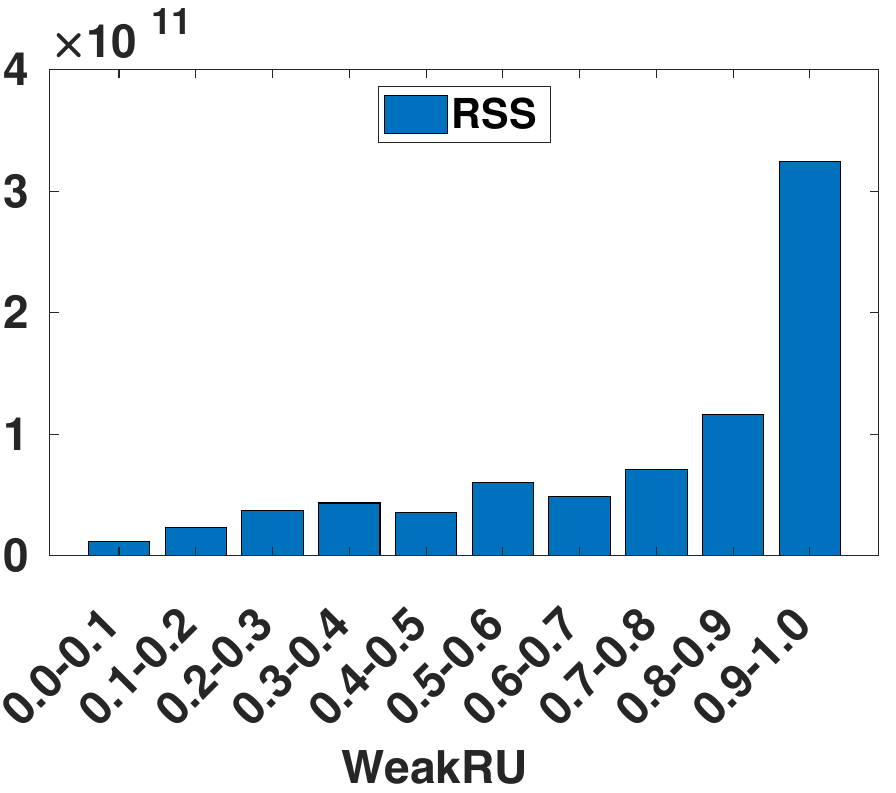}
        \vspace{-5mm}\caption{{{\it HS}, effectiveness of \wru on regression}}
        \label{fig:wdt_hs_uniform}
    \end{subfigure}\hfill
    \begin{subfigure}[t]{0.23\linewidth}
        \centering
        \includegraphics[width=\textwidth]{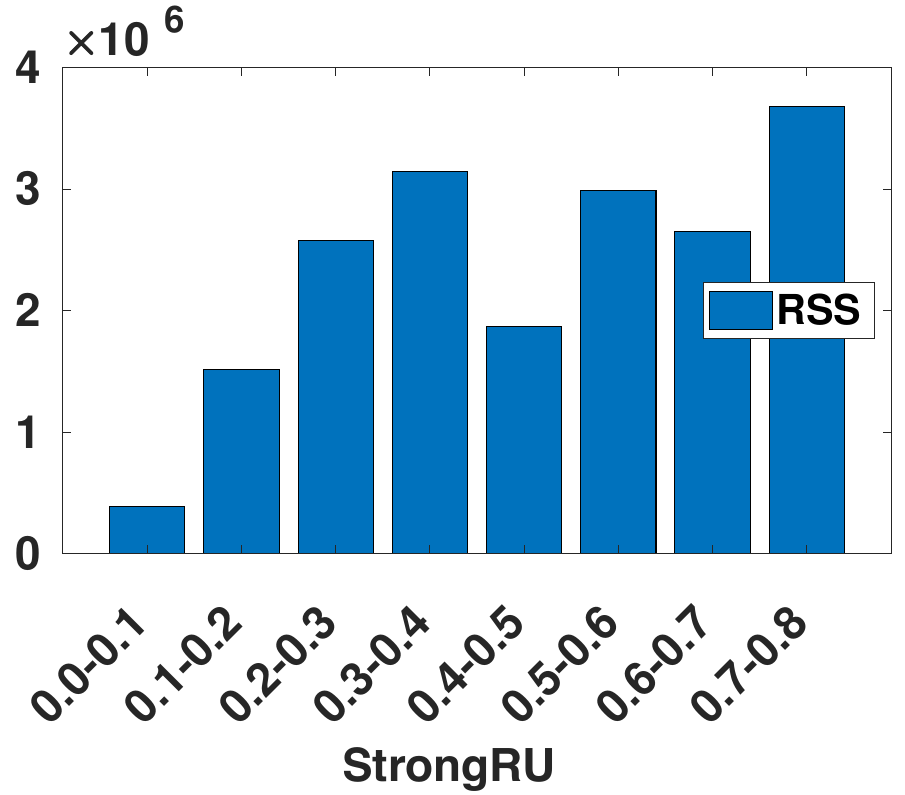}
        \vspace{-5mm}\caption{{{\it DI}, effectiveness of \sru on regression}}
        \label{fig:sdt_diamond_uniform}
    \end{subfigure}\hfill
    \begin{subfigure}[t]{0.23\linewidth}
        \centering
        \includegraphics[width=\textwidth]{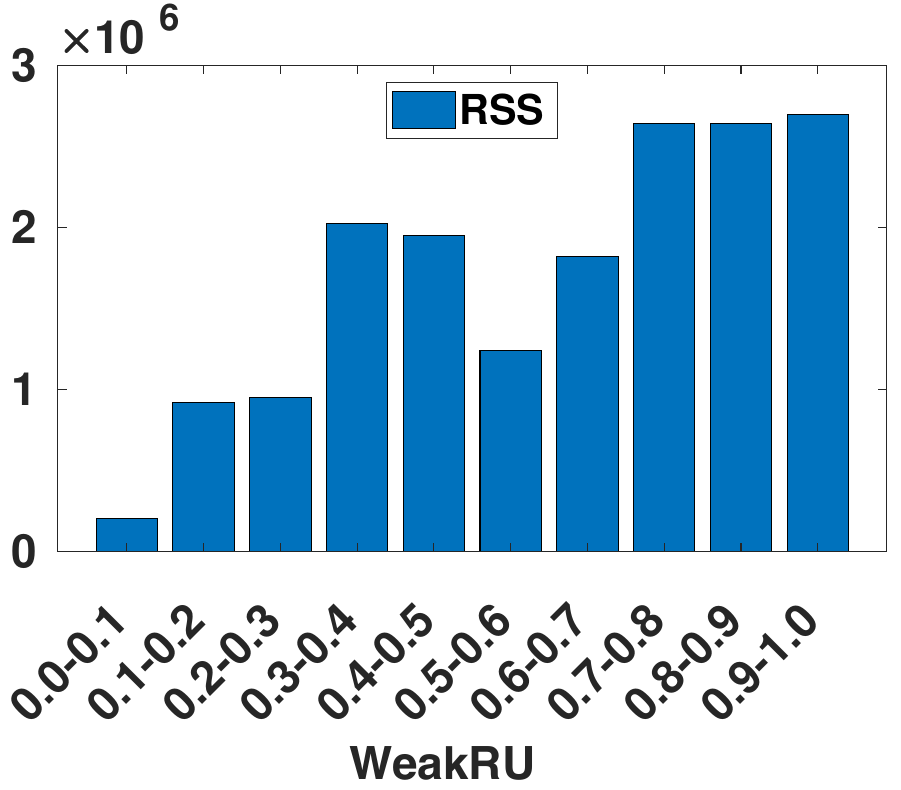}
        \vspace{-5mm}\caption{{{\it DI}, effectiveness of \wru on regression}}
        \label{fig:wdt_diamond_uniform}
    \end{subfigure}
\end{minipage}
\caption{{uniformly sampled training data and outliers removed from test set}}
\label{fig:uniformly_sampled_train_test}
\end{figure*}

\vspace{-2mm}
\subsubsection{Proof of Concept Experiments with Uniformity of Train/Test Samples}\label{exp:additional:train-test}
\vspace{-2mm}
In \S~\ref{sec:datasets}, to overcome the challenge of collecting enough samples to evaluate the effectiveness of \ru measures, we proposed sub-sampling from a large data set, removing the outliers and adding the outliers back to the test set to cover larger parts of the query space. 
However, as mentioned in \S~\ref{sec:datasets}, the downside of this approach is that it further reduces the presentation of points from under-represented regions in the training set, which may impact the model performance for those regions. 
Alternatively, 
in this experiment, we uniformly sample two sets from the underlying distribution. The first set serves as the training set and we did not remove the outliers from it.
The second set is used for creating the test set by finding its outliers and adding to the test set. 
We repeat this experiment for two classifications ({\it AD} and {\it DCC}) and two regression ({\it HS} and {\it DI}) data sets. The results are illustrated in Figure \ref{fig:uniformly_sampled_train_test}. In summary, the results are consistent with the previous observations, confirming the validity of our previous experiment and the minimal impact of removing the outliers from the training set to the results. 

\begin{figure*}[!tb]
    \scriptsize
    \centering
    \def\arraystretch{1.5}
    \begin{tabular}{||@{}c@{}|c|c|@{}c@{}|c|c|@{}c|@{}c@{}||}
        \hline
        \textbf{line of work}&\textbf{target}&\textbf{fidelity}&\textbf{output}&\textbf{task}&\textbf{components}&\textbf{model advocacy}&\textbf{data profiler} \\ [0.5ex]  \hline \hline 
        \makecell{\ru measures \\ (this work)}&data&local&probabilistic score&\makecell{classification\\ regression}&\makecell{lack of certainty \\ lack of representation}&challenge&yes\\ \hline
        \makecell{prediction probabilities \\ \cite{zadrozny2001obtaining,zadrozny2002transforming,platt1999probabilistic,niculescu2005predicting}}&model&global&probalistic score&classification&lack of certainty&challenge&no\\ \hline
        \makecell{prediction intervals \\ \cite{khosravi2010lower,pearce2018high,chatfield93predictionintervals}}&model&global&interval&regression&lack of certainty&challenge&no\\ \hline
        \makecell{conformal prediction \\ \cite{angelopoulos2021gentle,shafer2008tutorial}}&model&global&\makecell{interval \\set}&\makecell{classification \\ regression}&lack of certainty&challenge&no\\ \hline
        \makecell{uncertainty sampling \\ \cite{lewis1995sequential,sharma2013most}}&model&global&\makecell{continuous score}&\makecell{classification \\ regression}&lack of certainty&challenge&no\\ \hline
        \makecell{data coverage \\ \cite{asudeh2019assessing,asudeh2021identifying,lin2020identifying} }&data&local&binary signal&\makecell{classification \\ regression}&lack of representation&challenge&yes\\ \hline
        \makecell{local interpretation \\ \cite{lundberg2017unified,ribeiro2016should,datta2016algorithmic}}&model&local&\makecell{prediction probability \\ feature effect}&\makecell{classification \\ regression}&n/a&support&no\\ \hline
        \makecell{out-of-distribution\\ generalizability  \cite{carlini2019distribution} }&model&global&continuous score&\makecell{classification \\ regression}&lack of representation&challenge&yes\\ \hline
    \end{tabular}
    \vspace{-1mm}
    \caption{descriptive comparison of \ru measures and related work}
    \label{fig:related_works}
    \vspace{-5mm}
\end{figure*}

\vspace{-3mm}
\section{Related Work}\label{sec:related}
\submit{\vspace{-1mm}}
Responsible data science has become a timely topic, to which the data management community has extensively contributed~\cite{salimi2019interventional,salimi2020database,kuhlman2020rank,asudeh2019designing}.

In particular,~\cite {fariha2021} introduces a data profiling primitive {\it conformance constraint} to characterize whether inference over a tuple is untrustworthy. By the assumption that the {\it conformance constraints} always hold, they claim that they can use a tuple's deviation from the constraint as a proxy to trust a model's outcome for that tuple. 
Besides, extensive studies on different dimensions of trust in ML and AI have been presented in \cite{kentour2021analysis,liu2021trustworthy}.
It is also worth mentioning the body of work on the notion of trustworthiness of data sources that focuses on the correctness and legitimacy of data sources\cite{jayasinghe2017data,dong2015knowledge}, however despite the similar terminology, it is a different concept from our problem.

Related work also includes~\cite{blundell2015,pakdaman2015,zhang2020,abdar2021} that aim to estimate and quantify uncertainty in AI models, however, they have a different perspective on the issue as they extract the uncertainty from models, while our measures are data-centric.
Probabilistic classifiers predict a probability distribution over the set of classes for a given query point instead of simply returning the most likely class \cite{zadrozny2001obtaining,zadrozny2002transforming,platt1999probabilistic,niculescu2005predicting}. A given probability metric such as log loss or Brier score is calculated for each example to evaluate the predicted probabilities. Not all of the common classifiers are intrinsically probabilistic and some return distorted probabilities that need to be calibrated. Prediction probabilities are computed using the model trained for global performance and may not be accurate for the unrepresented regions.
Prediction Intervals (PIs) are a common practice for quantifying the uncertainty associated with a model's prediction of a query point in regression tasks \cite{chatfield93predictionintervals,pearce2018high,khosravi2010lower}. PIs consist of a lower and upper bound that contains a future observation with a specified level of confidence. 
Although PIs can be constructed in multiple ways, there is a negative correlation between the quality of the PI and the computational load associated with it \cite{khosravi2011comprehensive}.
Conformal Prediction (CP) is another standard way of quantifying uncertainty in both classification and regression problems returning confidence intervals and confidence sets respectively, guaranteeing a user-specified confidence level.
\techrep{ 
Benefiting from a heuristic notion of uncertainty in the model of choice, a scoring function $s(x,y)$ is defined that assigns uncertainty values to query point $x$ given target variable $y$ with larger values to the cases that $x$ and $y$ disagree more. Next, the $1-\alpha$ quantile ($\alpha$ being the user-specified confidence level) of the calibration set scores is calculated and used to form the prediction set for the new examples.
Uncertainty sampling~\cite{lewis1995sequential,sharma2013most} is also a related model-centric strategy used in active learning to select the most informative data points for annotation or labeling. The goal of active learning is to reduce the labeling cost by selecting the most valuable data instances to be labeled, rather than labeling all available data. Uncertainty sampling achieves this by selecting data points about which the model is uncertain or has low confidence in its predictions. The model typically measures uncertainty through metrics like entropy, margin, or least confidence using the probabilities assigned to different classes.}

It is important to note that all aforementioned model-centric approaches, including PI and CP, estimate intervals, probabilities, and scores using model(s) built
by maximizing the {\em expected performance} on {\em random} sample from the underlying distribution.
As a result, while they may provide accurate estimations for the dense regions of data (e.g. majority groups), their estimation accuracy is questionable for the poorly represented regions (e.g. minority groups).
In particular, \cite{angelopoulos2021gentle} recognizes the lack of guarantees in the performance of CP for such regions.
On the contrary, prediction outcomes are specifically unreliable for regions that are unlikely to be sampled. As a result, as we further discuss in \S~\ref{sec:measure:toy}, such approaches fail for cases that are not represented by the training data. This is consistent with our experimental evaluations.
\techrep{
PI and CP methods usually rely on techniques such as bootstrapping and constructing ensembles to elicit uncertainty, which regardless of the number of subsamples or ensembles created, fails to account for the regions that are not represented.
Contrarily, our proposed measures are computed {\em locally} around the query point (in form of lack of certainty and lack of representation) and therefore are equally accurate for different regions of data. 
Finally, while PI and CP return an interval or set for each query point, the results may
be too generic (e.g. including a large set) or lack a proper explanation for the user to make an informed decision. 
}

The notion of data {\it coverage} is a related topic that has been studied across different settings~\cite{jin2020mithracoverage,asudeh2019assessing,lin2020identifying,asudeh2021identifying,tae2021slice,accinelli2021impact,moskovitch2020countata,accinelli2020coverage}. For categorical data, uncovered regions are identified in form of value combinations (e.g. Hispanic Females) called patterns. A pattern is uncovered if there are not enough samples matching it~\cite{jin2020mithracoverage,asudeh2019assessing,lin2020identifying}.
{\it Coverage} on continuous space is studied in~\cite{asudeh2021identifying}. Accordingly, lack of {\it coverage} is identified as any point in the data space that doesn't have enough points in a fixed-radius neighborhood around it. 
\techrep{Although {coverage} does not provide a score for an arbitrary query point, following the idea of whether the point is covered or not, users can decide whether to trust the outcome of the model for that query point.}

Out-of-distribution generalizability is another related topic from the ML community that quantifies the degree to which a query point is an outlier in the underlying distribution.  
Specifically, \cite{carlini2019distribution} proposes five metrics for identifying well-represented examples. 
\techrep{These metrics are shown to be highly correlated, stable, and model-agnostic.}
The metrics rank examples based on different measures within ensembles, distance to the decision boundary, or prediction difference of two models for the same query point (holdout retraining). It is important to note that these techniques are model-agnostic in the sense that they have consistent results for different models and parameters, however, unlike our techniques that merely assess representation from the data, they still measure representation within model properties.

Another related topic is the body of work on local interpretation methods for explaining individual predictions \cite{molnar2020interpretable}. LIME provides local explanations for a model's prediction behavior on query points by substituting the original complex model with a locally interpretable surrogate model. 
\techrep{
Being a model-agnostic technique, to realize what parts of the input are involved in the prediction, LIME perturbs the query point by creating samples around its neighborhood and observes how the model performs for the perturbed samples. Next, the samples are weighted with regard to their proximity to the original query point, and an interpretable model is constructed on the new samples. The learned model should be locally a good approximation and be used to interpret the original model. 
}
We note that interpretation methods justify a model's reasoning for a particular behavior. Conversely, our measures raise warnings to cast doubt when the prediction outcome is not reliable for a specific case.

SHAP (Shapley Additive Explanations)~\cite{lundberg2017unified} is another model-agnostic framework for explaining individual predictions made by machine learning models. SHAP values are based on cooperative game theory concepts, specifically the Shapley value, which allocates a fair contribution to each feature in the prediction. SHAP assigns importance values to input features, indicating their contribution to a model's prediction. It also can provide both local explanations for individual predictions and global insights into overall model behavior.
Similar to SHAP, QII (Quantitative Input Influence)~\cite{datta2016algorithmic} also uses Shapley values to explain individual predictions, yet, instead of adopting the conditional approach used in SHAP, QII draws ideas from the causal inference and follows an interventional approach. The QII method addresses feature correlations by iteratively altering individual features and calculating the average impact of each change on the model's output, considering all features used in constructing the model.

Figure \ref{fig:related_works}, presents an extensive comparison between the related body of work and our proposed measures and demonstrates how our measures stand out in the skyline. The techniques are examined based on the following properties:
\submit{
\textit{target} specifies whether the technique is model-centric or data-centric. \textit{fidelity} shows if the technique evaluates trust locally or only provides global assurance (may fail for sparse regions in the data). \textit{output} specifies the outcome of the technique. \textit{task} specifies the learning problem. \textit{component} specifies the considered complications causing the trust problems. \textit{model advocacy} specifies whether the technique questions the outcome of the model or tries to justify it. \textit{data profiler} specifies whether or not the outcome of the technique is considered a property of the data.
}
\techrep{
\begin{itemize}[leftmargin=*]
    \item \textit{target} specifies whether the technique targets data or model. 
    \item \textit{fidelity} specifies whether the technique evaluates trust locally or only provides global assurance (may fail for sparse regions in the data).
    \item \textit{output} specifies the outcome of the technique.
    \item \textit{task} specifies the learning problem.
    \item \textit{component} specifies the considered complications causing the trust problems.
    \item \textit{model advocacy} specifies whether the technique questions the outcome of the model or tries to justify it.
    \item \textit{data profiler} specifies whether or not the outcome of the technique is considered as a property of the data.
\end{itemize}}
To the best of our knowledge, our paper is the first to provide data-centric \ru measures to identify the scope of use of data sets for individual predictions.
The techniques proposed in this paper rely on the extensive research and advanced algorithms for outlier detection ~\cite{ramaswamy2000,hautamaki2004,9006151,breunig2000lof} and uncertainty computing ~\cite{shannon1948mathematical,breiman2017classification,brier1950verification,gebel2009multivariate}.
\vspace{-5mm}
\section{Conclusion}\label{sec:conc}
Towards addressing the need for trustworthy AI, in this paper, we proposed \ru measures as the warning signals that limit datasets' scope of use for predicting future query points. These measures are valuable alongside other techniques for trustworthy AI.
We proposed novel ideas for the effective implementation of \ru measures and designed efficient algorithms that scale to very large, high-dimensional data.
Our comprehensive experiments on real-world and synthetic data sets validated our proposal and verified the scalability of our algorithms with sub-second run times. 

\bibliographystyle{spmpsci}
\bibliography{ref}

\end{document}